\documentclass[10pt]{article}

\usepackage{fullpage}
\usepackage{setspace}
\usepackage{parskip}
\usepackage{titlesec}
\usepackage[section]{placeins}
\usepackage{xcolor}
\usepackage{breakcites}
\usepackage{lineno}
\usepackage{hyphenat}

\PassOptionsToPackage{hyphens}{url}
\usepackage[colorlinks = true,
            linkcolor = blue,
            urlcolor  = blue,
            citecolor = blue,
            anchorcolor = blue]{hyperref}
\usepackage{etoolbox}

\renewenvironment{abstract}
  {{\bfseries\noindent{\abstractname}\par\nobreak}\footnotesize}
  {\bigskip}

\titlespacing{\section}{0pt}{*3}{*1}
\titlespacing{\subsection}{0pt}{*2}{*0.5}
\titlespacing{\subsubsection}{0pt}{*1.5}{0pt}

\usepackage{authblk}

\usepackage{graphicx}
\usepackage[space]{grffile}
\usepackage{latexsym}
\usepackage{textcomp}
\usepackage{longtable}
\usepackage{tabulary}
\usepackage{booktabs,array,multirow}
\usepackage{amsfonts,amsmath,amssymb}
\providecommand\citet{\cite}
\providecommand\citep{\cite}

\newif\iflatexml\latexmlfalse

\AtBeginDocument{\DeclareGraphicsExtensions{.pdf,.PDF,.eps,.EPS,.png,.PNG,.tif,.TIF,.jpg,.JPG,.jpeg,.JPEG}}

\usepackage[utf8]{inputenc}
\usepackage[english]{babel}

\usepackage{float}

\usepackage{bm}   
\usepackage{csquotes}

\RequirePackage[style=numeric-comp,sorting=none]{biblatex}
\usepackage{orcidlink}

\iflatexml

\else
\fi

\begin{document}

\title{
Topical Review: Extracting Molecular Frame Photoionization Dynamics from Experimental Data
}

\author[1]{Paul Hockett\thanks{paul@femtolab.ca} \orcidlink{0000-0001-9561-8433}}%
\author[2]{Varun Makhija\thanks{vmakhija@umw.edu} \orcidlink{0000-0002-4975-4888}}
\affil[1]{National Research Council of Canada, 100 Sussex Drive, Ottawa, ON, K1A 0R6, Canada}%
\affil[2]{Department of Chemistry and Physics, University of Mary Washington, 1301 College Avenue, Fredericksburg VA, 22401.}%

\vspace{-1em}

\date{\today}

\begingroup
\let\center\flushleft
\let\endcenter\endflushleft
\maketitle
\endgroup

\begin{abstract}Methods for experimental reconstruction of molecular frame (MF) photoionization dynamics, and related properties - specifically MF photoelectron angular distributions (PADs) and continuum density matrices - are outlined and discussed. General concepts are introduced for the non-expert reader, and experimental and theoretical techniques are further outlined in some depth. Particular focus is placed on a detailed example of numerical reconstruction techniques for matrix-element retrieval from time-domain experimental measurements making use of rotational-wavepackets (i.e. aligned frame measurements) - the ``bootstrapping to the MF" methodology - and a matrix-inversion technique for direct MF-PAD recovery. Ongoing resources for interested researchers are also introduced, including sample data, reconstruction codes (the \textit{Photoelectron Metrology Toolkit}, written in python, and associated \textit{Quantum Metrology with Photoelectrons} platform/ecosystem), and literature via online repositories; it is hoped these resources will be of ongoing use to the community.\end{abstract}
\tableofcontents

\section{Overview}

\subsection{Topical introduction}
The main aim of this topical review is to discuss the determination of molecular frame (MF) photoionization dynamics (and related properties) from laboratory frame (LF) measurements. This particular problem is a subset of the larger topic of determining quantum mechanical properties of molecules (and general quantum mechanical systems) which has been, of course, long at the heart of molecular physics, physical chemistry and related disciplines. Spectroscopy, in particular, has the underlying goal of the determination of atomic and molecular properties with high precision; the ``inverse” problem of transforming \textit{ab initio} (computational) results, which naturally start in the MF, to the LF (in order to compare with experimental measurements and phenomena), also has a long and storied history. In both cases the issue is, in very general terms, one of complexity and averaging over unobserved quantities/properties/degrees-of-freedom of the system; furthermore, in many cases, certain properties may be poorly understood, may fundamentally obscure other properties of interest, and/or may not be readily computed. These issues are particularly relevant for the specific case of photoionization dynamics, which is an inherently complicated scattering problem, and may be strongly coupled to other molecular properties, i.e. electronic, vibrational and rotational dynamics. 

While many of these issues are general in quantum state reconstruction problems~\cite{Fano1957}, photoionization dynamics is the focus of this review, and the determination of photoionization dynamics and correlated observables in the molecular frame the main topic of discussion. Historically this determination has been termed a ``complete" photoionization experiment, and has also recently been reframed in terms of quantum tomography and metrology, which has essentially the same aims of complete system reconstruction - in the photoionzation case the photoionization matrix elements fully describe the electron-molecule scattering event initiated by photo-absorption, hence the continuum state populated by photoionization or, equivalently, the continuum density matrix.

Herein, the problem of complexity is approached generally in terms of the dimensionality of the problem, and the information content of the measurement; examples are built-up and discussed in these terms. However, note that the aim here is not for a comprehensive review of the literature, but rather a topical introduction (Sect. \ref{sec:Framing}) and more detailed background (Sect. \ref{sec:Concepts}), including recent progress in this area, followed by a (reader-extensible) tutorial overview of concepts, grounded in numerical examples (Sect. \ref{sec:Recon}). Overall, the aims herein are to introduce new researchers to this interesting topic, present realistic case-studies, and (attempt to) build bridges between some historically disparate sub-topics/areas of the field, as well as emerging methods in related fields.


In order to try and fulfill these aims, and to make a useful contribution to the community, this review aims to provide a number of supplementary resources for researches, and engender discussion on this topic. The numerical work follows open-science principles, with corresponding data and open-source code available online; the underlying \textit{Quantum Metrology with Photoelectrons} platform is introduced in Sect. \ref{sec:numerics-intro}, and a full list of relevant online resources is given in Sect. \ref{sec:resources}.
In particular the analysis routines demonstrated in Sect. \ref{sec:bootstrapping} - along with relevant data - are available online, as a set of Jupyter computational notebooks backed by open-source python libraries (Sect. \ref{sec:numerics-intro}).
It is hoped that, in this way, this manuscript will become a living document, and a useful resource for interested researchers that will grow over time. It is also hoped that this manuscript, and especially the online discussions, will serve  to garner opinions from a cross-section of interested researchers, and help to bridge the gaps between the various, historically somewhat disparate, communities (e.g. spectroscopy, general AMO physics, quantum information etc.) interested in quantum state reconstruction in various cases. With this in mind, references are only given sparingly in the main text, and mainly provided as lists in Sect. \ref{sec:Appendix-A}; the full bibliography is also available online (via \href{https://www.zotero.org/groups/4733878/molecular_frame_pads_measurements_and_reconstruction}{a Zotero group} \cite{hockettZoteroGroupsMolecular}) - this is very far from a comprehensive list: it is again hoped that this will be used, and grown, by interested researchers.

\subsection{Outline}

This topical review is structured as follows:

\begin{itemize}
\item Sect. \ref{sec:Framing}: general introduction and discussion of the topic in broad terms, suitable for a general reader.
\item Sect. \ref{sec:Concepts}: discussion of experimental techniques (Sect. \ref{sec:experimentalTechniques}), and a more detailed introduction to theory, specifically a tensor formulation  (Sect. \ref{sec:theoretical-techniques}) and associated numerical methods for MF retrieval (Sect. \ref{sec:recon-techniques-intro}).
\item Sect. \ref{sec:Recon}: worked examples for MF reconstruction for two retrieval protocols. This forms the main substance of the review, in particular Sect. \ref{sec:bootstrapping} provides a detailed case-study; note that figures are interactive in HTML versions of the manuscript and also available in the data repository for the manuscript \cite{hockett2022MFreconFigshare}.
\item Sect. \ref{sec:summary-outlook}: Summary \& outlook.
\item Sect. \ref{sec:resources}: additional resources, including codes, data and notebooks pertaining to Sect. \ref{sec:Recon}.
\item Sect. \ref{sec:Appendix-A}: further reading.
\item Sect. \ref{sec:Appendix-B}: extended theoretical details.
\end{itemize}

\section{Framing the problem\label{sec:Framing}}

In this section some general introduction and discussion is provided, in rather broad terms, suitable for a general reader. For a more detailed discussion of photoionization problems see Sect. \ref{sec:Concepts}.

\subsection{Molecular properties}
A very general problem in molecular physics is the determination of intrinsic molecular properties from experimental measurements, and comparing such measurements with theoretical predictions. The key difficulty is, usually, the averaging or integration over unobserved degrees-of-freedom (DOFs) in the measurements. The exact nature of the DOFs will, naturally, be system and problem dependent, as will the coupling strength of the unobserved DOFs to the system properties of interest. From a spectroscopic perspective, one usually considers the DOFs in terms of a Born-Oppenheimer (BO) separation of the full molecular wavefunction, i.e. in terms of electronic, vibrational and rotational DOFs. These BO DOFs are useful, since they provide an intuitive separation of states, which can be regarded as uncoupled to a first approximation. These states may then approximately correlate with the choice of spectroscopic methodology applied (in terms of characteristic energy or time regimes), and often provide a good approximation to the fully-coupled system. One can also consider this issue in terms of a general quantum mechanical language of eigenstates, wavefunctions, wavepackets (more appropriate for dynamical systems), density matrices and so forth. Framing the problem in this language perhaps highlights the generality of the molecular case as a many-body quantum mechanical matter system, to which certain formal DOF separations are often applicable - but, regardless of the language, the problem remains that many-body systems are analytically unsolvable and can be computationally intractable. 
(For a tutorial-style introduction to some of these issues in the time-resolved case, see Ref. \cite{wu2011TimeresolvedPhotoelectronSpectroscopy} and references therein.) 

In favourable cases, careful experimental design can obviate the issue of DOF averaging in specific cases via, for instance, state-selection of the system prior to measurement to reduce the complexity of the problem at hand. This is typical in frequency-resolved spectroscopy, see, e.g., introductory spectroscopy textbooks \cite{bunkerMolSymm, herzberg1945molecular, hollasHighRes}. Similarly, preparing a specific ``zeroth-order" wavepacket, viewed as a superposition of BO states, is the typical target case in time-resolved spectroscopy experiments, see, e.g., Refs. \cite{Tannor2007,Stolow2008,wu2011TimeresolvedPhotoelectronSpectroscopy}. However, in many (perhaps most) cases this is not feasible due to the inherent complexity of the system, and/or coupling (non-separability) of states, and/or experimental issues or limitations. The problem becomes significantly worse for larger systems as the number of DOFs (hence density of states) increases, and/or if the DOFs are continuous - rather than quantised - properties, and/or if many states are populated. In general, then, this is a problem which must ideally be addressed at a high level, by a combination of experiment and theory. Happily, both are increasingly possible, and becoming more routine, as technology improves; of particular relevance to the photoionization case at hand is the advent of photoelectron imaging \cite{Whitaker2003}, advances in short laser pulses and control, and the ongoing march of available computational power and software.

In terms of photoionization studies, the aims can be viewed both in terms of control and in terms of measurement and reconstruction. For instance, a basic photoelectron imaging experiment may seek to measure photoelectron energy and lab-frame (LF) angular distributions from a given system. A more sophisticated experiment may seek to control these observables in some way, e.g. via state-preparation or ionizing pulse polarization, or may seek to use these measurements as a sensitive probe of some other DOF of interest, e.g. vibrational motion. A yet more sophisticated methodology may aim to directly obtain molecular frame information, or aim to obtain a ``complete" quantum mechanical description of the photoionization process from a set of measurements. This may be an end in itself, or serve as a more sensitive probe of DOFs of interest. (For further discussion of experimental techniques, see Sect. \ref{sec:experimentalTechniques}; for further general discussion along these lines, see for example Refs. \cite{hockett2018QMP1, kleinpoppen2013perfect, Reid2012, Stolow2008}.)
\subsection{Molecular frame observables\label{sec:MF-intro}}

A very general issue, of fundamental interest, which falls in the latter categories - and lacks a general solution - is the measurement of observables which depend on the orientation of the molecular frame (MF). In general, this can be regarded as a geometric problem in the spatial (rotational) degrees of freedom, in which the intrinsic MF properties are averaged (or smeared out) in a given measurement. For instance, in a typical gas phase molecular spectroscopy measurement, the observables in the molecular frame  are averaged over all possible molecular orientations in a laboratory frame (LF) measurement. The inverse problem is also difficult, i.e. simulating specific experimental observables starting from \textit{ab initio} calculations in the molecular frame. In this case, to simulate LF results, knowledge of the degree and type of geometric averaging is generally required. In more complex cases \textit{ab initio} calculations may need to be carried out for various molecular orientations, or coupled DOFs such as vibrational motions. Of particular interest in the current case will be the aligned frame (AF), which signifies a measurement frame with some degree of spatial preparation (alignment or orientation of the molecular axis ensemble) of the sample via, for instance, photon absorption (see Sect. \ref{sec:MF-control} for further discussion of the AF).

Despite, or perhaps because of, the inherent difficulties, obtaining  MF measurements remains a topic of great interest, and much progress in experimental methodologies has been made in recent decades \cite{Becker1998,Reid2003,Reid2012,kleinpoppen2013perfect,Yagishita2015,hockett2018QMP1}. 
These methodologies can be classified, approximately, as sitting somewhere on the spectrum between (1) relatively direct and (2) indirect methodologies. In the limit of (1), the methodologies are designed to enable control and reconstruction of the MF, hence measure MF observables (somewhat) directly, with minimal data processing requirements. Another avenue to MF observables is the limiting case of indirect methodologies (2), which involve more simulation and have less stringent (or at least different) experimental requirements. In the current context, this can be defined as techniques which make use of detailed theory and analysis procedures to reconstruct MF properties from LF measurements. Broadly speaking, one can also consider indirect methods as a post-processing or analysis-based approach, with a significant theoretical and computational requirement (akin to computational imaging techniques, as well as many traditional energy-domain spectroscopy techniques). In contrast, direct methods are, conceptually, closer to purely experimental methods, with more emphasis placed on detection techniques and capabilities, although significant numerical data analysis may still be required. 

There is, naturally, significant overlap between these extreme case definitions, and most methodologies and extant demonstrations fall somewhere on the spectrum between direct (``purely" experimental) and indirect. In particular, indirect approaches will typically still require some degree of (potentially sophisticated) experimental control, and a set of associated measurements, to be useful. For example, a set of LF measurements with different polarization geometries or with an optically-prepared aligned ensemble may be required. Direct measurements usually require specific molecular behaviour(s) and/or preparation in order to measure suitable observables, due to experimental restrictions. A typical requirement in multi-particle coincidence imaging techniques may be, for example, molecular fragmentation following photoionization, since it is measurement of the fragments that allows for MF reconstruction in these methodologies. Historically different methods have typically been pursued by experimental communities making use of either laser-based (LF, AF and control type schemes) or synchrotron-based (direct MF measurements) experimental methods. Whilst the underlying photoionization and molecular physics is shared, the different classes of experiment typically suit soft or hard photon energies, for valence photoionization and multi-photon techniques, or core ionization and dissociative photoionization based techniques, respectively, although the split is not rigorous. However, there is increasingly overlap between the communities, particularly in the last decade or two. In particular, the advent of time-resolved hard photon sources and strong-field and emergent attosecond laser techniques has pushed developments on the ``tabletop" side (for a broad review of techniques in attoscience, see Ref. \cite{Krausz2009}). Meanwhile, the advent of free-electron lasers (FELs), and the increasing availability of modern ultrafast lasers at synchrotron and FEL facilities, has enabled multi-source experiments in the time or frequency domain spanning optical and X-ray wavelengths and methods. (For recent perspectives covering many of these topics, including recent ultrafast and X-ray developments, see, for instance, Refs. \cite{Young2018,ueda2019RoadmapPhotonicElectronic}.)

It is also noteworthy that in the context of ultracold physics a number of sophisticated techniques have been developed to coherently populate individual molecular eigenstates \cite{mitra2022QuantumControlMolecules}. While these techniques have yet to be applied to molecular photoionization, recent examples exist of similar techniques applied to atomic photoionization \cite{desilva2021CircularDichroismAtomic}.


\begin{figure}[]
\begin{center}
\includegraphics[width=\textwidth,height=\dimexpr\textheight-4\baselineskip-\abovecaptionskip-\belowcaptionskip\relax,keepaspectratio]{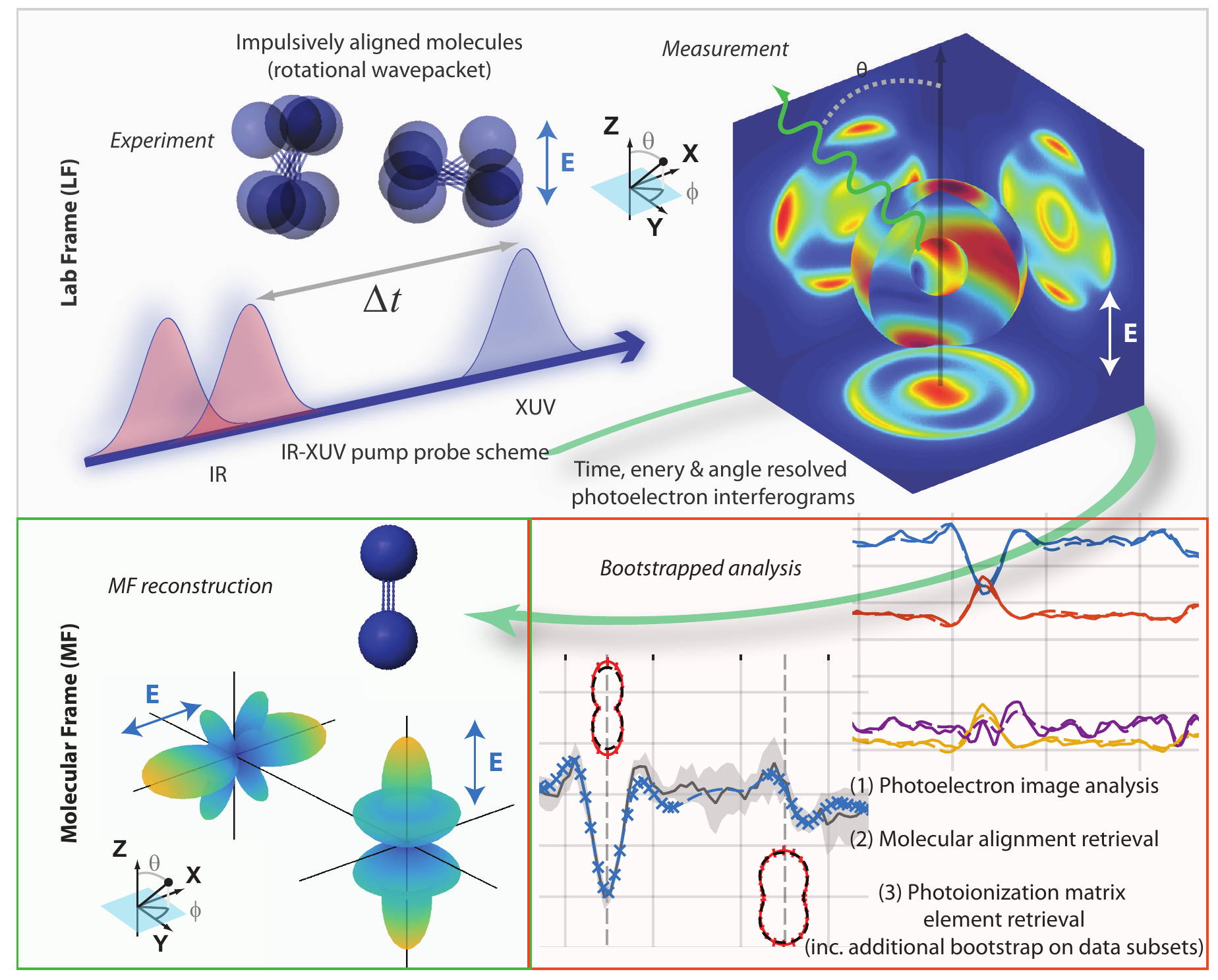}
\caption{Conceptual outline for retrieval of MF properties for ``bootstrapping to the MF", via a set of time-resolved measurements and suitable post-processing scheme. In the LF, a set of laser pulses creates and probes an aligned distribution of molecules, and photoelectron images are measured. The experimental data is analyzed through a multi-step ``bootstrap" protocol to obtain matrix elements, which constitute a complete description of the photoionization event. These can further be used to obtain MF observables, for any polarization geometry. Note the coordinates show will be used throughout the manuscript, with the LF z-axis defined relative to the laser field (\textbf{E}) polarization, and the MF defined with the molecular axis aligned along the z-axis (and polarization geometries for the laser field (\textbf{E}) referenced to this axis).\label{781808}}
\end{center}
\end{figure}


\subsection{Photoionization observables, properties and dimensionality\label{sec:Photo-into}}

Fundamentally, experimental observables depend on intrinsic molecular properties. In the  context of molecular physics, such properties are naturally expressed in the molecular frame - for instance: bond lengths, bond angles, polarisabilities, absorption cross-sections, dipole matrix elements and so on. Whilst it is often the case that such properties may be viewed as stationary, and possibly approximated semi-classical, ultimately the dynamics of the system are also of great interest. For instance, in a stationary state the equilibrium or average bond-lengths of a system are often viewed as a well-defined molecular property, but really it is the associated vibrational wavefunction that defines this. However, in a state-resolved experiment, the averaging over the vibrational DOF may be small (i.e. the wavefunction is localised), so this approximation holds. More generally this may not be the case, and will not be the case in a time-resolved experimental methodology wherein a superposition of vibrational states  - a wavepacket - may be prepared. In this sense, the observables in a given case may also be affected by experimental conditions and methods, and degrees-of-freedom or dynamics, all of which may affect what are viewed as fundamental molecular properties. 

As well as MF observables, one may also be interested in determining the underlying intrinsic molecular properties which govern the observable, and/or exploring extrinsic properties which modify or control the observable. For instance, the response to application of different laser field(s) (wavelength, intensity, polarization...), preparation of specific molecular states, or wavepacket dynamics in time-resolved studies. In some cases the mapping between observable(s) and property/ies of interest may be relatively direct, in other cases much less so!

To mitigate these issues, one generally seeks a ``maximum information" measurement to try and understand, untangle or even quantitatively determine the various effects at play. Clearly, the amount of information required will depend on the type of analysis, and complexity of the problem. 
This concept is expanded below in terms of a fundamental 1D example (Sect. \ref{sec:1D-case}), and the discussion is extended to the more complex (higher dimensionality) case of the measurement of photoelectron angular distributions in the remainder of this section (Sect. \ref{sec:flux-intro}). An important caveat to this is that, for analysis or retrieval of phases, the measurements must have some degree of phase-sensitivity and coherence - this may be inherent to the observables or DOFs chosen, or imparted experimentally via  interferometric (coherent multi-path) schemes.

\subsubsection{A brief discussion of rotation: 1D problems and beyond\label{sec:1D-case}}

A basic, traditional, example - at least for simple molecules - is the determination of bond lengths from LF measurements of rotational energy spectra. In this case, the intrinsic molecular properties (equilibrium bond lengths and geometry) map relatively cleanly to the observable energy levels; one can thus uniquely determine the properties of interest via a suitable data analysis methodology (for a general introduction, see Ref. \cite{hollasHighRes}). For instance, in the simplest case of a rigid homonuclear diatomic system, the problem can be regarded as 1-dimensional, with a single bond length to be determined from a 1D experimental measurement made in the LF. A measurement of the rotational energy spectrum is sufficient to provide bond-length information. In this case, there is no issue with orientational averaging, since the position of features in the energy spectrum are invariant to orientation, although their magnitudes are not. 
(In a more precise description, the observed spectrum is determined by the (rotational) transition moments, which are tied to the molecular symmetry axes. The magnitudes are the projection into the LF of these transition moments, but their energy eigenstates are invariant to orientation; an angle-integrated measurement will thus provide a sufficient information content.)
In this case, the intrinsic molecular properties can therefore be mapped to the LF observable relatively clearly, and the retrieval of these properties from a measurement is similarly direct. However, for more complicated systems, additional information may be required, either in terms of a richer ND (N-dimensional) measurement technique and/or a series of simple measurements with different ``control" parameters, which are chosen to affect the observable but not the intrinsic molecular properties.


More generally, one can regard a hierarchy of difficulty in the determination of molecular properties, based on the complexity or dimensionality of the system and observable(s).
\begin{itemize}
\item Simple systems, ``1D” methods, where a dataset of a single variable (e.g. rotational energy spectrum) is measured. For simple cases, e.g. rigid diatomic molecules, the determination of molecular properties - the rotational energy levels and, hence, the bond length - is relatively direct. (See, for example, Chpt. 5 in Ref. \cite{hollasHighRes}.) 
\item Intermediate cases, e.g. larger, but fairly rigid, molecules (for instance, formaldehyde ($H_2CO$) and aniline ($C_6NH_7$), as discussed in Ref. \cite{hollasHighRes}), possibly with congested spectra. In such cases the ``direct” determination of molecular properties may not be possible without more sophisticated modelling and \textit{ab initio} computations. Additional information from higher-dimensionality observables, e.g. polarization-dependent energy spectra, may also be required. An interesting example, and relevant to latter discussion herein, is the use of rotational coherence spectroscopy (RCS) in such cases \cite{Felker1987} - this can be considered as the narrow-wavepacket limit of more general (non-adiabatic) rotational wavepacket based methods (Sect. \ref{sec:MF-control}).
\item For the most complex cases, e.g. large, floppy molecules or complexes \cite{bunkerMolSymm,schmiedt2015SymmetryExtremelyFloppy}, determination of molecular properties in an absolute sense may be impossible or, rather, may be a poorly posed question since DOFs may become strongly coupled, and simplifying approximations such as fixed point-group symmetry and the BO separation break down \cite{bunkerMolSymm}. High level computational results (if possible/tractable) may be required to understand the spectrum, and associated dynamics, along with multi-dimensional observables (e.g. temperature dependent spectra, dependence on isotopically-substituted species and so on). Such cases remain at the cutting-edge of research in molecular spectroscopy \cite{schmiedt2015SymmetryExtremelyFloppy}.
\end{itemize}    

Whilst this discussion may seem a little arbitrary initially, aside from the general concepts, there is a strong link between traditional rotational spectroscopy and modern molecular alignment techniques (as will be discussed later, Sect. \ref{sec:MF-control}). Furthermore, control of the averaging over rotational (geometric) DOFs is a key to stepping into the MF in time-domain experiments.

\subsubsection{Photoelectron flux in the LF and MF\label{sec:flux-intro}}

A (generally) more complicated example is the topic of this article, i.e. the determination of MF photoelectron distributions, and the intrinsic molecular photoionization dynamics which underlie the observables.  The former may involve direct or indirect techniques; the latter is usually defined as the retrieval of the (complex valued) ionization dipole matrix elements, hence is indirect by definition (since phase information may not be directly measured). 

A broad conceptual overview for a time-dependent measurement scheme and associated observables is illustrated in Fig. \ref{781808}; although the figure shows a specific experimental scheme, the concepts are general. In this example - which forms the basis for the case study of Sect. \ref{sec:bootstrapping} - a multi-pulse experimental scheme is used to prepare an aligned molecular axis distribution, and AF photoionization measurements are made.

The preparation (or pump) step in this case is impulsive molecular alignment, in which one or more laser pulses, of short duration as compared to
the characteristic rotational timescale, are used to create a rotational wavepacket in the system. The evolution of this wavepacket corresponds to different ensemble alignments (LF projections), as a function of time. A probe-pulse of similar pulse-duration, and with controllable time-delay $\Delta t$, can then probe specific alignments as a function of time. In general, pulses in the femto-second regime are suitable for this type of experimental scheme, and the technique is quite widely applicable; further discussion can be found in Sect. \ref{sec:MF-control}.

The type of measurements shown are photoelectron images, which provide the angle and energy resolved photoelectron flux. Since the molecular axis distribution is time-dependent in this case,
a set of time-dependent measurements provide a set of observables with different spatial averaging at each measurement time $t$. These observables can be parameterised as a set of time-dependent parameters as shown in the lower right panel. The observables of interest - the photoelectron flux as a function of energy ($\epsilon$), ejection angles ($\theta,\phi$), and time ($t$) - can be written generally as an expansion in spherical harmonics:

\begin{equation}
\bar{I}(\epsilon,t,\theta,\phi)=\sum_{L=0}^{2n}\sum_{M=-L}^{L}\bar{\beta}_{L,M}(\epsilon,t)Y_{L,M}(\theta,\phi)\label{eq:AF-PAD-general}
\end{equation}

Here the flux in the laboratory frame (LF) or aligned frame (AF) is denoted $\bar{I}(\epsilon,t,\theta,\phi)$, with the bar signifying ensemble averaging, and the molecular frame (MF) flux by $I(\epsilon,t,\theta,\phi)$.  Similarly, the expansion parameters $\bar{\beta}_{L,M}(\epsilon,t)$ include a bar for the LF/AF case. These observables are generally termed photoelectron angular distributions (PADs), often with a prefix denoting the reference frame, e.g. LFPADs, MFPADs, and the associated expansion parameters $\bar{\beta}_{L,M}(\epsilon,t)$ are generically termed ``anisotropy" parameters. The polar coordinate system $(\theta,\phi)$ is referenced to an experimentally-defined axis in the LF/AF case (usually defined by the laser polarization), and the molecular symmetry axis in the MF, as indicated in the corresponding panels of Fig. \ref{781808}. 

The spherical harmonic rank and order of the observables, $(L,M)$, are constrained by experimental factors in the LF/AF, and $n$ is typically limited by the molecular alignment, which is correlated with the photon-order for gas phase experiments. Generally, this can be considered in terms of conservation of total angular momentum in the LF \cite{Yang1948}, and each photon imparts one unit of angular momentum. For basic cases these limits may be low: for instance, a simple 1-photon photoionization event ($n=1$) from an isotropic ensemble (zero net ensemble angular momentum) defines $L_{max}=2$; for cylindrically or axially symmetric cases (i.e. $D_{\infty h}$ symmetry) $M=0$ only. 

In the MF $n$ is constrained only by the maximum continuum electron angular momentum $n=l_{max}$ imparted by the scattering event \cite{Dill1976} (note lower-case $l$ for the electron angular momentum). For these cases, $l_{max}=4$ is often given as a reasonable rule-of-thumb for the continuum - hence $L_{max}=8$ - although in practice higher-$l$ may be populated. Further details are discussed below, with a realistic example case forming the basis of Sect. \ref{sec:bootstrapping}. (For further introductory discussion and examples of LF and MF PADs, see Refs. \cite{Reid2003,hockett2018QMP1}; a recent review article can be found in Ref. \cite{dowek2022TrendsAngleresolvedMolecular}.)


Returning to Fig. \ref{781808}, note, in particular that the LF measurements (top panel) involve averaging over an ensemble of molecules with different orientations, leading to averaging over the molecular frame observables. The $\bar{\beta}_{L,M}(\epsilon,t)$ parameters in this case are shown in the lower right panel (and in more detail in Fig. \ref{720080}), and are constrained to $L_{max}=6$ and $M=0$ (cylindrical or axial symmetry). Two examples of the corresponding AFPADs $\bar{I}(\epsilon,t,\theta,\phi)$ are also shown, as simple polar plots, in the panel. These AFPADs display fairly simple, albeit time-dependent, angular structure. 
The corresponding MFPADs (lower left panel) are highly structured,  time-independent, quantities $I(\epsilon,\theta,\phi)$ in this case. The difference in the complexity of the MFPADs and AFPADs is typical of spatial averaging, and indicative of why one wishes to obtain MF results if possible. Crucially, these types of process are coherent, and the PADs are sensitive to the (relative) phases of the continuum electrons. PADs may also respond to other phase contributions depending on the type of experiment, for example the spatial averaging in an AF measurement is a coherent process.
In cases - such as this example - where the time-dependence is purely geometric, and is separable from the ionization matrix elements, the total information content can be broadly viewed as the number of sets of $\{L,M\}$ at a given $\epsilon$. 
This represents a rich dataset for retrieving matrix elements and reconstruction of the MFPADs, and this fairly general case is explored in detail herein (Sect. \ref{sec:MF-recon-basic-intro} introduces the concepts, and Sect. \ref{sec:bootstrapping} presents a full numerical case-study).

\subsection{Photoionization dynamics\label{sec:dynamics-intro}} 




The core physics of photoionization has been covered extensively in the literature, and only a very brief overview is provided here with sufficient detail to introduce the MF reconstruction problem; the reader is referred to the literature listed in Appendix \ref{sec:theory-lit} for further details and general discussion. Technical details of the formalism applied for the reconstruction techniques discussed herein can be found in Sect. \ref{sec:tensor-formulation}.

Photoionization can be described by the coupling of an initial state of the system to a particular final state (photoion(s) plus free photoelectron(s)), coupled by an electric field/photon. Very generically, this can be written as a matrix element $\langle\Psi_i|\hat{\Gamma}(\boldsymbol{\mathbf{E}})|\Psi_f\rangle$, where $\hat{\Gamma}(\boldsymbol{\mathbf{E}})$ defines the light-matter coupling operator (depending on the electric field $\boldsymbol{\mathbf{E}}$), and $\Psi_i$, $\Psi_f$ the total wavefunctions of the initial and final states respectively. 

There are many flavours of this fundamental light-matter interaction, depending on system and coupling; the discussion here is confined to the simplest case of single-photon absorption, in the weak field (or perturbative), dipolar regime, resulting in a single photoelectron. (For more discussion of various approximations in photoionzation, see Refs. \cite{Seideman2002,Seideman2001}.)


Underlying the photoelecton observables is the photoelectron continuum state $\left|\mathbf{k}\right>$, prepared via photoionization.
The photoelectron momentum vector is denoted generally by $\boldsymbol{\mathbf{k}}=k\mathbf{\hat{k}}$, in the MF. 
The ionization matrix elements associated with this transition 
provide the set of quantum amplitudes completely defining the final continuum scattering state,
\begin{equation}
\left|\Psi_f\right> = \sum{\int{\left|\Psi_{+};\bf{k}\right>\left<\Psi_{+};\mathbf{k}|\Psi_f\right> d\bf{k}}},
\label{eq:cstate}
\end{equation}
where the sum is over states of the molecular ion $\left|\Psi_{+}\right>$. The number of ionic states accessed depends on the nature of the ionizing pulse and interaction. For the dipolar case,

\begin{equation}
\hat{\Gamma}(\boldsymbol{\mathbf{E}}) = \hat{\mathbf{\mu}}\cdot\boldsymbol{\mathbf{E}}
\end{equation}

Hence,

\begin{equation}
\left<\Psi_{+};\mathbf{k}|\Psi_f\right> =\langle\Psi_{+};\,\mathbf{k}|\hat{\mathbf{\mu}}\cdot\boldsymbol{\mathbf{E}}|\Psi_{i}\rangle
\label{eq:matE-dipole}
\end{equation}

Where the notation implies a perturbative photoionization event from an initial state $i$ to a particular ion plus electron state following absorption of a photon $h\nu$, 
$|\Psi_{i}\rangle+h\nu{\rightarrow}|\Psi_{+};\boldsymbol{\mathbf{k}}\rangle$, and $\hat{\mu}\cdot\boldsymbol{\mathbf{E}}$ is the usual dipole interaction term \cite{qOptics}, which includes a sum over all electrons $s$ defined in position space as $\mathbf{r_{s}}$:  

\begin{equation}
\hat{\mu}=-e\sum_{s}\mathbf{r_{s}}
\label{eq:dipole-operator}
\end{equation}

The position space photoelectron wavefunction is typically expressed in the ``partial waves" basis, expanded as (asymptotic) continuum eigenstates of orbital angular momentum, with angular momentum components $(l,m)$ (note lower case notation for the partial wave components),  


\begin{equation}
\Psi_\mathbf{k}(\mathbf{r})\equiv\left<\mathbf{r}|\mathbf{k}\right> = \sum_{lm}Y_{lm}(\mathbf{\hat{k}})\psi_{lm}(\mathbf{r},k)
\label{eq:elwf}
\end{equation}

where $\mathbf{r}$ are MF electronic coordinates and $Y_{lm}(\mathbf{\hat{k}})$ are the spherical harmonics.

Similarly, the ionization dipole matrix elements can be separated generally into radial (energy-dependent or `dynamical' terms) and geometric (angular momentum) parts (this separation is essentially the Wigner-Eckart Theorem, see Ref. \cite{zareAngMom} for general discussion), and written generally as (using notation similar to \cite{Reid1991}): 

\begin{equation}
\langle\Psi_{+};\,\mathbf{k}|\hat{\mathbf{\mu}}\cdot\boldsymbol{\mathbf{E}}|\Psi_{i}\rangle = \sum_{lm}\gamma_{l,m}\mathbf{r}_{k,l,m}
\label{eq:r-kllam}
\end{equation}

Provided that the geometric part of the matrix elements $\gamma_{l,m}$ are known, knowledge of the so-called radial (or reduced) dipole matrix elements, at a given 
$k$, 
thus equates to a full description of the system photoionization dynamics (and, hence, the observables). The $\gamma_{l,m}$ includes the geometric rotations  into the LF arising from the dot product in Eqn.~\ref{eq:r-kllam}, as well as all other angular-momentum coupling terms.



For the simplest treatment, the radial matrix element can be approximated as a 1-electron integral involving the initial electronic state (orbital), and final continuum photoelectron wavefunction:


\begin{equation}
\mathbf{r}_{k,l,m}=\int\psi_{lm}(\mathbf{r},k)\mathbf{r}\Psi_{i}(\mathbf{r})d\mathbf{r}
\label{eq:r-kllam-integral}
\end{equation}

As noted above, the geometric terms $\gamma_{l,m}$ are analytical functions which can be computed for a given case - minimally requiring knowledge of the molecular symmetry and polarization geometry, although other factors may also play a role (see Sect. \ref{sec:full-tensor-expansion} for details). 


The photoelectron angular distribution (PAD) at a given $(\epsilon,t)$ can then be determined by the squared projection of $\left|\Psi_f\right>$ onto a specific state $\left|\Psi_{+};\bf{k}\right>$ (see Sect. \ref{sec:theoretical-techniques}), and therefore the amplitudes in Eqn.~\ref{eq:r-kllam} also determine the observable anisotropy parameters $\beta_{L,M}(\epsilon,t)$ (Eqn. \ref{eq:AF-PAD-general}). (Note that the photoelectron energy $\epsilon$ and (scalar) momentum $k$ are used somewhat interchangeably herein, with the former usually preferred in reference to observables.) Note, also, that in the treatment above there is no time-dependence incorporated in the notation; however, a time-dependent treatment readily follows, and may be incorporated either as explicit time-dependent modulations in the expansion of the wavefunctions for a given case, or implicitly in the radial matrix elements. Examples of the former include a rotational or vibrational wavepacket, or a time-dependent laser field. The rotational wavepacket case is discussed herein (see Sect. \ref{sec:full-tensor-expansion}). The radial matrix elements are a sensitive function of molecular geometry and electronic configuration in general, hence may be considered to be responsive to molecular dynamics, although they are formally time-independent in a Born-Oppenheimer basis. For further general discussion and examples see Ref. \cite{wu2011TimeresolvedPhotoelectronSpectroscopy}. Discussions of more complex cases with electronic and nuclear dynamics can be found in Refs.  \cite{arasaki2000ProbingWavepacketDynamics,Seideman2001, Suzuki2001,Stolow2008}.

Typically, for reconstruction experiments, a given measurement will be selected to simplify this as much as possible by, e.g., populating only a single ionic state (or states for which the corresponding observables are experimentally energetically-resolvable), and with a bandwidth $d\bf{k}$ which is small enough such that the matrix elements can be assumed constant. Importantly, the angle-resolved observables are sensitive to the magnitudes and (relative) phases of these matrix elements, and can be considered as angular interferograms (Fig. \ref{781808} top right).

\subsection{MF photoionization measurements and matrix element retrieval: approaching the problem and accounting for complexity\label{sec:MF-recon-basic-intro}}

As discussed in Sect. \ref{sec:MF-intro}, MF observables may be sought via (1) direct or (2) indirect methods. Following Sect. \ref{sec:dynamics-intro} the difficulty of both methods may begin to become apparent - both require sophisticated measurements and data analysis, and the underlying photoionization dynamics may be very rich. For (1) the aim is to obtain highly-structured $\beta_{L,M}(\epsilon,t)$ from ``fixed-in-space" molecules, whilst for (2) sufficient measurements must be made to either reconstruct these observables and/or the underlying matrix elements from $\bar{\beta}_{L,M}(\epsilon,t)$ measurements. In both cases experimental measurements are designed, ideally, to avoid averaging over any correlated DOFs which map into the observables (or otherwise account for them in some manner) - specific methods are discussed further in Sect. \ref{sec:experimentalTechniques}.

For indirect methods, the MF observables and/or matrix elements are reconstructed or retrieved from the LF measurements, via inversion or fitting methodologies, and these techniques are the main focus of Sect. \ref{sec:Recon}. The difficulty in matrix element reconstruction in general arises from the fact that there are typically many component partial waves (matrix elements) for even a simple system, and that determination of both magnitudes and phases is required. Hence this can be viewed as a form of quantum tomography, or a specific class of (quantum) phase-retrieval problems. In terms of the MF observable, these properties result in a quantity that may be highly structured and, hence, is (in general) particularly susceptible to orientational averaging (Fig. \ref{781808}). Furthermore, it may be particularly sensitive to averaging over other DOFs (e.g. vibronic states) and/or molecular dynamics, due to inherent sensitivity of the scattering process to molecular structure. Whilst a number of direct and indirect techniques have been used to obtain the relevant MF observables and/or dipole matrix elements, many outstanding questions remain, and this is an ongoing, interesting and challenging area of research \cite{hockett2018QMP1, hockett2018QMP2}.

Although the core physics is complicated, relatively high-dimensionality observables are possible for photoelectron measurements (Sect. \ref{sec:info-content}), hence experimental progress can be made to understand these light-matter interactions. The PAD is the key observable, which may be measured in the LF or MF (see Sect. \ref{sec:Photo-into}), and additionally interrogated as a function of other experimental parameters: of particular interest (and readily amenable to experimental control) are the ionizing field properties (polarization, intensity, wavelength, duration), the axis distribution of the molecular ensemble in the LF (alignment), and orientation in the MF. 

As a reasonable first approximation, photoionization can be treated as a single active electron (SAE) problem, and one in which the remainder of the system is static during the photoionization process (impulsive, or sudden, approximation): this allows the problem to be defined in terms terms of three key components: 

\begin{enumerate}
\item the initial (ionizing) state (electronic) wavefunction $\Psi_i$ (and the final ion state is assumed to be identical in character, minus the ionizing electron $\Psi_+ = \Psi_i(N-1)$, hence the hole created has the same orbital structure as the ionizing electron),
\item the structure of the continuum (free electron) wavefunction 
$\Psi_{k}(\mathbf{r})$,
\item the dipole matrix elements coupling these (single electron) wavefunctions.
\end{enumerate}


However, this problem still remains rather complicated (hence interesting), since the structure of the initial and continuum states depends sensitively on the molecular geometry (atomic positions and electron distribution, i.e. the full vibronic wavefunction) of the ionizing system. Nonetheless, significant progress can be made in both experimental analysis and \textit{ab initio} theory in this reduced case, and the approximations are valid for many interesting real cases (e.g. small, relatively rigid, polyatomics). Naturally, this zero-order treatment also provides a framework within which other effects can be recognised and understood, in terms of which physical assumptions are broken. Certain types of experiment may be sensitive to certain effects and DOFs - for instance, time-resolved observables may map the vibrational dependence, and pulse-intensity studies may indicate if and when the weak-field approximation breaks down.



\section{Concepts \& techniques\label{sec:Concepts}}

In this section a more detailed discussion of photoionization is provided, which underpins the quantum state and MF reconstruction presented in Sect. \ref{sec:Recon}.

\subsection{Experimental techniques\label{sec:experimentalTechniques}}

Following the above discussions, experimental methodologies for the determination of MF observables can be viewed from the perspective of direct and indirect techniques. 

In the direct case, access to the MF is sought via essentially one of two schemes:

\begin{enumerate}
\item Alignment/orientation of the molecular frame (Sect. \ref{sec:MF-control}).
\item Post-hoc reconstruction of the molecular frame from a suitable measurement (Sect. \ref{sec:fixed-in-space}).
\end{enumerate}

Indirect techniques are the main focus of this manuscript, and experimental implications are briefly introduced in Sect. \ref{sec:MF-recon-expt}.



\subsubsection{Molecular alignment and control: conforming the MF to the LF\label{sec:MF-control}}

The first category covers a range of techniques. For gas phase experiments, the most common methods involve creating some form of alignment or orientation \footnote{In the technical sense, alignment retains 
inversion symmetry in the LF, while orientation typically implies reduction of the LF symmetry to match the molecular point group symmetry. The term "orientation" is used herein as synonymous with the MF for an arbitrary molecular system, but in some cases - e.g. homonuclear diatomics - alignment may be sufficient for observation of MF observables.} 
in the gas phase molecular ensemble, which defines a relationship between the LF and MF. In general, measurements made from such an ensemble can be termed as corresponding to ``the aligned frame (AF)", and may still involve averaging over some DOFs; in the 
classical limit of perfect orientation, the AF and MF are conformal/indistinguishable. 

Perhaps the simplest AF technique is the creation of alignment via a single-photon pump process (as used in many resonance-enhanced mulit-photon ionization (REMPI) type experimental schemes, which may even be rotational-state selected); in this case a parallel or perpendicular transition moment will create a $\cos^2(\theta)$ or $\sin^2(\theta)$ distribution, respectively, of the corresponding molecular axis. Any such axis distribution, in which there is a defined arrangement of axes created in the LF, can be discussed, and characterised, in terms of the axis distribution moments (ADMs). ADMs are coefficients in a multipole expansion, in terms of Wigner D-Matrix Elements (see Sect. \ref{sec:full-tensor-expansion}), of the molecular axis probability distribution. These are spherical tensors, equivalent to density matrix elements \cite{BlumDensityMat}. Many authors have address aspects of this problem in the past in frequency-domain work, see, for instance, the textbooks of Zare \cite{zareAngMom} and Blum \cite{BlumDensityMat}, treatments for various experimental cases in Refs. \cite{Docker1988,Dubs1989,Greene1983}, and application in complete photoionization experiments in Refs. \cite{Leahy1991,hockett2009RotationallyResolvedPhotoelectron}.

Further control can be gained via a single, or sequence of, N-photon transitions, or strong-field mediated techniques. Of the latter, adiabatic and non-adiabatic alignment methods are particularly powerful, and make use of a strong, slowly-varying or impulsive laser field respectively. (Here the ``slow" and ``impulsive" time-scales are defined in relation to molecular rotations, roughly on the ps time-scale, with ns and fs laser fields corresponding to the typical slow and fast control fields.) In the former case, the molecular axis, or axes, will gradually align along the electric-field vector(s) while the field is present. In the latter, impulsive case, a broad rotational wavepacket (RWP) can be created, initiating complex rotational dynamics including field-free revivals of ensemble alignment. Both techniques are powerful, but multiple laser fields are typically required in order to control more than one molecular axis, leading to relatively complex experimental requirements. The absolute degree of alignment obtained in a given case is also dependent on a number of intrinsic and experimental properties, including the molecular polarisability 
and moment of inertia tensors, rotational temperature and separability of the rotational degrees of freedom from other DOFs (loosely speaking, this can be considered in terms of the stiffness of the molecule). Recent studies of molecules embedded in Helium droplets have addressed some of these issues, achieving stronger and longer lived 3D alignment. These studies also examined several complications associated with coupling between molecular and droplet DOFs.   Therefore, although general in principle, in practice not all molecular targets are amenable to ``good" (i.e. a high degree of) alignment. For more general details, see, for example, Refs. \cite{koch2019QuantumControlMolecular,Stapelfeldt2003,nielsen2022Helium}, and for applications in photoionization see Sect. \ref{sec:theory-lit}.

Whilst gas phase alignment experiments can become rather complex, multi-pulse affairs, they are increasingly popular in the AMO community for a number of possible reasons. Conceptually and experimentally, they are a relatively tractable extension to existing techniques. They are interesting experiments in their own right, and, practically, they are usually feasible with existing high-power pulsed laser sources in the ns to fs regime. Alignment techniques have been combined with a range of different probes, including non-linear and high-harmonic optical probes, as well as photoionization-based methods - for recent reviews see \cite{hasegawa2015NonadiabaticMolecularAlignment,koch2019QuantumControlMolecular}. 

An alternative, very different, technique of orientational control is via embedding the target species in a matrix, or via deposition on a surface, which defines a spatial orientation. This approach has been taken primarily by the surface science community, and the required methods may often be readily applied to a range of targets using existing experimental apparatus and techniques (hence their ready adoption, analogous to the ready adoption of multi-pulse laser schemes in the gas phase community), combined with suitably-prepared surfaces. Although not the topic of this manuscript, there is certainly work in this vein conceptually related to the discussions herein, including ARPES (angle-resolved photoemission spectroscopy) and SERS (surface-enhanced, coherent anti-Stokes Raman scattering) studies. In these techniques, molecular orientation is well-defined, but at the expense of interactions with the bulk - although the latter may also be of interest and/or probed by the measurement. 
Examples include ARPES studies in which orbital densities of adsorbed species are reconstructed (``orbital tomography") \cite{Puschnig2009a,dauth2014AngleResolvedPhotoemission}, and SERS work making use of functionalised nano-particles for ``single molecule" fluorescence studies of vibrational wavepackets \cite{Yampolsky2014}. 

\subsubsection{Fixed-in-space molecules: MF via axis reconstruction\label{sec:fixed-in-space}}

The second category covers methodologies which make use of experimental information to reconstruct, post-facto, molecular alignment at the time of a light-matter interaction. This usually involves making a ``kinematically complete" class of measurement, which provides the full energy or momentum partitioning of the products of a light-matter interaction. In order for the alignment of a given axis to be defined in this case, there must be a clear energy partitioning defining it - typically dissociation is required, although some axes in a given problem may be inferred from related or proxy measurements. The simplest example is the dissociation of a diatomic molecule, in this case measuring the momentum of just one product atom/ion will enable the original orientation of the molecule to be determined 
provided the molecule does not rotate while dissociating (i.e. photodisociation is also sudden/impulsive, or equivalently geometrically-uncoupled - otherwise this remains a DOF to be averaged over, resulting in ``recoil-frame" (RF) measurements!). Combined with measurement of an electron, in coincidence, the MF photoelectron distribution can be recovered from a set of such measurements. Recent discussions and reviews of this area can be found in Refs. \cite{Yagishita2005,Reid2012,dowek2012PhotoionizationDynamicsPhotoemission,Yagishita2015,jahnke2022PhotoelectronDiffraction}, and some (representative) examples from the literature are given in Sect. \ref{appendix:MF-expt}.

Further extensions to such a measurement can probe additional dynamics in the MF, for instance electron-electron correlation effects in double ionization \cite{Akoury2007}. 
For larger molecules this becomes more complicated, and additionally requires that axial-recoil conditions are fulfilled (i.e. energy is not partitioned into other DOFs during dissociation). Such measurements are, therefore, well-suited to diatomics, and small polyatomics, and light-matter interactions involving core-ionization(s) events. For valence studies these techniques are less directly applicable, since dissociative events are less common (and may be slow/complex), although potentially can be applied in Coulomb-explosion imaging (CEI) type scenarios. In CEI methods, intense fields are used to strip multiple electrons from the target system, causing the molecule to (Coulombically) explode. Provided the intense pulse is short (relative to time-scales of atomic motion), measurement of multiple ionic fragments yields a map of the geometry of the system at the instant of the interaction \cite{stapelfeldt1998TimeresolvedCoulombExplosion,Underwood2015,Slater2015}. 

Another caveat for this class of measurement is the requirement for coincidence (or covariance) data collection, which typically limits count rates significantly, as well as the total number of products which may be feasibly measured - although the requirements may be somewhat relaxed for covariance studies. For these various reasons, amongst others, these experiments have typically been performed at synchrotrons (and, recently, at FELs) with high repetition rates (high KHz, MHz) and at hard photon energies. However, laser-based experiments are also relatively common, particularly in the strong-field community, and may become more so as sources with high-repetition rates and high peak intensities become commercially available. 

Recent state-of-the-art MF measurements have successfully measured 4-fold coincidences, and 5-particle covariance maps, in order to map polyatomic molecules and vibrational dynamics. 
For a recent example, illustrating the power of such a technique for CEI imaging of iodopyridine and iodopyrazine, see Ref.   \cite{boll2022XrayMultiphotoninducedCoulomb}. However, to date these techniques have mainly been used as structural (nuclear) probes. In terms of MFPAD measurements, these techniques remain very challenging, since the CEI schemes required produce many secondary electrons which cannot typically be distinguished.


\subsubsection{Post-processing \& ``complete" photoionization studies: MF via reconstruction\label{sec:MF-recon-expt}}


A significant issue with ``direct" experimental approaches to the MF is the difficulty of the measurement, and the degree of MF fidelity obtained, particularly in more complex systems. (Related to this is the issue of whether the free system is measured, or whether it is perturbed in some way, e.g. by an alignment laser field or a coupled system.) A complementary approach is to employ a post-processing approach in which underlying MF properties, possibly even the full set of photoionization matrix elements, are sought from LF or AF measurements, as already introduced in Sect. \ref{sec:MF-recon-basic-intro}. Such schemes are potentially demanding and complex in terms of the computational effort required to post-process the experimental data, but may also be significantly less demanding experimentally than direct MF measurements. Such schemes additionally have the potential to provide more fundamental information on photoionization dynamics.

Experimental techniques to generate suitable datasets for analysis are many and varied. These include the direct analysis of MF measurements to obtain the underlying photoionization dynamics, the use of frequency-resolved methods, and the use of molecular alignment techniques. Some representative examples of such ``complete" photoionization studies from the literature are given in Sect. \ref{sec:CompleteLit}. 
The main focus of Sect. \ref{sec:Recon} is the analysis of time-resolved data from an aligned system with a prepared RWP, although retrieval from the MF is also investigated herein (Sect. \ref{sec:recon-from-MFPADs}). Whilst the RWP case is experimentally similar to the ``direct" MF measurement case outlined above, the requirements on the degree of alignment are much reduced, since the fidelity arises from the analysis, rather than the absolute maximum degree of alignment obtained. Specifically, the fidelity of the reconstruction can be considered as a result of the total information content of the time-domain measurement (related to the number of distinct molecular axis distributions probed), as distinct from the ``direct" case which is constrained by a single measurement at the best alignment or orientation obtained. In practice this alignment needs to be extremely good to truly approach MF information in general, and this may not be possible in many cases. It will also be constrained in general by the symmetry of the problem. For an example case study, see Ref. \cite{reid2018AccessingMolecularFramea}, which suggests $\langle\cos^2(\theta)\rangle>0.9$ as a \textit{minimum} requirement, for measurements of relatively simple, cylindrically-symmetric, MFPADs; another illustrative example of the loss of information in the MF to LF transformation can be found in Ref. \cite{Underwood2000}.

This implies that reconstruction methods may be rather more general, and the RWP case in particular is expected to be applicable to any molecular system, although outstanding questions on the required and obtainable information content remain (see Sect. \ref{sec:info-content}). Questions of fidelity of reconstruction are also a matter of ongoing research. This will, again, depend on both the type and nature of the experimental measurements, and the reconstruction methodology, both of which may involve or assume certain additional DOFs or physical behaviours which are required for tractable theory but may not hold in practice. Examples include cases with multiple conformers, floppy systems and the presence of additional (e.g. vibrational) dynamics. In all cases progress may be possible, but will require additional experimental and/or computational effort to control, isolate, simulate or reconstruct the additional DOFs. (See Ref. \cite{Takatsuka2000} for a ``basic" theoretical vibrational wavepacket example in $Na_2$, and for a more complex case in $CS_2$ Ref. \cite{wang2017MonitoringNonadiabaticDynamics}.)

A benchmark example is the retrieval of the photoionization dynamics of $N_2$, since this is a simple, fairly rigid, system amenable to experimental control ($NO$ has also been investigated by a number of authors, see Sect. \ref{sec:CompleteLit}). The authors of this manuscript, with a number of collaborators, demonstrated that photoionization matrix elements can be retrieved for one-photon ionisation of $N_2$ by time resolved measurements of LFPADs from a rotational wavepacket. The experiments did achieve a relatively high degree of alignment via a two-pulse pump scheme, with a maximum $\langle\cos^2(\theta)\rangle$ of $\sim 0.8$, and 11 temporal data-points (obtained over the half and full RWP revivals) were found to be sufficient for full matrix element retrieval and MFPAD reconstruction. This methodology, ``bootstrapping to the molecular frame", is the focus of Sect. \ref{sec:bootstrapping} below, with additional details and results provided for the case of radial matrix element extraction for N$_2$.  In follow up work, it was shown that for molecules with $D_{nh}$ point group symmetry the retrieval of the MFPAD is possible directly via a matrix inversion methodology, bypassing the difficulty of extracting the radial matrix elements, and this is discussed in Sect. \ref{sec:Matrix-inversion-example}.


\subsubsection{Technology and outlook}

A final note on experimental methods is the possibilities afforded by technological (rather than conceptual) developments. Naturally, the techniques above rely on a certain degree of experimental and technological sophistication, and this, of course, generally increases over time as a technique matures and ``enabling" technologies develop. Examples, in this context, are the development of, and gradual improvements in, particle imaging detectors. Such progress allows for more sophisticated experiments with, e.g. multi-particle coincidence detection, higher detection rates, energy-multiplexed measurements and so on. In short, the experimental dimensionality or information content can be increased. For developing general methods, which can be applied in complex cases, this becomes increasingly important (as do related technological capabilities, such as data storage and processing, to handle the enhanced complexity). 

Historically, photoionization measurement capabilities have gone from 1D (flux) and 2D (flux and kinetic energy) in the early days, advanced to sequential or parallel measurements at different angles for various flavours of angle-resolved studies, and to full 3D or 4D capabilities in modern ``imaging" type systems (flux and 2D or 3D kinetic energy vector resolution). This is expanded by the possibility of additional experimental measurement dimensions such as time, polarisation, laser power, wavelength and so forth. Advances in experimental methods, in particular the development of high count-rate 3D detectors, is ongoing (see, for examples and historical context, \cite{Parker1997,Dorner1997a,Continetti2001,Vallance2013,chandler2017PerspectiveAdvancedParticle}). 

As well as detector technology, general developments in experimental methods, system integration, computational power and so forth further enable novel techniques and/or fusion of existing methodologies. For instance, it is of note that with the advent of more sophisticated laser-based techniques, and short-pulse FELs, the combination of laser alignment techniques with dissociative photoionization measurements is also now increasingly common. Other examples include increasingly sophisticated multi-pulse measurements and measurements with shaped laser pulses. An illustrative example of the rich data available from a modern, sophisticated, experimental scheme - CEI with a pixel-imaging mass-spectroscopy (PImMS) camera and covariance analysis - see \cite{Slater2015}. Further examples of experimental developments from the literature, with respect to matrix element retrieval, can be found in Sect. \ref{sec:CompleteLit}.




\subsection{Theoretical techniques\label{sec:theoretical-techniques}}



\subsubsection{Tensor formulation of photoionization\label{sec:tensor-formulation}}

A number of authors have treated MFPADs and related problems, see Appendix \ref{sec:theory-lit} for some examples. Herein, a geometric tensor based formalism is developed, which is close in spirit to the treatments given by Underwood and co-workers \cite{Reid2000, Stolow2008, Underwood2000}, but further separates various sets of physical parameters into dedicated tensors. This allows for a unified theoretical and numerical treatment, where the latter computes properties as tensor variables which can be further manipulated and investigated. 
Furthermore, the tensors can readily be converted to a density matrix representation \cite{BlumDensityMat, zareAngMom}, which is more natural for some quantities, and also emphasizes the link to quantum state tomography and other quantum information techniques. Much of the theoretical background, as well as application to aspects of the current problem, can be found in the textbooks of Blum \cite{BlumDensityMat} and Zare \cite{zareAngMom}.

Within this treatment, the observables can be defined in a series of simplified forms, emphasizing the quantities of interest for a given problem. Some details are defined in the following subsections, and further detailed in Appendix \ref{appendix:formalism}.

\subsubsection{Channel functions\label{sec:channel-funcs}}

A simple form of the equations, amenable to fitting, is to write the observables in terms of ``channel functions", which define the ionization continuum for a given case and set of parameters $u$ (e.g. defined for the MF, or defined for a specific experimental configuration). Here we briefly introduce the notation, which expresses the observables $\beta_{L,M}^{u}$ (Eqn. \ref{eq:AF-PAD-general}) as a tensor product of channel functions ($\varUpsilon_{L,M}^{u,\zeta\zeta'}$), which collect all of the geometric parameters, and ionization matrix elements $\mathbb{I}^{\zeta\zeta'}$. At a basic level, the channel functions can be viewed simply as defining all of the analytical parameters in a given case; further theoretical details are unpacked below (Sect. \ref{sec:full-tensor-expansion} and Appendix \ref{appendix:formalism}), and specific cases in Sect. \ref{sec:bootstrapping-info-sensitivity}.

Making use of the channel functions, the observables can be written as:

\begin{equation}
\beta_{L,M}^{u}=\sum_{\zeta,\zeta'}\varUpsilon_{L,M}^{u,\zeta\zeta'}\mathbb{I}^{\zeta\zeta'}\label{eqn:channel-fns}
\end{equation}


where $\zeta,\zeta'$ collect all the required quantum numbers, and define all (coherent) pairs of components. The term $\mathbb{I}^{\zeta\zeta'}$ denotes the coherent square of the ionization matrix elements:

\begin{equation}
\mathbb{I}^{\zeta,\zeta}=I^{\zeta}(\epsilon)I^{\zeta'*}(\epsilon)
\label{eqn:I-zeta}
\end{equation}

This is effectively a convolution equation (cf. Refs. \cite{Reid2000,gregory2021MolecularFramePhotoelectron}) with channel functions, for a given ``experiment'' $u$, summed over all terms $\zeta,\zeta'$. Aside from the change in notation (which is here chosen to match the formalism of Refs. \cite{Gianturco1994, Lucchese1986, Natalense1999}), see also Sect. \ref{sec:mat-ele-conventions}), these matrix elements are essentially identical to the simplified (radial) forms $\mathbf{r}_{k,l,m}$ defined in Eqn. \ref{eq:r-kllam}, in the case where $\zeta=k,l,m$. These complex matrix elements can also be equivalently defined in a magnitude, phase form:

\begin{equation}
I^{\zeta}(\epsilon)\equiv\mathbf{r}_{\zeta}\equiv r_{\zeta}e^{i\phi_{\zeta}}
\end{equation}

This tensorial form is numerically implemented in the ePSproc codebase \cite{ePSprocGithub}. This is in contradistinction to standard numerical routines in which the requisite terms are usually computed from vectorial and/or nested summations, and typically implement the full computation of the observables in one computational routine. The standard approach can be somewhat opaque to detailed interpretation.  The PEMtk codebase \cite{hockett2021PEMtkGithub} implements matrix element retrieval based on this formalism, with pre-computation of all the geometric tensor components (channel functions) prior to a fitting protocol for matrix element analysis. This is essentially a fit to Eqn. \ref{eqn:channel-fns} given a set of $\beta_{L,M}^{u}$, with $I^{\zeta}(\epsilon)$ as the unknowns (in magnitude, phase form). The main computational cost of a tensor-based approach is that more RAM is required to store the full set of tensor variables.

\subsubsection{Density matrix representation\label{sec:density-mat-basic}}
The density operator associated with the continuum state in Eqn.~\ref{eq:cstate} is easily written as $\hat{\rho}=|\Psi_c\rangle\langle\Psi_c|$. In the channel function basis, this leads to a density matrix given by the radial matrix elements:

\begin{equation}
\mathbf{\rho}^{\zeta\zeta'} = \mathbb{I}^{\zeta,\zeta'}
\label{eqn:radial-density-mat}
\end{equation}

Since the matrix elements characterise the scattering event, the density matrix provides an equivalent characterisation of the scattering event.
An example case is discussed in Sect. \ref{sec:den-mat-N2} (see Fig. \ref{998904}); for more details, and further discussion, see Sect. \ref{sec:density-mat-full}. Further discussion can also be found in the literature, see, e.g., Ref. \cite{BlumDensityMat} for general discussion, Ref. \cite{Reid1991} for application in pump-probe schemes.

\subsubsection{Full tensor expansion\label{sec:full-tensor-expansion}}

In more detail, the channel functions can be given as a set of tensors, defining each aspect of the problem.

For the MF:

\begin{eqnarray}
\beta_{L,-M}^{\mu_{i},\mu_{f}}(\epsilon) & = & (-1)^{M}\sum_{P,R',R}(2P+1)^{\frac{1}{2}}{E_{P-R}(\hat{e};\mu_{0})}\\
 & \times &\sum_{l,m,\mu}\sum_{l',m',\mu'}(-1)^{(\mu'-\mu_{0})}{\Lambda_{R',R}(R_{\hat{n}};\mu,P,R,R')B_{L,-M}(l,l',m,m')}\\
 & \times & I_{l,m,\mu}^{p_{i}\mu_{i},p_{f}\mu_{f}}(\epsilon)I_{l',m',\mu'}^{p_{i}\mu_{i},p_{f}\mu_{f}*}(\epsilon)\label{eq:BLM-tensor-MF}
\end{eqnarray}

And the LF/AF as:

\begin{eqnarray}
\bar{\beta}_{L,-M}^{\mu_{i},\mu_{f}}(E,t) & = & (-1)^{M}\sum_{P,R',R}{[P]^{\frac{1}{2}}}{E_{P-R}(\hat{e};\mu_{0})}\\
 & \times &\sum_{l,m,\mu}\sum_{l',m',\mu'}(-1)^{(\mu'-\mu_{0})}{\Lambda_{R'}(\mu,P,R')B_{L,S-R'}(l,l',m,m')}\\
 & \times &I_{l,m,\mu}^{p_{i}\mu_{i},p_{f}\mu_{f}}(\epsilon)I_{l',m',\mu'}^{p_{i}\mu_{i},p_{f}\mu_{f}*}(\epsilon)\sum_{K,Q,S}\Delta_{L,M}(K,Q,S)A_{Q,S}^{K}(t)\label{eq:BLM-tensor-AF}
\end{eqnarray}

In both cases a set of geometric tensor terms are required, which are fully defined in Appendix \ref{appendix:formalism}; these terms provide details of:

\begin{itemize}
\item ${E_{P-R}(\hat{e};\mu_{0})}$: polarization geometry \& coupling with the electric field.
\item $B_{L,S-R'}(l,l',m,m')$: geometric coupling of the partial waves into the $\beta_{L,M}$ terms (spherical tensors).
\item $\Lambda_{R'}(\mu,P,R')$: frame couplings and rotations.
\item $\Delta_{L,M}(K,Q,S)$: alignment frame coupling.
\item $A_{Q,S}^{K}(t)$: ensemble alignment described as a set of axis distribution moments (ADMs).
\end{itemize}

And \(I_{l,m,\mu}^{p_{i}\mu_{i},p_{f}\mu_{f}}(\epsilon)\) are the (radial) dipole ionization matrix elements, as a function of energy \(\epsilon\). These matrix elements are essentially identical to the simplified forms $r_{k,l,m}$ defined in Eqn. \ref{eq:r-kllam}, except with additional indices to label symmetry and polarization components
defined by a set of partial-waves \(\{l,m\}\), for polarization component \(\mu\) (denoting the photon angular momentum components) and channels (symmetries) labelled by initial and final state indexes \({p_{i}\mu_{i},p_{f}\mu_{f}}\). The notation here follows that used by ePolyScat \cite{Gianturco1994, Lucchese1986, Natalense1999}, and these matrix elements again represent the quantities  to be obtained numerically from data analysis, or from an \href{https://epsproc.readthedocs.io/en/latest/ePS_ePSproc_tutorial/ePS_tutorial_080520.html\#Theoretical-background}{ePolyScat (or similar) calculation}. 

Note that, in this case as given, time-dependence arises purely from the \(A_{Q,S}^{K}(t)\) terms in the AF case, and the electric field term currently describes only the photon angular momentum coupling,
although can in principle also describe time-dependent/shaped fields. Similarly, a time-dependent initial state (e.g. a vibrational wavepacket) could also describe a time-dependent MF case.

It should be emphasized, however, that the underlying physical quantities are essentially identical in all the theoretical approaches, with a set of coupled angular-momenta defining the geometrical part of the photoionization problem, despite these differences in the details of the theory and notation.

\subsubsection{Information content\label{sec:info-content}}

As discussed in Ref. \cite{hockett2018QMP2}, the information content of a single observable might be regarded as simply the number of contributing $\beta_{L,M}$ parameters. In set notation:

\begin{equation}
M=\mathrm{n}\{\beta_{L,M}\}
\end{equation}

where $M$ is the information content of the measurement, defined
as $\mathrm{n}\{...\}$ the cardinality (number of elements) of the
set of contributing parameters. A set of measurements, made for some
experimental variable $u$, will then have a total information content:

\begin{equation}
M_{u}=\sum_{u}\mathrm{n}\{\beta_{L,M}^{u}\}
\end{equation}

In the case where a single measurement contains multiple $\beta_{L,M}$, e.g. as a function of energy $\epsilon$ or time $t$, the information content will naturally be larger:

\begin{eqnarray}
M_{u,\epsilon,t} & = & \sum_{u,\epsilon,t}\mathrm{n}\{\beta_{L,M}^{u}(\epsilon,t)\}\\
 & = & M_{u}\times M_{\epsilon,t}
\end{eqnarray}

where the second line pertains if each measurement has the same native
information content $M_{\epsilon,t}$ per energy and time point, independent of any other experimental parameters $u$. It may be that the variable
$\epsilon$ is continuous (e.g. photoelectron energy), but in practice it
will usually be discretized in some fashion by the measurement.

In terms of purely experimental methodologies, a larger $M_{u}$ clearly defines a richer experimental measurement which explores more of the total measurement space spanned by the full set of $\{\beta_{L,M}^{u}(k,t)\}$. However, in this basic definition a larger $M_{u}$ does not necessarily indicate a higher information content for quantum retrieval applications.
The reason for this is simply down to the complexity of the problem
(cf. Eqn. \ref{eqn:channel-fns}), in which many couplings define
the sensitivity of the observable to the underlying system properties
of interest. In this sense, more measurements, and larger $M$, may
only add redundancy, rather than new information.

A more complete accounting of information content would, therefore,
also include the channel couplings, i.e. sensitivity/dependence of the observable to a given system property, in some manner. For the case of a time-dependent measurement, arising from a rotational wavepacket, this can be written as:

\begin{equation}
M_{u}=\mathrm{n}\{\varUpsilon_{L,M}^{u}(\epsilon,t)\}
\end{equation}

In this case, each $(\epsilon,t)$ is treated as an independent measurement with unique information content, although there may be redundancy as a function of $t$ depending on the nature of the rotational wavepacket and channel functions. This is explored further in Sect. \ref{sec:bootstrapping-info-sensitivity}. (Note this is in distinction to previously demonstrated cases where the time-dependence was created from a shaped laser-field, and was integrated over in the measurements, which provided a coherently-multiplexed case, see Refs. \cite{hockett2014CompletePhotoionizationExperiments, hockett2015CompletePhotoionizationExperiments,hockett2015CoherentControlPhotoelectron} for details.)




\subsection{Retrieval \& reconstruction techniques\label{sec:recon-techniques-intro}}


Following the tensor notation presented above, a ``complete" photoionization experiment can be characterized as recovery of the matrix elements $I^{\zeta}(\epsilon)$ from the experimental measurements or, equivalently, the density matrix $\mathbf{\rho}^{\zeta\zeta'}$. (For further discussion, see Refs. \cite{Reid2003,kleinpoppen2013perfect,hockett2018QMP1}.) This may be possible provided the channel functions are known, and the information content of the measurements is sufficient. (Note here that the matrix elements are assumed to be time-independent, although that may not be the case for the most complicated examples including vibronic dynamics \cite{hockett2018QMP2}.) 

Additionally, for schemes making use of molecular alignment, the molecular axis distributions must also be characterised. For the rotational wavepacket case, this is discussed in Sect. \ref{sec:RWPs}. This can actually be considered as a reduced-dimensionality MF signal retrieval problem, and also forms the first step in both the generalised ``bootstrapping" method (Sect. \ref{sec:bootstrapping}) and matrix inversion techniques (Sect. \ref{sec:matrix-inv-intro}).

Of particular import for matrix element retrieval is the phase-sensitive nature of the observables, which is required in order to obtain partial wave phase information. PADs can also be considered as angular interferograms, and reconstruction can be considered conceptually similar to other phase-retrieval problems, e.g. optical field recovery with techniques such as FROG \cite{trebino2000FrequencyResolvedOpticalGating}, and general quantum tomography \cite{MauroDAriano2003}.

\subsubsection{Freely rotating molecules: MF via time evolution\label{sec:RWPs}}

The efforts to align and orient molecules discussed in the previous sections necessarily led to detailed studies of the rotational dynamics of molecules after interaction with a non-resonant femtosecond laser pulse. A significant outcome of these studies has been the development of a reliable model capable of accurate simulations of rotational wavepacket dynamics that quantitatively agree with experimental results. By measurement of a signal from a time evolving rotational wavepacket, this ability to accurately simulate the wavepacket dynamics can be used to reconstruct the measured signal in the molecular frame. Since in this case the time resolved measurement constitutes a set of measurements of the same quantity from a variety of molecular axes distributions, it is reasonable to conclude that if the axes distributions are known, and provided a large enough space of orientations is explored by the molecule over the experimental time window, the molecular frame signal should be extractable. 

This is relatively straight forward for a signal that is a single number (scalar) in the MF for a given polarization of the light, such as the photoionization yield. Such a signal may, in general, be expressed as an expansion,
\begin{equation}
S(\theta,\chi)=\sum_{jk}C_{jk}D^{j}_{0k}(\theta,\chi),
\label{eq:mfrealsig}
\end{equation}
where $\theta$ and $\chi$ are the MF spherical polar and azimuthal angles of the linearly polarized electric field vector generating the signal; $C_{jk}$ are unknown expansion coefficients; and $D^{j}_{0k}$ are the Wigner D-Matrix elements, a basis on the space of orientations. A time resolved measurement of $S$ from a rotational wavepacket is the quantum expectation value of this expression,
\begin{equation}
\langle S \rangle(t) = \sum_{jk}C_{jk}\langle D^{j}_{0k} \rangle (t).
\label{eq:St-Cjk}
\end{equation}
Since the rotational wavepacket can be accurately simulated, the $\langle D^{j}_{0k} \rangle (t)$ are considered known. The time resolved signal $\langle S \rangle(t)$ being measured, the unknown coefficients $C_{jk}$ can be determined by linear regression, and the molecular frame signal in Eqn.~\ref{eq:mfrealsig} constructed. In this form the method was initially applied to strong field ionization and dubbed Orientation Reconstruction through Rotational Coherence Spectroscopy (ORRCS) \cite{makhija2016ORRCS,wang2017ORRCS}.
It has since been applied to strong field ionization of various molecules \cite{sandor2018ORRCS,sandor2019ORRCS,wangjam2021ORRCS},
strong field dissociation \cite{lam2020ORRCS} 
and few-photon ionization \cite{lam2022ORRCS}. 
(As hinted at in Sect. \ref{sec:MF-control}, a large range of other experimental methods have also addressed alignment and orientation dependence and retrieval, other recent examples include Coulomb-explosion imaging \cite{Underwood2015}, high-harmonic spectroscopy \cite{he2018RealTimeObservationMolecular, he2020MeasuringRotationalTemperature}, optical imaging \cite{Loriot2008} and rotational echo spectroscopy \cite{wang2022RotationalEchoSpectroscopy}, see Refs. \cite{Ramakrishna2013,koch2019QuantumControlMolecular} for further discussion.)


The case of PADs is a more challenging one, since they are not generally described by Eqn.~\ref{eq:mfrealsig}. Instead, both LFPADs and MFPADs are determined by the radial dipole matrix elements as described above (Sects. \ref{sec:dynamics-intro}, \ref{sec:channel-funcs}). However, the correspondence of the problem with an equation of the form of Eqn. \ref{eq:St-Cjk} - essentially a convolution - can be made. This is discussed in detail in Ref. \cite{Underwood2000}. In the current case Eqns. \ref{eqn:channel-fns}, \ref{eq:BLM-tensor-AF} can be rewritten in a similar form to Eqn. \ref{eq:St-Cjk} by explicitly separating out the axis distribution moments $A_{Q,S}^{K}(t)$ and collapsing all other terms. The case of photoionization from a time-dependent ensemble can then be reparameterized as:




\begin{equation}
\bar{\beta}_{L,M}^{u}(\epsilon,t)=\sum_{K,Q,S}\bar{C}_{KQS}^{LM}(\epsilon)A_{Q,S}^{K}(t)
\label{eqn:beta-convolution-C}
\end{equation}


Here the set of axis distribution moments can thus be viewed as modulating all observables $\beta_{L,M}^{u}(t)$. The unknowns, $\bar{C}_{KQS}^{LM}$ and axis distribution moments $A_{Q,S}^{K}(t)$, can be retrieved in a similar manner to that discussed for the simpler scalar observable case above, i.e. via linear regression with simulated rotational wavepackets. 

In practice this equates to (accurately) simulating rotational wavepackets, hence obtaining the corresponding $A_{Q,S}^{K}(t)$ parameters (expectation values), as a function of laser fluence and rotational temperature. Given experimental data, a 2D uncertainty (or error) surface in these two fundamental quantities can then be obtained from a linear regression for each set of $A_{Q,S}^{K}(t)$. The closest set of parameters to the experimental case is then determined by selection of the best results (smallest uncertainty) from such a parameter-space mapping, which constitutes determination of both the rotational wavepacket (hence $A_{Q,S}^{K}(t)$) and $\bar{C}_{KQS}^{LM}(\epsilon)$. Optimally, the corresponding physical properties can be cross-checked with other experimental estimates for additional confirmation of the fidelity of the protocol, although this may not always be possible. Note that, in this case, the photoionization dynamics are phenomenologically described by the real parameters $\bar{C}_{KQS}^{LM}$, but details of the matrix elements are not obtained directly; however, these parameters can be further used for the matrix inversion method (Sect. \ref{sec:matrix-inv-intro}), and are formally defined therein (Eqn. \ref{eq:C-AF}).


\subsubsection{Fitting methodologies\label{sec:fitting-intro}}

The nature of the photoionization problem suggests that a fitting approach can work, in general, which can be expressed (for example) in the standard way as a (non-linear) least-squares minimization problem:

\begin{equation}
\chi^{2}(\mathbb{I}^{\zeta\zeta'})=\sum_{u}\left[\beta^{u}_{L,M}(\epsilon,t;\mathbb{I}^{\zeta\zeta'})-\beta^{u}_{L,M}(\epsilon,t)\right]^{2}\label{eq:chi2-I}
\end{equation}

where $\beta^{u}_{L,M}(\epsilon,t;\mathbb{I}^{\zeta\zeta'})$ denotes  the values from a model function, computed for a given set of (complex) matrix elements $\mathbb{I}^{\zeta\zeta'}$, and $\beta^{u}_{L,M}(\epsilon,t)$ the experimentally-measured parameters, for a given configuration $u$. Implicit in the notation is that the matrix elements are independent of $u$ (or otherwise averaged over $u$). Once the matrix elements are obtained in this manner then MF observables, for any arbitrary $u$, can be calculated. An example of such a protocol - specifically one based on time-domain measurements and making use of a rotational wavepacket - is shown in Fig. \ref{781808}, 
and the practical realisation of such a methodology is the topic of Sect. \ref{sec:bootstrapping} (see also Refs. \cite{hockett2018QMP2,marceau2017MolecularFrameReconstruction} for further discussion). As discussed in Sect. \ref{sec:MF-recon-expt}, other choices of experimental measurements may also be made, for instance direct MF measurements or frequency-domain measurements, some representative examples from the literature are given in Sect. \ref{sec:CompleteLit}. 

Although in principle a very general approach, outstanding questions with such protocols remain, in particular fit uniqueness and reproducibility, the optimal measurement space $u$ - or associated information content $M_u$ - for any given case or measurement schema, and how well they will scale to larger problems (more matrix elements/partial waves). (Again, see Refs. \cite{hockett2018QMP2,marceau2017MolecularFrameReconstruction} for further discussion.)

\subsubsection{Matrix inversion methodologies\label{sec:matrix-inv-intro}}

An alternative methodology has recently been demonstrated, in which the MF observables are determined via a matrix inversion protocol \cite{gregory2021MolecularFramePhotoelectron}. This method does not require - potentially time-consuming - numerical fitting, although still requires knowledge of the channel functions. A full outline of the matrix inversion method is given in Appendix \ref{app:mat-inversion}, and a brief overview below.

For the matrix-inversion approach, the relationship between the LF/AF and MF is considered in terms of a matrix transform:

\begin{equation}
\mathbf{C}^{mol}=\mathbf{G}\mathbf{C}^{lab},\label{eq:basic}
\end{equation}

where $\mathbf{C}$ are a set of coefficients that can be used to construct the $\beta_{LM}$ in the LF and MF. Explicitly, 


\begin{equation}
C_{PR}^{LM}(\epsilon,\Delta q)=\sum_{\zeta\zeta'}\mathbb{I}^{\zeta\zeta'}(\epsilon)\Gamma_{PR\Delta q}^{\zeta\zeta'LM}
\end{equation}

for the MF ($\mathbf{C}^{mol}$), and:

\begin{equation}
\bar{C}_{KQS}^{LM}(\epsilon)=\sum_{\zeta\zeta'}\mathbb{I}^{\zeta\zeta'}(\epsilon)\Gamma_{KQS}^{\zeta\zeta'LM}
\label{eq:C-AF}
\end{equation}

for the LF ($\mathbf{C}^{lab}$). We refer to $\mathbf{\Gamma}$ as the ``reduced" channel functions - similar to the channel functions defined previously (Sect. \ref{sec:channel-funcs}), but \textit{without} the inclusion of alignment ($\mathbf{C}^{lab}$) or frame rotation effects ($\mathbf{C}^{mol}$). These are explicitly indexed by all required quantum numbers for the LF and MF definitions (as previously, $\zeta$ denotes all other required indices). Again, full details can be found in Appendix \ref{app:mat-inversion}. Given these, it can be shown for molecules with $D_{nh}$ point group symmetry that a transformation matrix can be written as:

\begin{equation}
\mathbf{G}_{L'M'KS}^{LMP\Delta q}=\mathbf{\Gamma}_{P0\Delta q}^{\zeta\zeta'LM}(\mathbf{\Gamma}_{K0S}^{\zeta\zeta'L^{\prime}M^{\prime}})^{+}
\label{eq:MPinversion}
\end{equation}

Here $()^{+}$ indicates the Moore-Penrose inverse matrix of a reduced channel function, which can be computed numerically. Significantly, the matrix elements are not required for inversion, provided that $\mathbf{C}^{lab}$ is known (e.g. from a measurement), and that the reduced channel functions are computed. Therefore, this method does not provide a route to reconstruction of a full set of matrix elements, but can be used to obtain MFPADs, and has been demonstrated to work for linear ($N_2$, $D_{\infty h}$ symmetry) and 
asymmetric top ($C_{2}H_{4}$, $D_{2h}$ symmetry) examples. Although not as powerful as a complete experiment methodology, this technique is expected to scale more readily to larger problems (more matrix elements), for which a complete matrix element retrieval via fitting may be impossible.

\subsubsection{Numerical implementations\label{sec:numerics-intro}}

\begin{figure}[]
\begin{center}
\includegraphics[width=\textwidth,height=\dimexpr\textheight-4\baselineskip-\abovecaptionskip-\belowcaptionskip\relax,keepaspectratio]{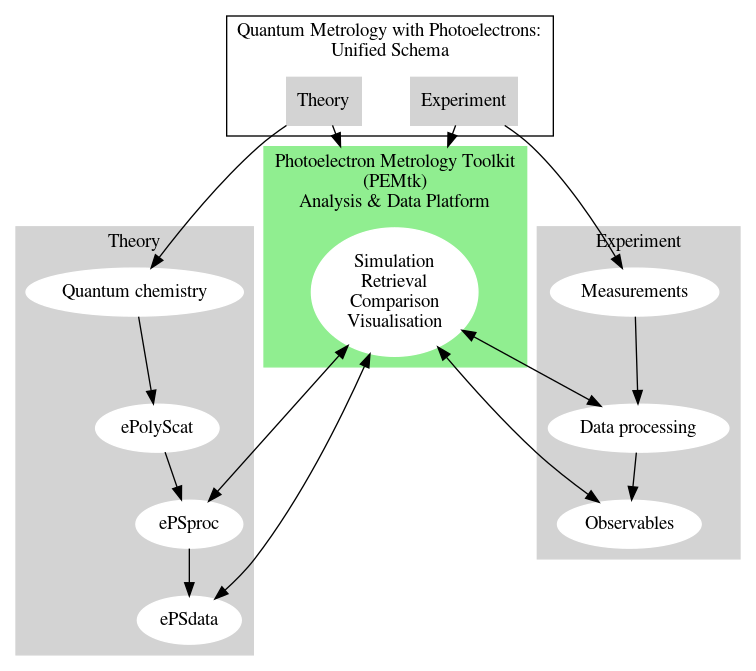}
\caption{Quantum metrology with photoelectrons platform outline.\label{239231}}
\end{center}
\end{figure}

The numerical implementation of the methodologies defined above has been variously implemented in the past, including code in Fortran, C and Matlab, often for specific cases only. Recently a unified Python codebase/ecosystem/platform has been in development to tackle various aspects of photoionization problems, including \textit{ab initio} computations and experimental data handling, and (generalised) matrix element retrieval. The ``Quantum Metrology with Photoelectrons"  platform is briefly introduced here, and is used for the analysis in Sects. \ref{sec:bootstrapping} and \ref{sec:recon-from-MFPADs}. Fig. \ref{239231} shows some of the main tools and tasks/layers. 

The two main components of the platform used herein are:

\begin{itemize}
\item The Photoelectron Metrology Toolkit (PEMtk) codebase \cite{hockett2021PEMtkDocs,hockett2021PEMtkGithub} aims to provide various general data handling routines for photoionization problems. At the time of writing, simulation of observables and fitting routines are implemented, along with some basic utility functions. Further implementation details can be found in Sect. \ref{sec:numerical-notes}, and the \href{https://pemtk.readthedocs.io/en/latest/about.html}{PEMtk documentation} \cite{hockett2021PEMtkDocs}.
\item The ePSproc codebase \cite{ePSprocAuthorea,ePSprocGithub,ePSprocDocs} aims to provide methods for post-processing with \textit{ab initio} radial dipole matrix elements from ePolyScat \cite{Lucchese1986,Gianturco1994,Natalense1999,luccheseEPolyScatUserManual}, or equivalent matrix elements from other sources, including computation of AF and MF observables. Manual computation without known matrix elements is also possible, e.g. for investigating limiting cases, or data analysis and fitting. These routines also provide the backend functionality for PEMtk fitting routines. See Sect. \ref{sec:numerical-notes} for additional notes.
\end{itemize}

Note that, at the time of writing, rotational wavepacket simulation is not yet implemented in the PEMtk suite, and these must be obtained via other codes. 

A Docker-based distribution of various codes for tackling photoionization problems is also available from the \href{https://github.com/phockett/open-photoionization-docker-stacks}{Open Photoionization Docker Stacks} project, which aims to make a range of these tools more accessible to interested researchers \cite{hockettOpenPhotoionizationDocker}.





\section{Reconstruction examples \& recent developments in time-domain measurements\label{sec:Recon}}

\begin{figure}[]
\begin{center}
\includegraphics[width=\textwidth,height=\dimexpr\textheight-4\baselineskip-\abovecaptionskip-\belowcaptionskip\relax,keepaspectratio]{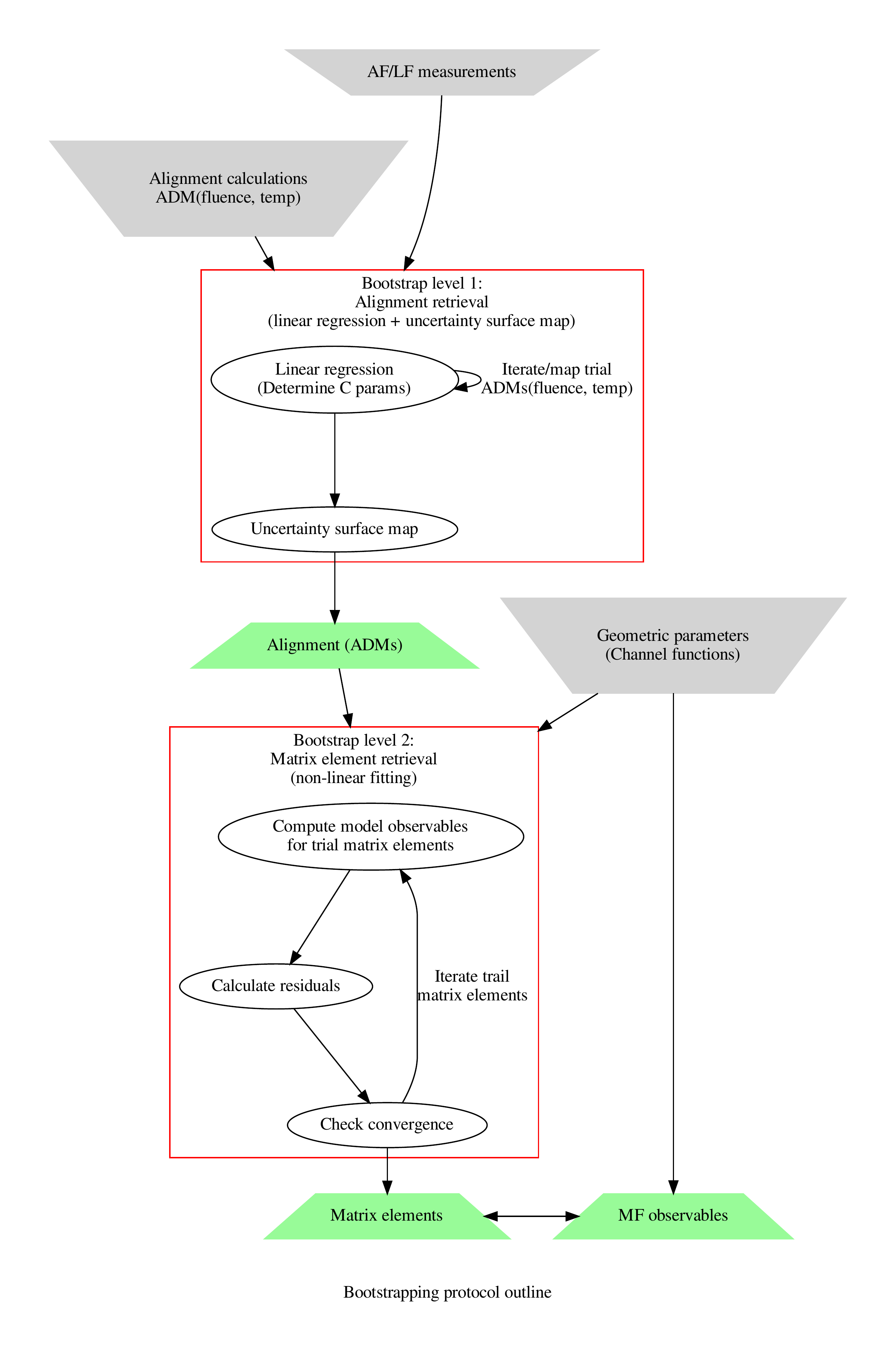}
\caption{Basic bootstrap protocol outline. (See also Fig. \ref{781808} .) Filled shapes indicate (upper) required inputs to the fitting stages (experimental measurements, trial ADMs, and geometric parameters), (lower) the final outputs (ADMs, matrix elements and MF observables).\label{807606}}
\end{center}
\end{figure}

\begin{figure}[]
\begin{center}
\includegraphics[width=\textwidth,height=\dimexpr\textheight-4\baselineskip-\abovecaptionskip-\belowcaptionskip\relax,keepaspectratio]{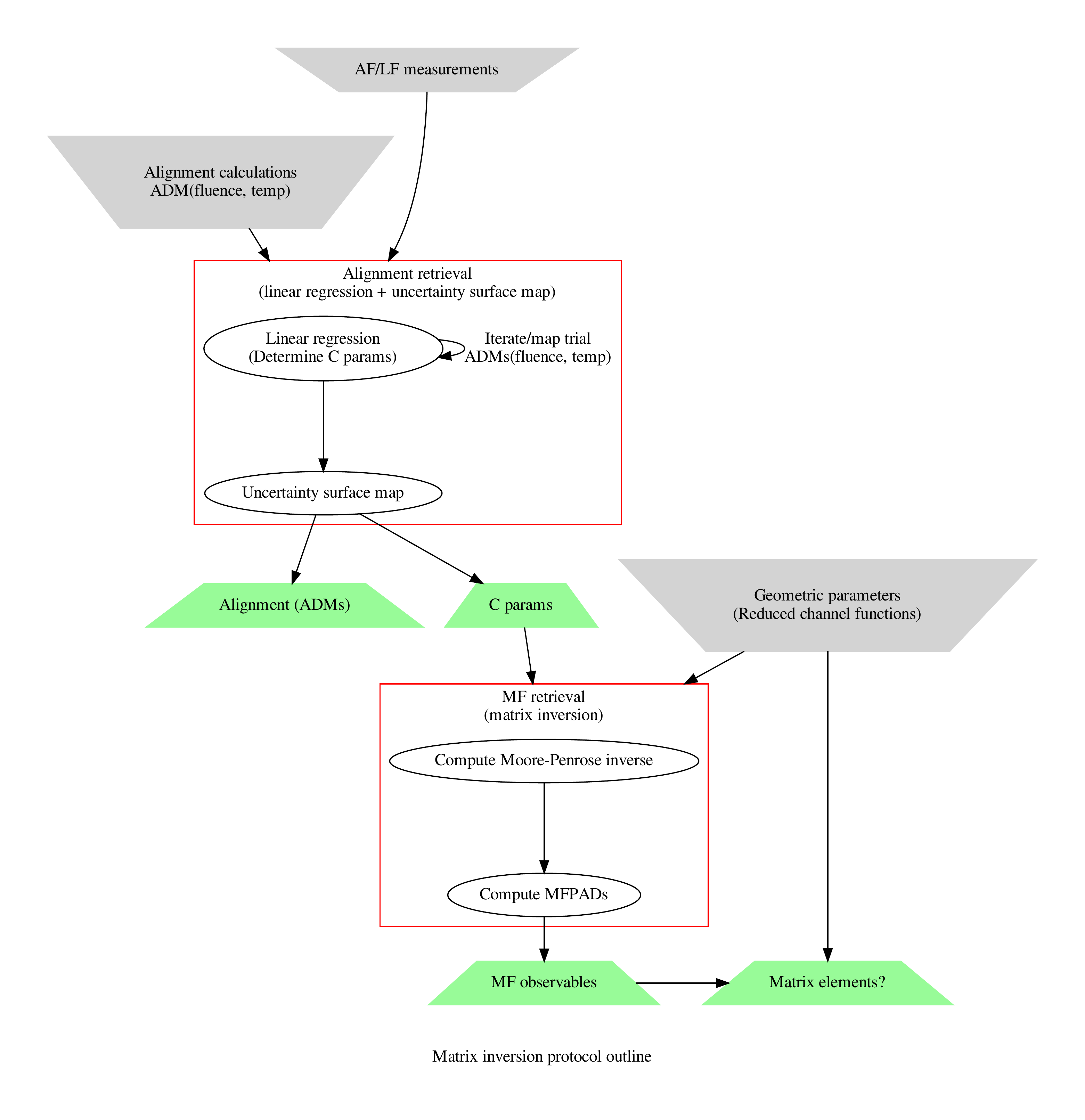}
\caption{Matrix inversion protocol outline. Filled shapes indicate (upper) required inputs to the fitting stages (experimental measurements, trial ADMs, and geometric parameters), (lower) the final outputs (ADMs, and MF observables), matrix elements may also be obtained via further treatment of the MF observables (Sect. \ref{sec:recon-from-MFPADs} ).    Note that the ADM retrieval step is the same as the level 1 bootstrapping stage (Fig. \ref{807606} ).\label{731792}}
\end{center}
\end{figure}

In this section/the remainder of the manuscript, reconstruction of MFPADs from time-domain measurements are considered using two methodologies:

\begin{enumerate}
\item ``Bootstrapping" to the MF \cite{hockett2018QMP1,hockett2018QMP2,marceau2017MolecularFrameReconstruction}. The protocol outline is shown in Fig. \ref{807606}, and discussed in Sect. \ref{sec:bootstrapping}.
\item Matrix reconstruction with Moore-Penrose inversion \cite{gregory2021MolecularFramePhotoelectron}. The protocol outline is shown in Fig. \ref{731792}, and discussed in Sect. \ref{sec:Matrix-inversion-example}.
\end{enumerate}

The techniques are closely related, and both make use of rotational wavepackets (geometrical coherences) to mediate the LF/AF information content of a set of measurements in the time-domain, but differ in the ``directness" of the reconstruction. In the former case, the aim is full matrix element retrieval (Sect. \ref{sec:bootstrap-fidelity}) or, equivalently, continuum density matrix reconstruction (Sect. \ref{sec:den-mat-N2}), and the MFPADs can then be computed from these ``complete" results (Sect. \ref{sec:bootstrap-MFPADs}). In the latter case of matrix reconstruction the MFPADs are determined, essentially, via a transformation matrix (Sect. \ref{sec:Matrix-inversion-example}), and full matrix element retrieval is not necessary. The former is therefore a full quantum state retrieval or quantum tomography, whilst the latter represents a reconstruction of the MF observables which is more akin to classical tomographic methods, albeit with some phase information retained. 
However, the matrix reconstruction method does not require time-consuming data fitting, and should also scale more readily to larger systems (with some caveats), so should be advantageous in problems where only the MFPADs are sought.

\begin{figure}[]
\begin{center}
\includegraphics[width=\textwidth,height=\dimexpr\textheight-4\baselineskip-\abovecaptionskip-\belowcaptionskip\relax,keepaspectratio]{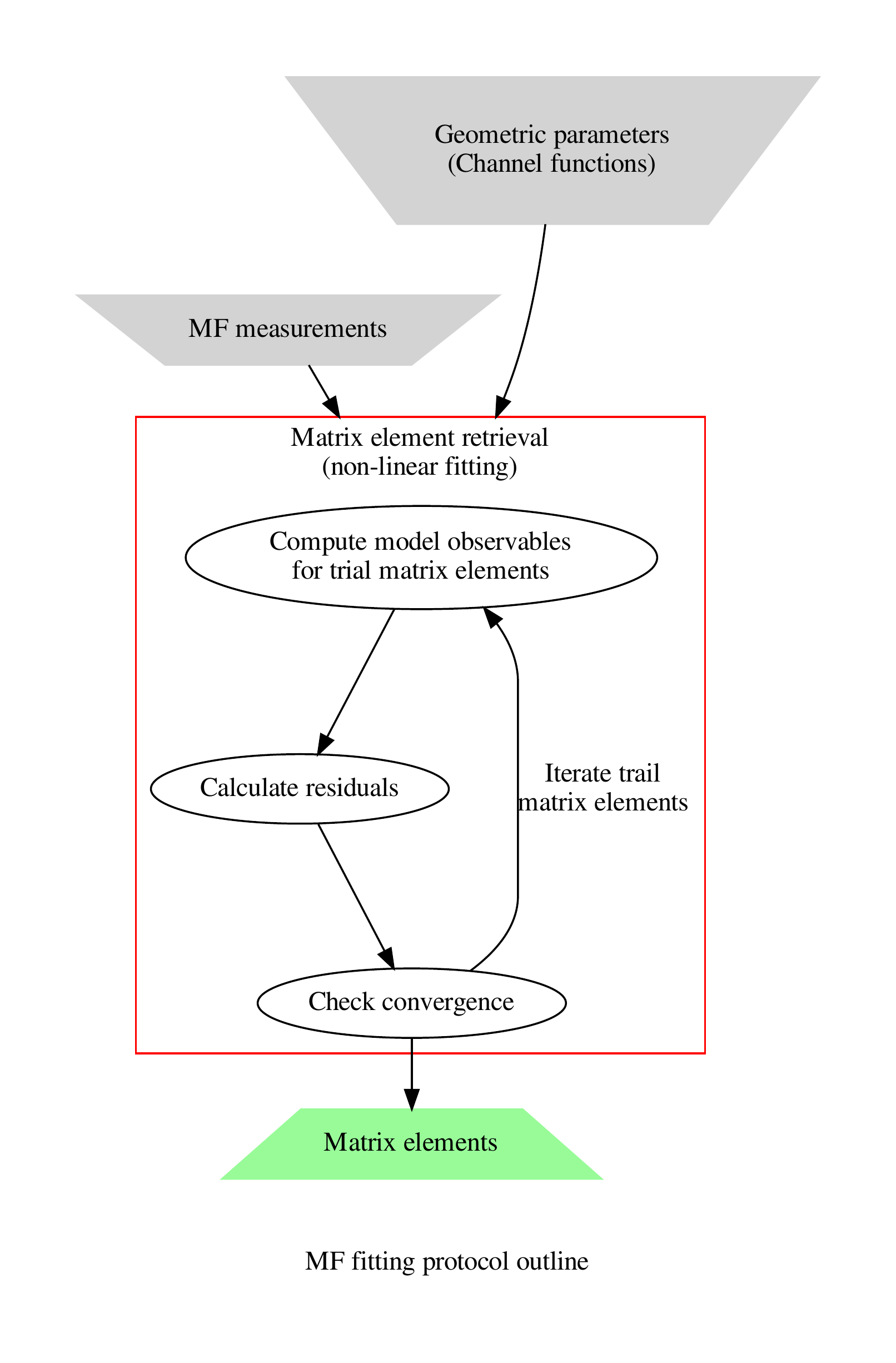}
\caption{MF matrix element retrieval protocol outline. Filled shapes indicate (upper) required inputs to the fitting stages (experimental measurements and geometric parameters), (lower) the final outputs (matrix elements).    Note that this is essentially identical to the level 2 bootstrapping protocol (Fig. \ref{807606} ), except with MF observables as the input. This can also be viewed as a generic model protocol, subject to the given set of measurements and associated channel functions.\label{671760}}
\end{center}
\end{figure}

Additionally, matrix element retrieval from MF observables is briefly addressed, the protocol is outlined in Fig. \ref{671760}, and discussed in Sect. \ref{sec:recon-from-MFPADs}. This protocol is essentially identical to the level 2 bootstrapping case, but with different geometric parameters and input dataset. This provides a route to quantum state reconstruction from direct MF measurements, \textit{or reconstructed MF observables}, hence provides a protocol which can be used to extend the matrix inversion method if desired, albeit with associated information content restrictions.

Finally, it is of note that the MF matrix element retrieval protocol of Fig. \ref{671760} is rather generic, and also forms the basis for other similar methods, e.g. cases where LF measurements are obtained as a function of polarization geometry. The difference in general is the exact form and information content of the input dataset, and the geometric parameters required for the given case (model).


The PEMtk python package \cite{hockett2021PEMtkDocs, hockett2021PEMtkGithub} currently implements the level 2 bootstrapping routines (Fig. \ref{807606}), and MF retrieval (Fig. \ref{671760}), and full computational notebooks, including source data and numerics, for these examples are available online \cite{hockett2022MFreconFigshare}. The matrix inversion technique will be implemented soon, examples shown herein are reproduced from Ref. \cite{gregory2021MolecularFramePhotoelectron}. It is hoped that interested readers will make use of these materials, and that this presentation will encourage other readers to try (and build upon) what has - up until quite recently - been a rather challenging and involved numerical analysis task. A full list of resources is given in Sect. \ref{sec:resources}, and additional numerical notes in Sect. \ref{sec:numerical-notes}.

\subsection{Bootstrapping to the MF example case\label{sec:bootstrapping}}


The ``generalised bootstrapping" methodology proceeds via a multi-stage fitting protocol to obtain both the AF axis distribution (level 1 of the bootstrap, linear fit), and the photoionization matrix elements (level 2 of the bootstrap, non-linear fit). A general overview is given in Fig. \ref{781808}, and a more detailed schematic overview is given in Fig. \ref{807606}. In the simplest case (as illustrated), this proceeds in a direct manner, and a final set of results are obtained. However, further bootstrapping - in the statistical sense - can also be employed to probe the fidelity of the reconstruction with different samples from the source data (typically sub-sets from the measured time-steps and/or $\beta_{LM}$ parameter sub-sets), to further test the robustness of the results, and estimate uncertainties and/or improve upon them. Some of these possibilities are explored below, and further details may also be found in Refs. \cite{hockett2018QMP1,hockett2018QMP2,marceau2017MolecularFrameReconstruction} (and references therein).

\subsubsection{Bootstrapping basics: dataset and setup}

For the example case, synthetic data was used, although the dataset was inspired by Ref. \cite{marceau2017MolecularFrameReconstruction}. In that case, experimental results were obtained via a (dual) pump-probe scheme, where the IR pump pulses prepared a rotational wavepacket in $N_2$, and a time-delayed XUV probe pulse ionized the sample - this scheme is illustrated in Fig. \ref{781808}. Multiple channels were probed in this manner, for a range of time-delays, to provide a dataset of AF-$\beta_{LM}(E,t)$ parameters. Here the focus is on the photoionization step, information content and further investigation of the retrieval routines; this is facilitated by the following choices and data:

\begin{enumerate}
\item The same rotational wavepacket \& molecular axis distribution as obtained experimentally is assumed. Consequently, the 1st (linear) stage of the bootstrapping protocol is not investigated herein. Although this may seem like a drastic omission, in general this stage of the methodology is expected to be quite robust, and this has been demonstrated in various investigations of rotational wavepackets and molecular alignment techniques, see discussion in Sect. \ref{sec:RWPs}.
\item To simulate the observables, photoionization matrix elements from ePolyScat \cite{Lucchese1986, Gianturco1994, Natalense1999, luccheseEPolyScatUserManual} calculations were used. The calculations made use of $N_2$ electronic structure input, computed with Gamess \cite{gamess, Gordon} (RHF/MP2/6-311G, bond length $1.07$~\AA). In the example illustrated herein, matrix elements for $E_{ke}=1~eV$ were used, and are given explicitly in Table \ref{tab:inputMatE}. This, naturally, also provides a means for direct comparison and fidelity analysis of the retrieval protocol (Sect. \ref{sec:bootstrap-fidelity}); the density matrix can also be used for analysis (Sect. \ref{sec:den-mat-N2}). In the current study only ionization of the $3\sigma_g$ orbital (HOMO) is investigated, corresponding to formation of  $X^2\Sigma_{g}^{+}$ ionic ground-state, i.e. $N_2(X^{1}\Sigma^{+}_{g}) \rightarrow N^+_2(X^{2}\Sigma^{+}_{g})$, generically denoted as the $X$-channel, or via the ionizing orbital as $N_2(3\sigma_g^{-1})$, for a more compact notation. 
\item To investigate limitations of the numerical routines, noise and other artifacts can be added to the simulated data. Different sub-sets of the data can also be readily analysed and compared.
\item Finally, it is of note that the numerical implementations are structured as a set of tensors, as close as possible to the formalism given above (Sect. \ref{sec:tensor-formulation}). This provides a means to further investigate the information content of various parts of the problem, and investigate their influence on the retrieval. In particular, these can indicate aspects of the data which may be most sensitive to particular matrix elements, most susceptible to noise and so forth. This is explored in Sect. \ref{sec:bootstrapping-info-sensitivity}
\end{enumerate}

\begin{figure}[]
\begin{center}
\includegraphics[width=\textwidth,height=\dimexpr\textheight-4\baselineskip-\abovecaptionskip-\belowcaptionskip\relax,keepaspectratio]{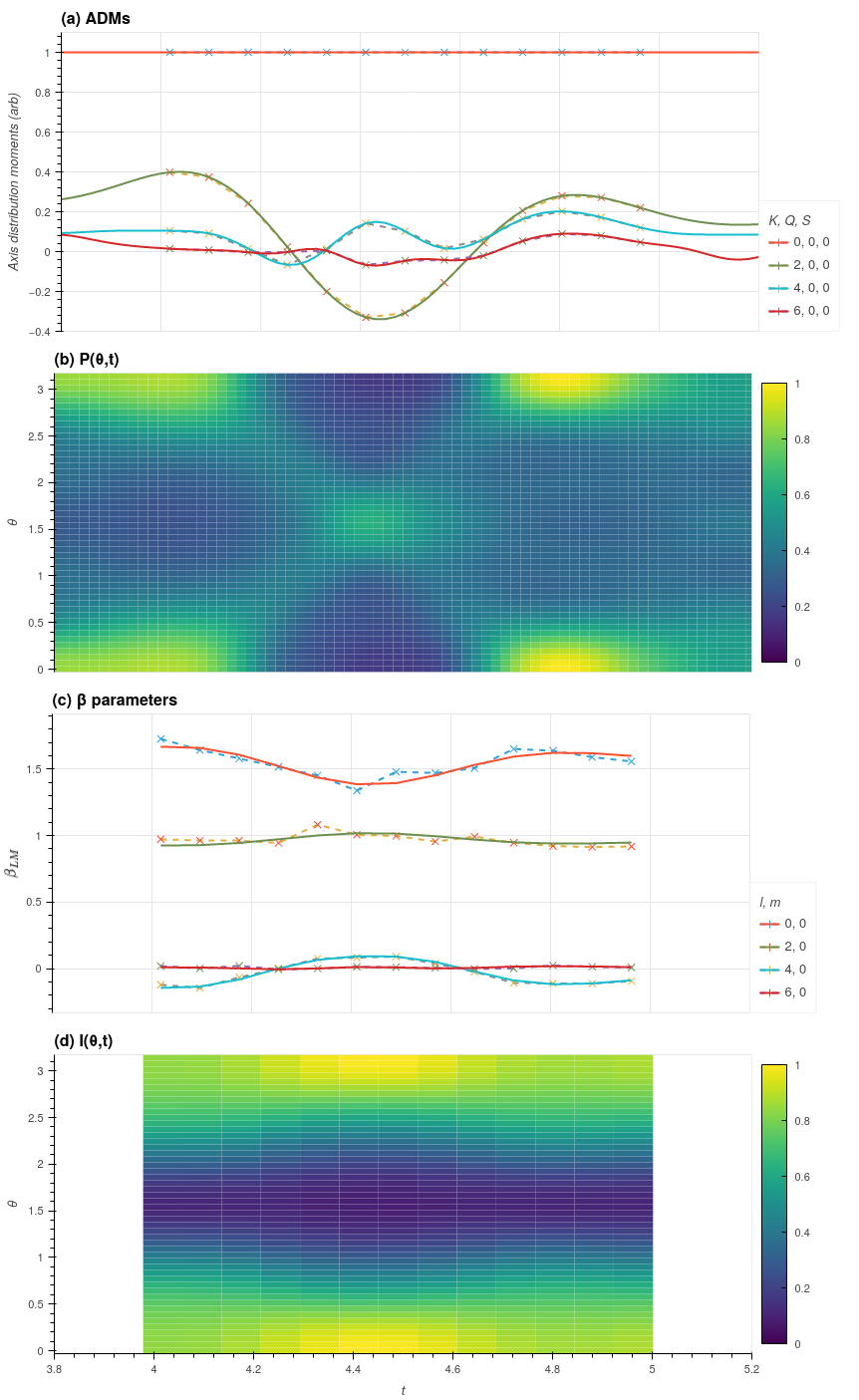}
\caption{Simulated data for N2: (Upper pair) (a) Axis distribution moments (ADMs, \(A^{K}_{Q,S}(t)\) ) and (b) corresponding axis distribution maps \(P(\theta, t)\) . (Lower pair) (c) Calculated observables, \(\beta_{L,M}\) parameters (solid lines) and sub-selected data points (dashed lines) with random noise added (up to 10\%) as used for the reconstruction protocol, (d)  full theoretical \(I(\theta,t)\) maps (ideal case with no added noise).\label{720080}}
\end{center}
\end{figure}

The sample dataset used for the results presented herein is illustrated in Figure \ref{720080} and, as mentioned previously, full numerical data can be accessed online (Sect. \ref{sec:resources}). In this case only a 1~ps subset of the simulation data was used, over the main revival feature. Figure \ref{720080} shows both the full simulation results and the sub-selected points (13 data points) with random noise added (up to 10\%), the latter is used as the input dataset for matrix element retrieval in the following sections. 

\begin{figure}[]
\begin{center}
\includegraphics[width=\textwidth,height=\dimexpr\textheight-4\baselineskip-\abovecaptionskip-\belowcaptionskip\relax,keepaspectratio]{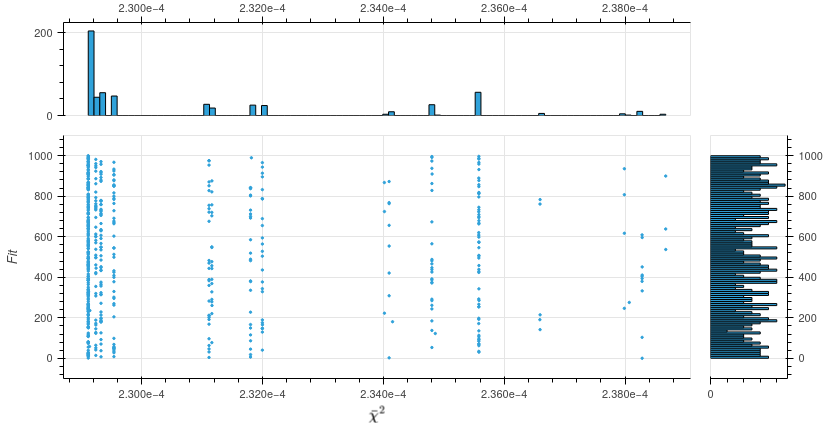}
\caption{Sample results from 1000 fits with noisy data, showing all fits with \(\bar{\chi}^2 < 2.4\times10^{-4}\) .\label{509194}}
\end{center}
\end{figure}

\begin{figure}[]
\begin{center}
\includegraphics[width=\textwidth,height=\dimexpr\textheight-4\baselineskip-\abovecaptionskip-\belowcaptionskip\relax,keepaspectratio]{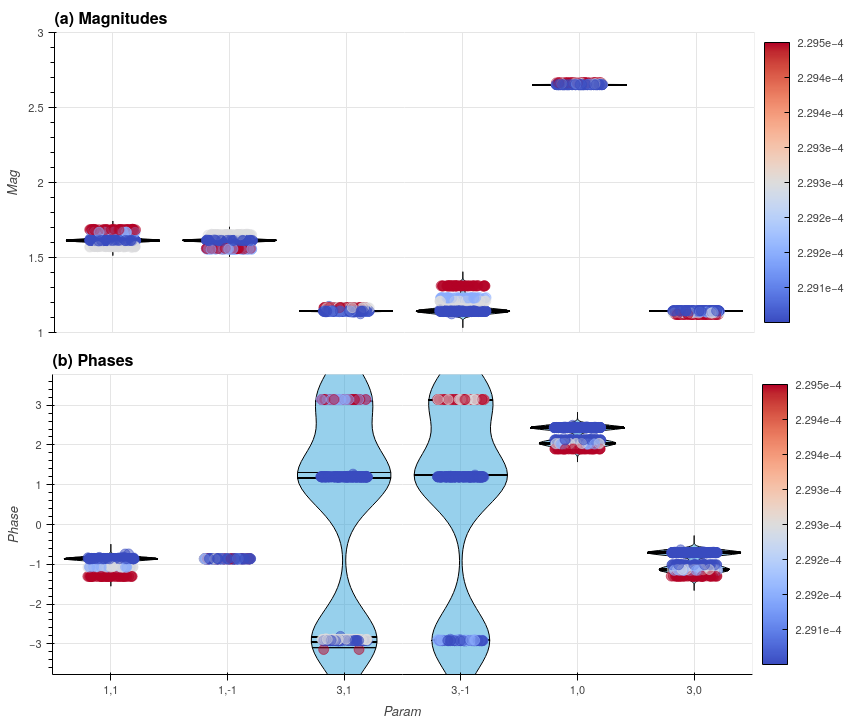}
\caption{Statistics for the 349 best fit results ( \(\bar{\chi}^2 < 2.3 \times 10^{-4}\) ). Plots show magnitudes (top) and phase (bottom) parameters, labelled by \((l,m)\) , with spread proportional to the number of results and colour-mapped by \(\bar{\chi}^2\) . Note the clustering by \(\bar{\chi}^2\) , where each cluster constitutes a viable set of matrix elements (particularly in \(\phi_{3,\pm1}\) terms) , and the lack of variation for \(\phi_{1,-1}\) , which was fixed as a reference.\label{494229}}
\end{center}
\end{figure}

\begin{figure}[]
\begin{center}
\includegraphics[width=\textwidth,height=\dimexpr\textheight-4\baselineskip-\abovecaptionskip-\belowcaptionskip\relax,keepaspectratio]{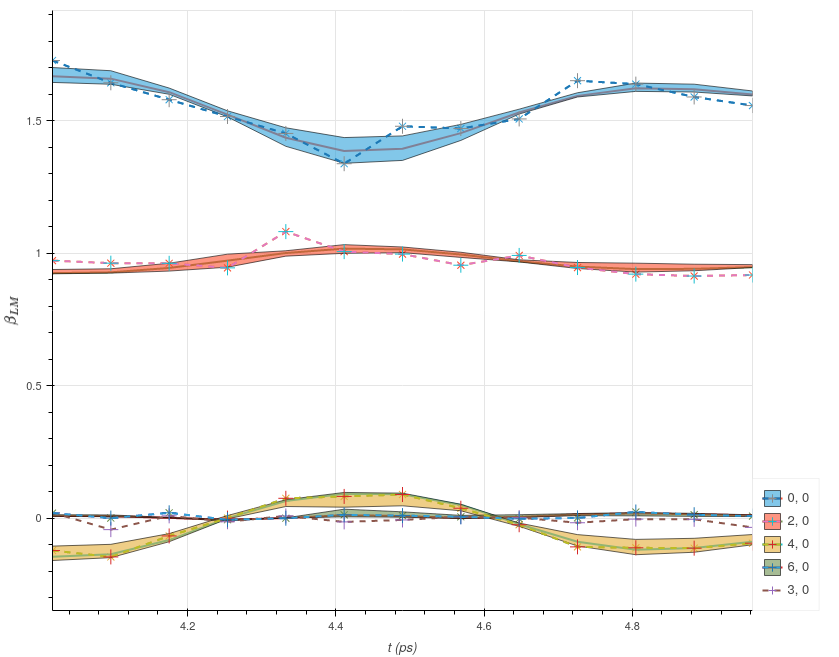}
\caption{Best fit results (minimum \(\bar{\chi}^2\) only) with uncertainty estimation. Plots show original simulation (solid lines), simulation with noise used for fitting (dashed lines with `+' markers) and mean values \(\pm1\sigma\) (filled bands). Note that (spurious) \(\beta_{3,0}\) data is present in the input noisy data, but not in the outputs, as defined by symmetry.\label{743962}}
\end{center}
\end{figure}

\subsubsection{Bootstrapping basics: fit setup}

To fit the simulated data, a set of fitting parameters were defined in (magnitude, phase) form. In this case, the parameter set was simply defined from the input matrix elements, resulting in 6 complex radial matrix elements defining the problem. This equates to $2n$ parameters to determine in magnitude-phase form, or, if only relative phases are to be obtained, $2n-1$ parameters, where the remaining phase is defined as a reference, either pre or post fit. Similarly, if only relative magnitudes are obtained, the total number of parameters is reduced to $2n-2$. In general this step in the protocol may require manual configuration for the problem at hand, based on the symmetry of the system. The input matrix elements are listed in Table \ref{tab:inputMatE}. Of particular note in this case is that only $l=1,3$ partial-waves are present, and the two continuum symmetries - $\sigma_u$ and $\pi_u$ - which correlate with $m=0$ and $m=\pm1$ respectively, and correspond to parallel and perpendicular polarization geometries in the molecular frame. Furthermore, the $m=\pm1$ pairs are identical in this case. Because the $l,\pm m$ indices are unique, they will be used generally as short-hand labels for the matrix elements, e.g. $|1,\pm1\rangle$ corresponds to the $|\zeta\rangle = |\pi_u,1,\pm 1,\mp 1 \rangle$ continuum component (full matrix element), and $\phi_{1,\pm1}$ corresponds to the phase of this component.

In the current demonstration case:

\begin{itemize}
\item A standard Levenberg-Marquardt least-squares minimization was used.
\item 11 of the 12 parameters were allowed to float freely (i.e. no symmetry relations were imposed a priori), with phases bounded to $-\pi\leq\phi\leq\pi$. 
\item $\phi_{1,-1}$ was set as a reference phase. 
\item To gain insight into the efficiency of the fitting routine, the validity of the methodology and the uniqueness of the fit results (Sect. \ref{sec:bootstrap-fidelity}), 1000 fits were performed on the same dataset, each seeded with randomised parameter values. 
\item Convergence criteria were set as a minimum gradient reduction of $1\times10^{-8}$, 
but fits were also capped by the total number of function evaluations. Whilst this prevents run-away cases, it also constitutes a second convergence criteria.
\item Results herein are categorized by reduced $\bar{\chi}^2$ values ($\bar{\chi}^2=\chi^2/(N-N_{vars})$, where $N$ is the number of data points, and $N_{vars}$ the number of parameters) unless otherwise noted.
\item Further details of the codebase and numerical implementation can be found in Sect. \ref{sec:numerical-notes}.
\end{itemize}

This methodology amounts to a statistical sampling of the solution hyperspace, which may be expected to contain some local minima in general high-dimensional cases.  Running in parallel on 20 (logical) cores of an AMD Threadripper 2950X based workstation, this required ~5GB of RAM and took on the order of 2 hours (note that the time per fit cycle had large variance, since convergence time depends on the start parameters); further benchmarks for the current codebase can be found in \href{https://pemtk.readthedocs.io/en/latest/index.html}{the PEMtk documentation} \cite{hockett2021PEMtkDocs}.


\subsubsection{Bootstrapping basics: fit results}

Results are summarised broadly in Fig. \ref{509194}, which shows the best fit results over the fits. A few (tentative) conclusions can be quickly drawn from this:

\begin{enumerate}
\item The results appear to indicate a well-defined, single global minima was found, with $\bar{\chi}^2 = 2.291\times10^{-3}$, and 203 fits (20\%) obtained this value.
\item There are additional close-lying bands with $\bar{\chi}^2 < 2.3\times10^{-3}$. In the test case, 349 fits (35\%) obtained this region of the fitting space. These results may be due to local minima, or simply fits which didn't quite converge for numerical reasons (e.g. reached the upper limit of iterations in fitting protocol); further exploration to identify whether these sub-sets of parameters are significantly different is therefore required.
\item There are multiple bands of fits which converged to higher $\bar{\chi}^2$; since there is a large step to the first of these ($\bar{\chi}^2 = 2.312\times10^{-3}$), and they appear with much lower frequency, these are most likely indicative of local minima.

\end{enumerate}

\subsubsection{Bootstrapping basics: retrieved parameter sets \& fidelity\label{sec:bootstrap-fidelity}}

To drill down into the results, the values and clustering of the retrieved parameter sets can be investigated, as well as the overall quality of the fitting. Fig. \ref{494229} illustrates the spread in retrieved parameters over the best fit region, and Fig. \ref{743962} shows the overall quality of the fits in this region. For the purposes of discussion, each band observed in Fig. \ref{509194} will be described as a $\bar{\chi}^2$ cluster or group, assumed to equate to a viable set of retrieved parameters, and the best such set should equate to the "true" values, i.e. the photoionization matrix elements, if the methodology is successful and there is a well-defined global minima. Again, some tentative conclusions can be drawn:

\begin{enumerate}
\item The results for the magnitudes show very little difference within each $\bar{\chi}^2$ cluster, and closer inspection reveals that values between clusters are typically within 5\%, with the exception of the $|3,\pm1\rangle$ case which shows a larger spread. This indicates a good agreement in all results.
\item The phase results broadly fall into two classes, (a) cases which show good agreement within a $\bar{\chi}^2$ cluster, and (b) cases which have two values per $\bar{\chi}^2$ cluster. Additionally there is a larger spread in the values between clusters in phase space, and this is again most apparent for the $|3,\pm1\rangle$ case, which spans the full $-\pi\leq\phi\leq\pi$ phase range. Note that there is no spread in the $\phi_{1,-1}$ (phase) parameter, this was fixed as a reference phase in the current case, with $\phi=-0.861$.
\item The well-defined grouping here indicates that the retrieved parameter sets are unique for each $\bar{\chi}^2$ cluster, and each case can be regarded as a distinct, viable, parameter set to be further investigated; a priori it is expected that the lowest $\bar{\chi}^2$ should be the true result, but in practice this may not always hold depending on the data.
\item The parameters with split/pairs of values are indicative of cases where there is an insensitivity in the fit. In this case, the $|3,\pm1\rangle$ case shows an insensitivity to the sign of the phase, which appears as a splitting due to the $\mod(\pi)$ nature of the result. There is a slight bias in the results towards one of the pair in each case, though this may be a statistical artifact. (This type of effect will be discussed further below.)
\end{enumerate}

\begin{figure}[]
\begin{center}
\includegraphics[width=\textwidth,height=\dimexpr\textheight-4\baselineskip-\abovecaptionskip-\belowcaptionskip\relax,keepaspectratio]{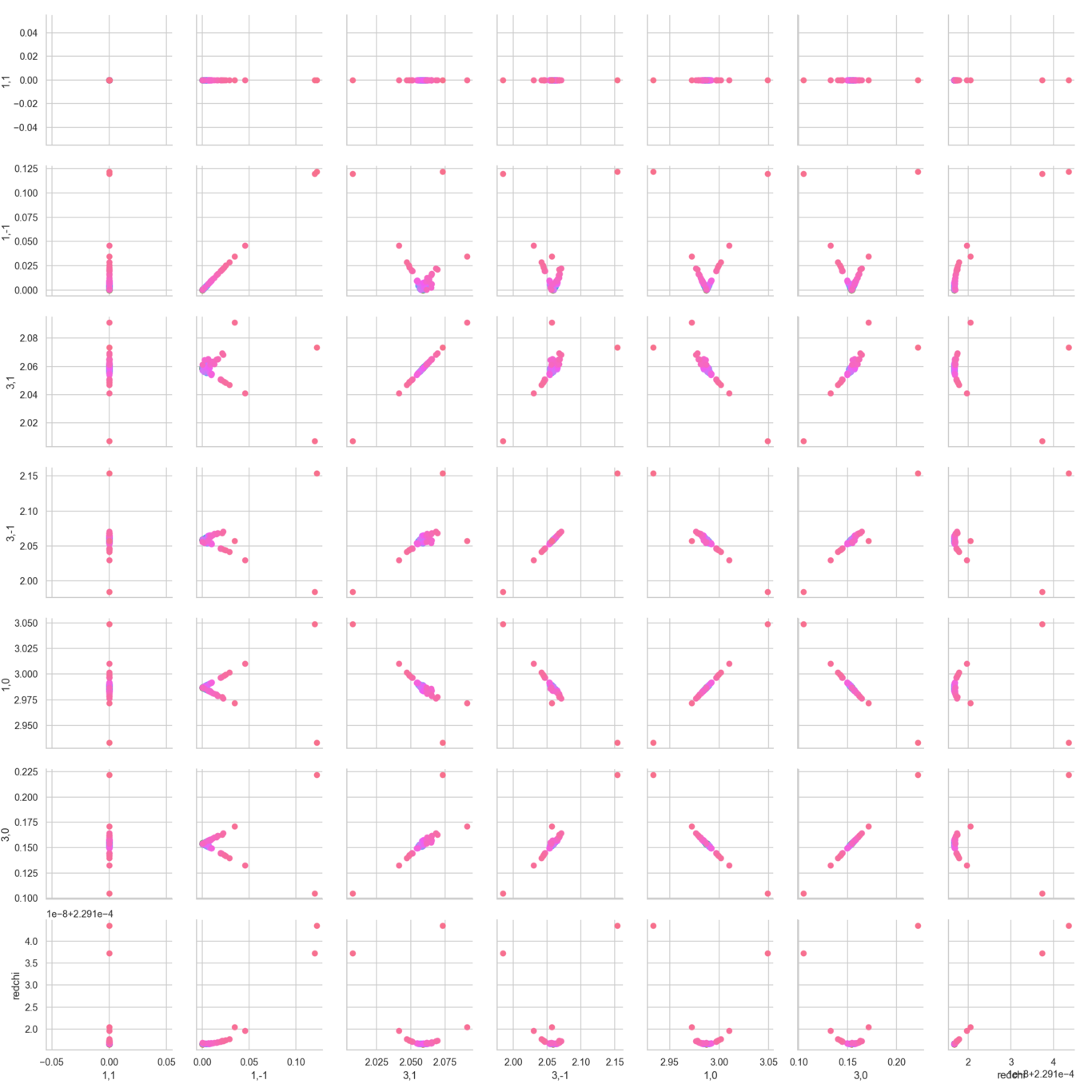}
\caption{Example correlation (pair) plot for retrieved phases. Note in this case the reference phase (top left panel) has been set to 0 to illustrate general patterns in the retrieved parameter sets. In general, the curvature indicates how well-defined a given parameter is, whilst V-patterns indicate cases with \(\pm\phi\) pairs. The bottom row and final column on the right show patterns of the parameters with respect to \(\bar{\chi}^2\) .\label{888108}}
\end{center}
\end{figure}

To further investigate the parameter sets, a correlation pair/matrix representation can be used. For the phases, an example is shown in Fig. \ref{888108}. In this plot each parameter is shown as a function of all other parameters for each set (fit), and coloured by $\bar{\chi}^2$. This results in a rather complex, but informative, representation. In this example the phases were ``corrected", with the reference value set to zero in order to illustrate general trends (i.e. as would typically be the case when an absolute reference value is not known a priori), which results in a zero spread for this parameter (top left panel in Fig. \ref{888108}). Additionally, the ranges are wrapped to $\mod(\pi)$, keeping positive values only, thus collapsing the $\pm$ pairs observed in Fig. \ref{494229} for $\phi_{3,\pm1}$. The bottom row and final column on the right of Fig. \ref{888108} show patterns of the parameters with $\bar{\chi}^2$ value. Of particular note here are the curves visible in most cases: these indicate the curvature of the $\bar{\chi}^2$ hypersurface with respect to the given coordinate (parameter), where a steeper curve indicates how well defined a given parameter is. The remaining correlation plots indicate the spread of individual parameters (diagonal panels), and parameter-parameter correlations (off-diagonal panels). In general good clustering is observed for the lowest $\bar{\chi}^2$ values, 
consistent with the expectations from Fig. \ref{494229}, with more spread apparent in higher-valued results. For the phases with $\pm$ pair splittings - for example the 2nd column in Fig. \ref{888108} - characteristic V shapes are observed, fanning out from the best-fit cluster as a function of $\bar{\chi}^2$; these correspond to cases where the $\pm$ pairs are not wrapped (corrected) to the same value, hence indicate larger uncertainties in these parameter sets. Some outlier parameter sets can also be seen in this plot, for instance the single points visible to the right of the panels in the 2nd column of Fig. \ref{888108}, which are correlated with larger $\bar{\chi}^2$ values (bottom panel of the column).

To investigate the fidelity of the results in this test case the best parameter set(s) can be compared with the input matrix elements, and results are presented in Table \ref{tab:matE}. In this comparison, uncertainties on the retrieved parameters were estimated statistically, based on the standard deviation in the best fit results for the lowest $\bar{\chi}^2$ group; in this case it is of note that no ``experimental" uncertainties were included in the fitting, so the estimates herein indicate purely the error in the retrieval procedure with noisy data (without noise, essentially perfect results are found in this case). 

\begin{itemize}
\item The final results are generally quite close to the inputs. In this case, the 10\% random noise essentially translates to a similar retrieval uncertainty in the magnitudes, although there is also a notable decrease in accuracy with the $|3,\pm1\rangle$ case, and significant increase in accuracy for the $\sigma_u$ cases.
\item The absolute phase values appear to be quite far off in some cases, but the phase corrected values typically look quite reasonable; this is consistent with a lack of sensitivity in the test dataset to the sign of the phases (as discussed elsewhere), but the remapped phases ($0:\pi$) are generally precise. Additionally, the ``absolute" phases given are averaged over all $\pm$ pairs, hence end up quite close to zero, whilst the ``corrected" phases are remapped before averaging. In general the former is, of course, not desirable, but serves to illustrated the type of issue that may arise in a retrieval protocol.
\item With the exception of $|3,0\rangle$ and $|1-1\rangle$, all remapped parameters are within 10\% of the reference values.
\item For the $\phi_{3,0}$ it appears that the value is not well-defined in the data, and is different from the reference value by $\sim\pi/4$.
\item For $\phi_{1-1}$ the absolute error is, indeed, small, but since this is close to zero, and should be defined as zero in the phase corrected case, the percentage error appears large, although the absolute value is close to the true value.
\item The doubly-degenerate $|\pi_u,l,\pm1\rangle$ cases are found to have approximately the same magnitudes and phases in the free fit. In this case this is expected from the input values, hence indicates a good fit result. In general, if known \textit{a priori}, such constraints can also be included in the fitting protocol, and should result in faster and more precise results where there are significant symmetry constraints on the results.
\item The differences between the data and reference values are much larger than the standard deviation of the fits. This is indicative of a good (close/singular) batch of fits, with a unique global minimum, but reveals that the results obtained are not perfect - as expected for noisy data. Adding more data-points to the fit, and/or using higher fidelity data would help in this case.
\item Alternative uncertainty estimations can be obtained from the individual parameter set results, based on the curvature of the $\chi^2$ hyperspace w.r.t. each parameter, or w.r.t. to all other parameters \cite{HockettThesis,Bevington1992}. In testing, only the first approach was considered, making use of values returned by the \href{https://lmfit.github.io/lmfit-py/fitting.html#uncertainties-in-variable-parameters-and-their-correlations}{lmfit library routines} \cite{LMFITDocumentation} from inversion of the Hessian matrix (see Sect. \ref{sec:numerical-notes}); however, this value was not found to be useful in many cases here, with relative errors into the thousands of \%, likely due to the strongly-correlated nature of the fit. The 2nd approach has been used previously, but is significantly more time-consuming (each test value necessitates a refitting of all other parameters), hence statistical uncertainties may represent the best approach for fast and robust estimations.
\item As noted above, although noise was added to the simulated data prior to fitting, experimental/data uncertainties were not included in the fitting. In general these may be expected to be relatively small for $\beta$ parameters obtained via imaging-type experiments with good S:N (for instance, the results of Refs. \cite{HockettThesis,marceau2017MolecularFrameReconstruction}), but may remain significant for absolute count-rates (i.e. $\beta_{0,0}$), and consequently are expected to map to larger uncertainties in the absolute magnitudes. However, fitting to count-rate normalised data (angular distributions only) can mitigate this effect, as was explored in the original demonstration of the technique \cite{marceau2017MolecularFrameReconstruction}, which returned accurate \textit{relative} magnitudes and MFPADs for the $X$-state with experimental uncertainties typically $<10\%$.
\end{itemize}

\begin{table}
\centering
\begin{tabular}{llllrrrl}
\toprule
   &   &    &    &          comp &     m &      p & labels \\
Cont & l & m & $\mu$ &               &       &        &        \\
\midrule
PU & 1 & -1 &  1 &  1.163-1.354j & 1.785 & -0.861 &   1,-1 \\
   &   &  1 & -1 &  1.163-1.354j & 1.785 & -0.861 &    1,1 \\
   & 3 & -1 &  1 & -0.803-0.017j & 0.803 & -3.120 &   3,-1 \\
   &   &  1 & -1 & -0.803-0.017j & 0.803 & -3.120 &    3,1 \\
SU & 1 &  0 &  0 & -2.317+1.359j & 2.686 &  2.611 &    1,0 \\
   & 3 &  0 &  0 &  1.106-0.087j & 1.109 & -0.079 &    3,0 \\
\bottomrule
\end{tabular}

\caption{\label{tab:inputMatE}Input matrix elements for the simulated data, from ePolyScat calculations for $N_2$ at 1~eV, (comp)lex values, and in (m)agnitude and (p)hase form. Note that the continuum symmetry is split into PU ($\pi_u$) and SU ($\sigma_u$) components, correlated with $m=\pm1,\mu=\mp1$ and $m=0,\mu=0$ respectively, i.e. perpendicular and parallel MF components; for brevity the matrix elements will be labelled simply by $|l,m\rangle$ in the text in general.}
\end{table}
\begin{table}
\centering
\begin{tabular}{lllrrrl}
\toprule
   &   & Type &     m &      p &    pc & labels \\
Cont & l & m &       &        &       &        \\
\midrule
PU & 1 & -1 & 1.614 & -0.859 & 0.000 &   1,-1 \\
   &   &  1 & 1.614 & -0.861 & 0.005 &    1,1 \\
   & 3 & -1 & 1.143 &  0.144 & 2.059 &   3,-1 \\
   &   &  1 & 1.142 &  0.143 & 2.059 &    3,1 \\
SU & 1 &  0 & 2.653 &  2.357 & 2.987 &    1,0 \\
   & 3 &  0 & 1.144 & -0.785 & 0.154 &    3,0 \\
\bottomrule
\end{tabular}

\caption{\label{tab:aggMatE}Retrieved values for (m)agnitudes, (p)hases and ``corrected" phases (pc).}
\end{table}
\begin{table}
\centering

\begin{tabular}{lllrrrrrr}
\toprule
   &          & l,m &     1,-1 &      1,1 &      3,-1 &       3,1 &       1,0 &       3,0 \\
Type & Source & dType &          &          &           &           &           &           \\
\midrule
m & mean & num &    1.614 &    1.614 &     1.143 &     1.142 &     2.653 &     1.144 \\
   & ref & num &    1.785 &    1.785 &     0.803 &     0.803 &     2.686 &     1.109 \\
   & diff & \% &   10.581 &   10.600 &    29.733 &    29.715 &     1.264 &     3.030 \\
   &          & num &   -0.171 &   -0.171 &     0.340 &     0.339 &    -0.034 &     0.035 \\
   & std & \% &    0.129 &    0.136 &     0.267 &     0.265 &     0.002 &     0.007 \\
   &          & num &    0.002 &    0.002 &     0.003 &     0.003 &     0.000 &     0.000 \\
   & diff/std & \% & 8190.232 & 7766.406 & 11153.884 & 11193.713 & 79089.050 & 44706.225 \\
p & mean & num &   -0.859 &   -0.861 &     0.144 &     0.143 &     2.357 &    -0.785 \\
   & ref & num &   -0.861 &   -0.861 &    -3.120 &    -3.120 &     2.611 &    -0.079 \\
   & diff & \% &    0.193 &    0.000 &  2264.893 &  2274.558 &    10.797 &    89.979 \\
   &          & num &    0.002 &    0.000 &     3.265 &     3.264 &    -0.254 &    -0.706 \\
   & std & \% &    1.636 &    0.000 &  1249.292 &  1255.151 &     5.709 &    17.101 \\
   &          & num &    0.014 &    0.000 &     1.801 &     1.801 &     0.135 &     0.134 \\
   & diff/std & \% &   11.803 &      inf &   181.294 &   181.218 &   189.126 &   526.167 \\
pc & mean & num &    0.000 &    0.005 &     2.059 &     2.059 &     2.987 &     0.154 \\
   & ref & num &    0.000 &    0.000 &     2.259 &     2.259 &     2.811 &     0.782 \\
   & diff & \% &      nan &  100.000 &     9.740 &     9.737 &     5.893 &   407.124 \\
   &          & num &    0.000 &    0.005 &    -0.201 &    -0.200 &     0.176 &    -0.628 \\
   & std & \% &      nan &  243.707 &     0.257 &     0.461 &     0.227 &     4.433 \\
   &          & num &    0.000 &    0.013 &     0.005 &     0.010 &     0.007 &     0.007 \\
   & diff/std & \% &      nan &   41.033 &  3791.665 &  2109.989 &  2599.980 &  9183.815 \\
\bottomrule
\end{tabular}

\caption{\label{tab:matE}Retrieved values and statistics for (m)agnitudes, (p)hases and ``corrected" phases (pc). For each type and parameter the mean result from the fits is given, along with the standard deviation, and a comparison to the (ref)erence computational data, with absolute numeric (`num') and percentage differences as defined by the `dType' column. Finally, the value of the difference/standard deviation is given as a rough metric for veracity.}
\end{table}

\begin{figure}[]
\begin{center}
\includegraphics[width=\textwidth,height=\dimexpr\textheight-4\baselineskip-\abovecaptionskip-\belowcaptionskip\relax,keepaspectratio]{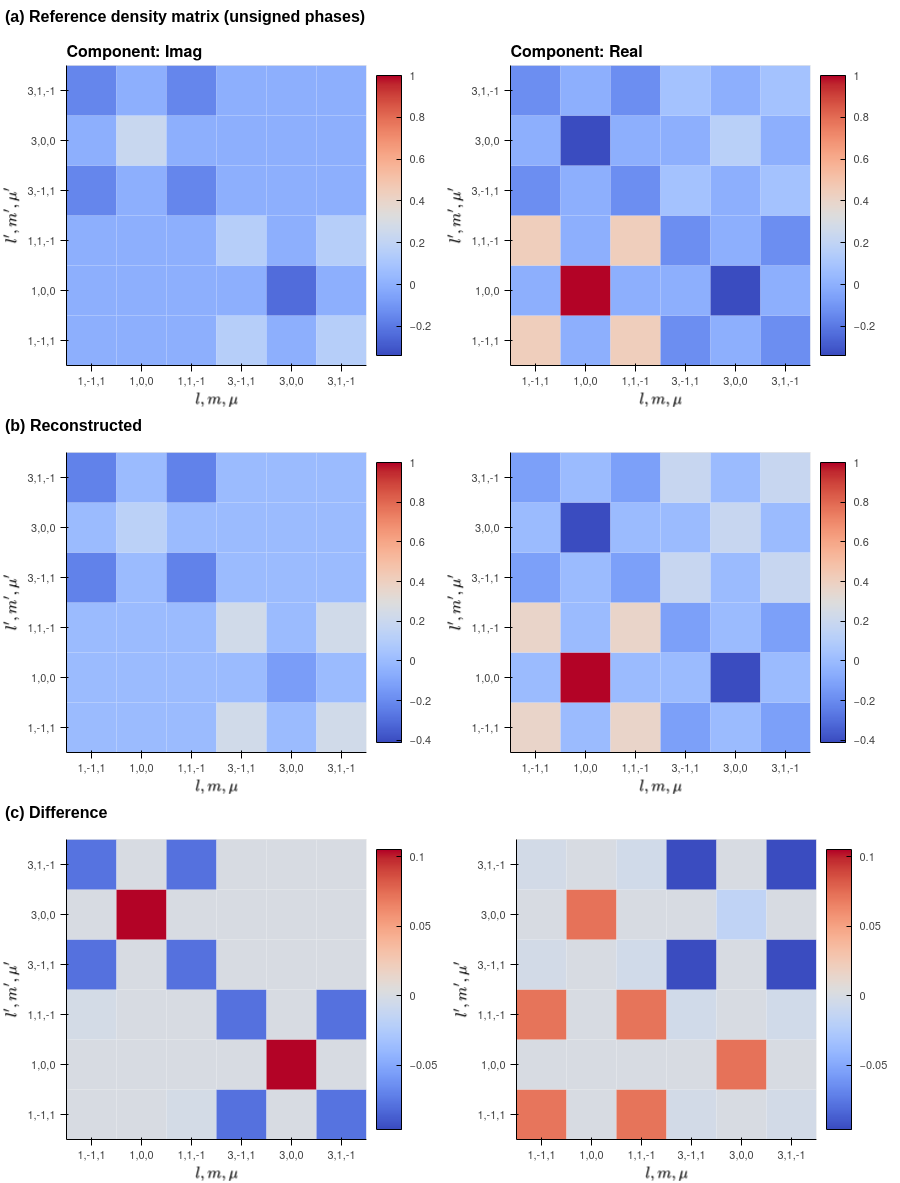}
\caption{Density matrix representations for the continuum. (top row) From reference matrix elements, (middle row) from the retrieved matrix elements, (bottom row) differences. Each row shows the elements for the (imag)inary and real components. Density matrix values (complex) were renormalised to unity by the maximum before plotting ( \(|1,0,0\rangle \langle1,0,0|\) element), and these representations show the case for unsigned phases; note also the rescaled colourmap for the difference plots.\label{998904}}
\end{center}
\end{figure}

\subsubsection{Bootstrapping basics: density matrix representation\label{sec:den-mat-N2}}

An alternative test of fidelity can be investigated via a density matrix representation of the results, this is shown in Fig. \ref{998904}. These plots show the real and imaginary components of the density matrices for the ionization continuum (as defined in Sect. \ref{sec:density-mat-basic}, Eqn. \ref{eqn:radial-density-mat}), normalised to the maximum (complex) value. Overall the agreement is good for the real values (right column of Fig. \ref{998904}), as expected from the values in Table \ref{tab:matE}, and the differences (bottom row) are typically $<10~\%$. Similar results are observed for the imaginary values (left column), which are shown for the unsigned phase case only (corresponding to phases set `pc' as per Table \ref{tab:matE}). For the signed phase case (not shown, but corresponding to phases set `p' as per Table \ref{tab:matE}) the loss of signs leads to larger differences in the off-diagonal density terms, and the possibility of phase flips, which 
can result in inverted patterns visible in the imaginary part of the density matrix. However, as discussed elsewhere (e.g. Sect. \ref{sec:bootstrap-fidelity}), symmetry constraints may render the continuum insensitive to the sign of the phases, and the retrieved matrix elements may therefore still present a high-fidelity reconstruction of the continuum.

Perhaps more interesting/useful in the density matrix representation is the visualisation of the phase relations between the photoionization matrix elements (the off-diagonal density matrix elements), and the ability to quickly check the overall pattern of the elements, hence confirm that no phase-relations are missing and orthogonality relations are fulfilled, or that very different patterns/sets of matrix elements were retrieved. Furthermore, the density matrix elements also provide a complete description of the photoionization event, and makes clear the equivalence of the ``complete" photoionization experiments (and associated continuum reconstruction methods) with quantum tomography schemes \cite{MauroDAriano2003}. It can be used as the starting point for further analysis based on density matrix techniques - this is discussed, for instance, in Ref. \cite{BlumDensityMat}, and can also be viewed as a bridge between traditional methods in spectroscopy and AMO physics, and more recent concepts in the quantum information sciences (e.g. Refs.  \cite{Tichy2011a,Yuen-Zhou2014}).


\subsubsection{Bootstrapping basics: obtaining MFPADs\label{sec:bootstrap-MFPADs}}

\begin{figure}[]
\begin{center}
\includegraphics[width=\textwidth,height=\dimexpr\textheight-4\baselineskip-\abovecaptionskip-\belowcaptionskip\relax,keepaspectratio]{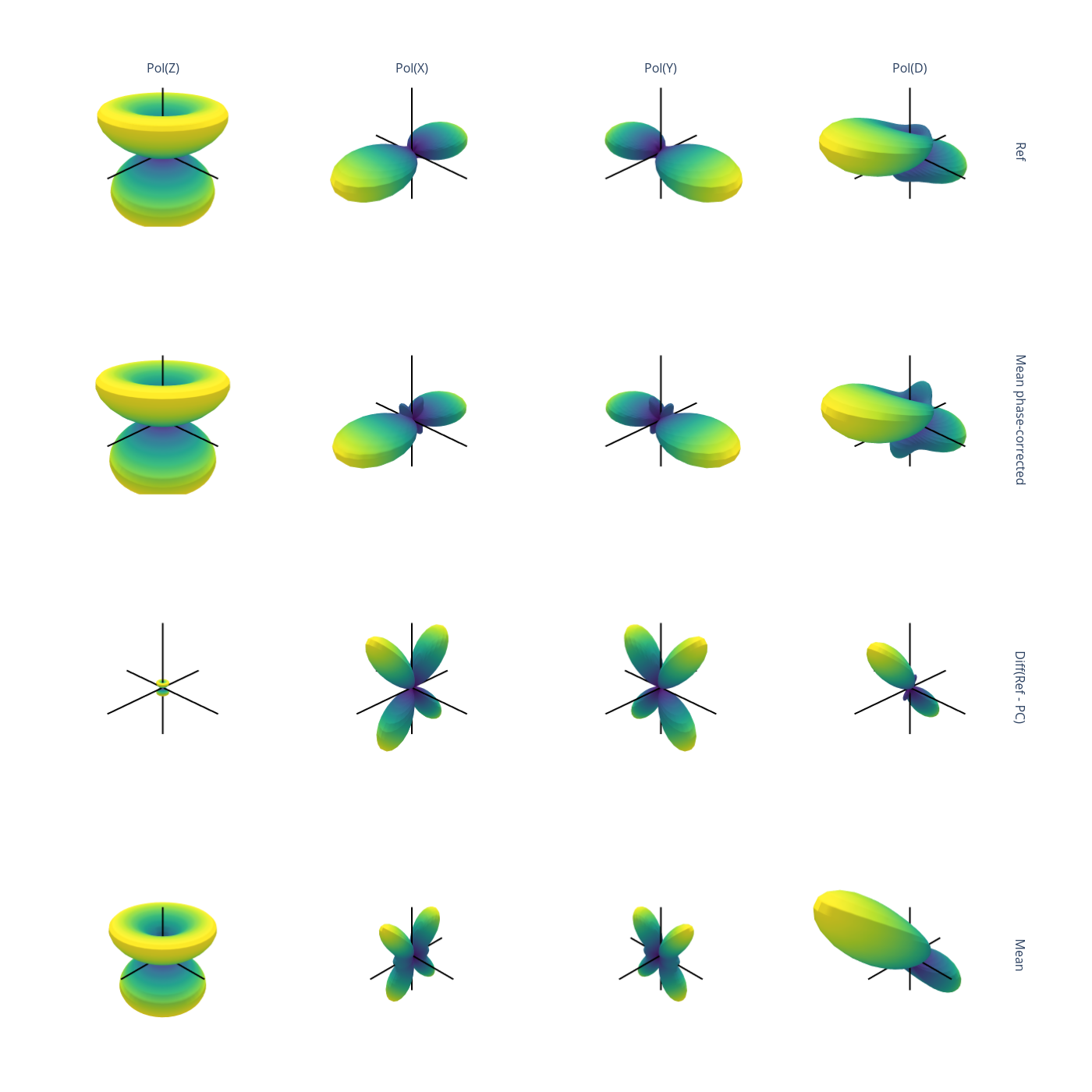}
\caption{Reconstructed MFPADs for N2. Columns show results for polarization geometries \((z,x,y,d)\) and rows the results for difference sets of matrix elements: retrieved mean case, phase-corrected case, reference (computational) results and difference between phase-correct and reference case. Note the colour-scale is rescaled per row, thus emphasizing the changes in form of the MFPADs (but obfuscating relative magnitudes), particularly for the difference case. See main text for further details.\label{454268}}
\end{center}
\end{figure}

With the matrix elements to hand, the MFPADs can also be computed, for any arbitrary molecular alignment and polarization. Some examples are shown in Fig. \ref{454268}, which shows results for the two sets of retrieved matrix elements discussed above (raw, and with phase-correction), along with the reference computational results (as used to originally generate the simulated data) and the differences between the phase-corrected case and the reference results.

A number of points are illustrated by the figure:

\begin{itemize}
\item The original/input matrix elements produce the MFPADs shown in the top row of the figure, which provide a reference for the fidelity of the reconstructed MFPADs. 
\item Polarization geometries are illustrated for four cases (see Fig. \ref{781808} for the MF reference coords), labelled $z,x,y,d$, where $z$ is parallel to the bond axis, $x,y$ perpendicular and $d$ ``diagonal" relative to the $z$-axis ($\theta_{z}=\pi/4$). In this case symmetry dictates that $z$ ($\sigma_u$ continuum) is cylindrically symmetric, the $x,y$ cases are a conjugate $\pi_u$ pair, and that the diagonal case mixes these components, hence breaks the symmetry and is sensitive to the relative phase between the continua.
\item The 2nd row illustrates the results from the phase-corrected parameters. These show generally good agreement with the reference case, apart from slightly exaggerated off-diagonal lobes for the $x,y$ results, consistent with the larger magnitudes for $|3,\pm1\rangle$ in the reconstructed matrix elements.
\item The 3rd row shows difference plots between the reference and phase-corrected results. This emphasizes the differences in the $|3,\pm1\rangle$ component magnitudes. The $z$ case shows very good agreement with the reference case (note the smaller magnitudes), whilst the $d$ case indicates some differences, again attributable to the differences in $|3,\pm1\rangle$ component magnitudes.
\item The bottom row illustrates the case of the phase-averaged (but not ``corrected") parameters. This serves to highlight the sensitivity of the PADs to the relative phases of the matrix elements - here the $x,y$ results show poor agreement with the reference case due to averaging over the $\pm\phi$ pairs, and this also affects the $d$ case.
\end{itemize}

Overall, with careful treatment of the phases, it is clear that high-fidelity MFPADs can be recovered in the current case. Although the difference plots indicate some differences between the reconstructed and reference case, the absolute changes in the form of the MFPADs are fairly minor. The lack of an absolute phase is not an issue in general for MFPAD recovery, although this does constitute a loss of relative phase in the continuum wavefunctions as a function of energy. 


\subsubsection{Further bootstrapping: information content \& sensitivity\label{sec:bootstrapping-info-sensitivity}}

As well as considering the results from full fits of the data, the inherent sensitivity of various aspects of the problem can also be investigated. In general, this will depend on the details of the problem at hand (symmetry, ADMs etc.), but can in essence be considered independently of the matrix elements themselves via ``channel functions" or equivalent (see Sect. \ref{sec:channel-funcs} and Sects. \ref{appendix:formalism}, \ref{sec:channel-funs-full} for full definitions). In the PEMtk routines, the various component tensors are computed and packaged as a basis set prior to fitting, and can be further examined independently. (For full details see \href{https://pemtk.readthedocs.io/en/latest/fitting/PEMtk_fitting_basis-set_demo_050621-full.html}{the PEMtk docs}.)

\begin{figure}[]
\begin{center}
\includegraphics[width=\textwidth,height=\dimexpr\textheight-4\baselineskip-\abovecaptionskip-\belowcaptionskip\relax,keepaspectratio]{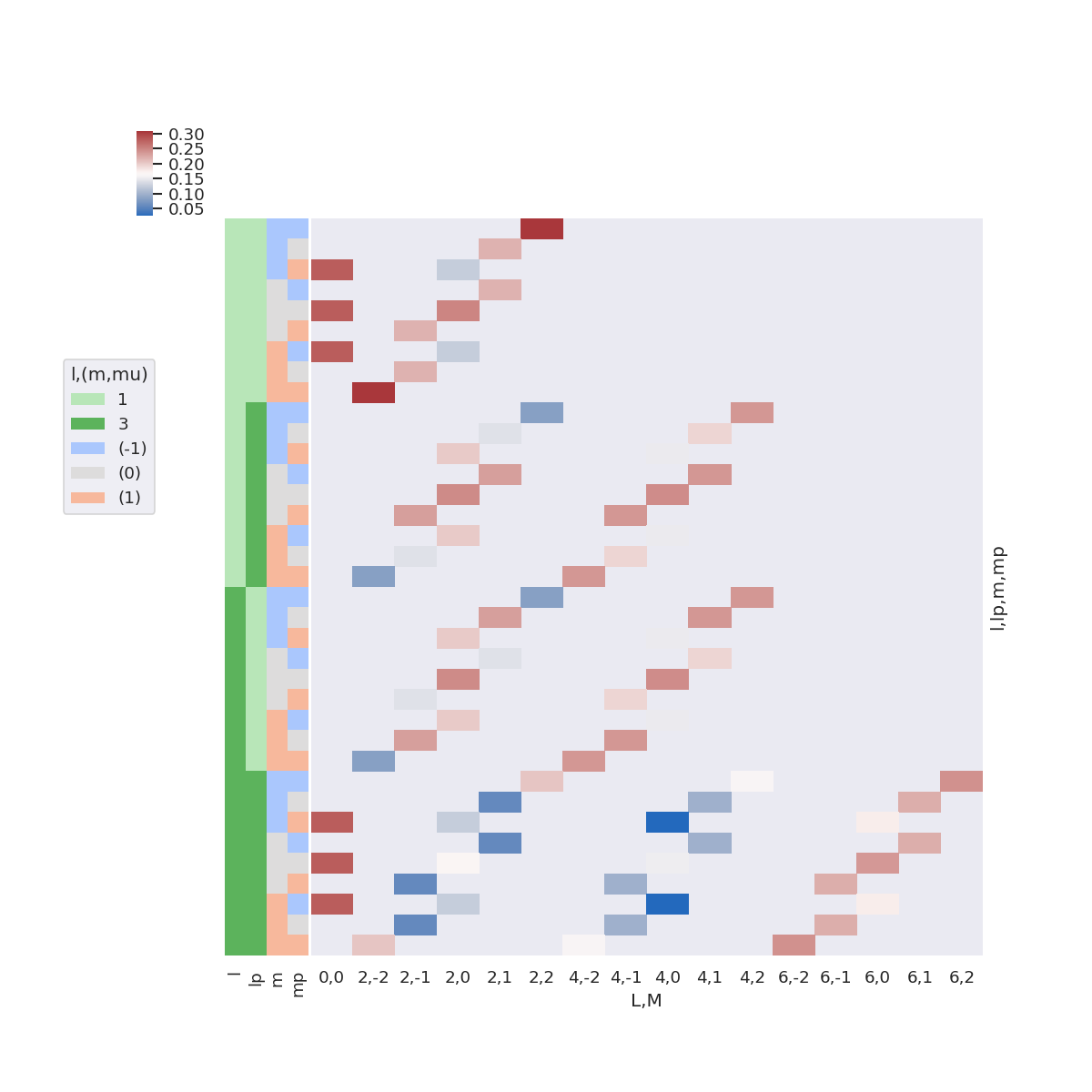}
\caption{Matrix element geometric coupling term \(B_{L,M}\). (Terms \(\ge0.001\) only.)\label{776753}}
\end{center}
\end{figure}

Fig. \ref{776753} illustrates the geometric coupling term $B_{L,M}$ (Eqn. \ref{eq:BLM-func-defn}), hence the sensitivity of different (L,M) terms to the matrix element products. This term incorporates the coupling of the partial wave pairs, $|l,m\rangle$ and $|l',m'\rangle$, into the term $B_{L,M}$, where $\{L,M\}$ are observable total angular momenta, 
hence indicates which terms are allowed for a given set of partial waves - in the current test case, $l=1,3$ only (as defined by the known matrix elements). This is essentially a way to visualize the general selection rules into the observable: for instance, only terms $l=l'$ and $m=-m'$ contribute to the overall photoionization cross-section term ($L=0, M=0$). However, since these terms are fairly simply followed algebraically in this case, via the rules inherent in the 3j product, this is not particularly insightful. These visualizations will become more useful when dealing with real sets of matrix elements, and specific polarization geometries, which will further modulate the $B_{L,M}$ terms. For example, in the AF $M$ is further restricted by $M = S-R_{p}$ (Eqn. \ref{eq:delta-func-defn}); in the current case with $S=0, R_p=0$ (this is typically the case for a cylindrically-symmetric experimental configuration), only $M=0$ terms will contribute.

\begin{figure}[]
\begin{center}
\includegraphics[width=\textwidth,height=\dimexpr\textheight-4\baselineskip-\abovecaptionskip-\belowcaptionskip\relax,keepaspectratio]{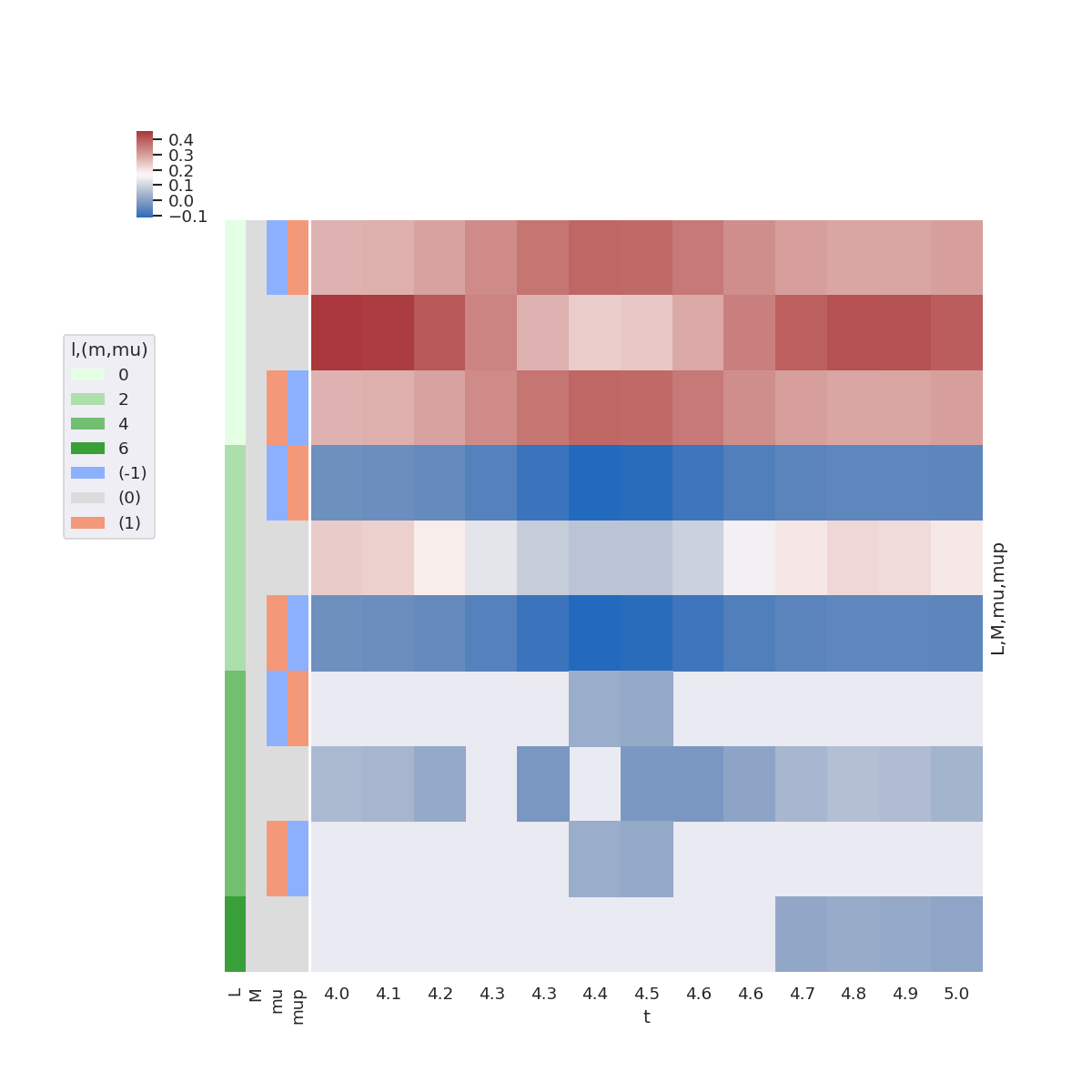}
\caption{Polarisation \& ADM product term, tensor product \(\Lambda_{R}\otimes E_{PR}(\hat{e})\otimes\Delta_{L,M}(K,Q,S)\otimes
A^{K}_{Q,S}(t)\).     (Terms \(\ge0.001\) only.)\label{652406}}
\end{center}
\end{figure}

Fig. \ref{652406} is more complicated, and illustrates the tensor product $\Lambda_{R}\otimes E_{PR}(\hat{e})\otimes \Delta_{L,M}(K,Q,S)\otimes A^{K}_{Q,S}(t)$, expanded over all quantum numbers.
This term, therefore, incorporates all of the dependence (or response) of the AF-$\beta_{LM}$s on the polarisation state, and the axis distribution. In this case, it's clear that there's a significant response to the alignment in the $L=0,2$ terms, some response in $L=4$ and - for the most part - no significant contribution from higher-order terms (threshold of 0.01), for the selected set of ADMs. This visualisation is potentially useful for planning measurements sensitive to certain properties, for example, in this case the $L=6$ term is significant only over a small $t$-range, so this region could be targeted experimentally to obtain data more sensitive to higher-order $l$-wave term couplings (per Fig. \ref{776753}). Conversely, the $L=0,2$ response terms are quite symmetric over the half-revival, so making experimental measurements at $t$-points symmetrically over this feature will provide redundant, but not additional, information content to the dataset for matrix-element retrieval. Also of note is the opposite response of the $\mu=\mu'=0$ terms and $\mu\neq\mu'$ (with $\mu=\pm1$, $\mu'=\mp1$) to the alignment, indicating different response and sensitivities in the polarization projections. This is particularly apparent in the $L=6$ term, where the $\mu\neq\mu'$ terms drop below threshold.

\begin{figure}[]
\begin{center}
\includegraphics[width=\textwidth,height=\dimexpr\textheight-4\baselineskip-\abovecaptionskip-\belowcaptionskip\relax,keepaspectratio]{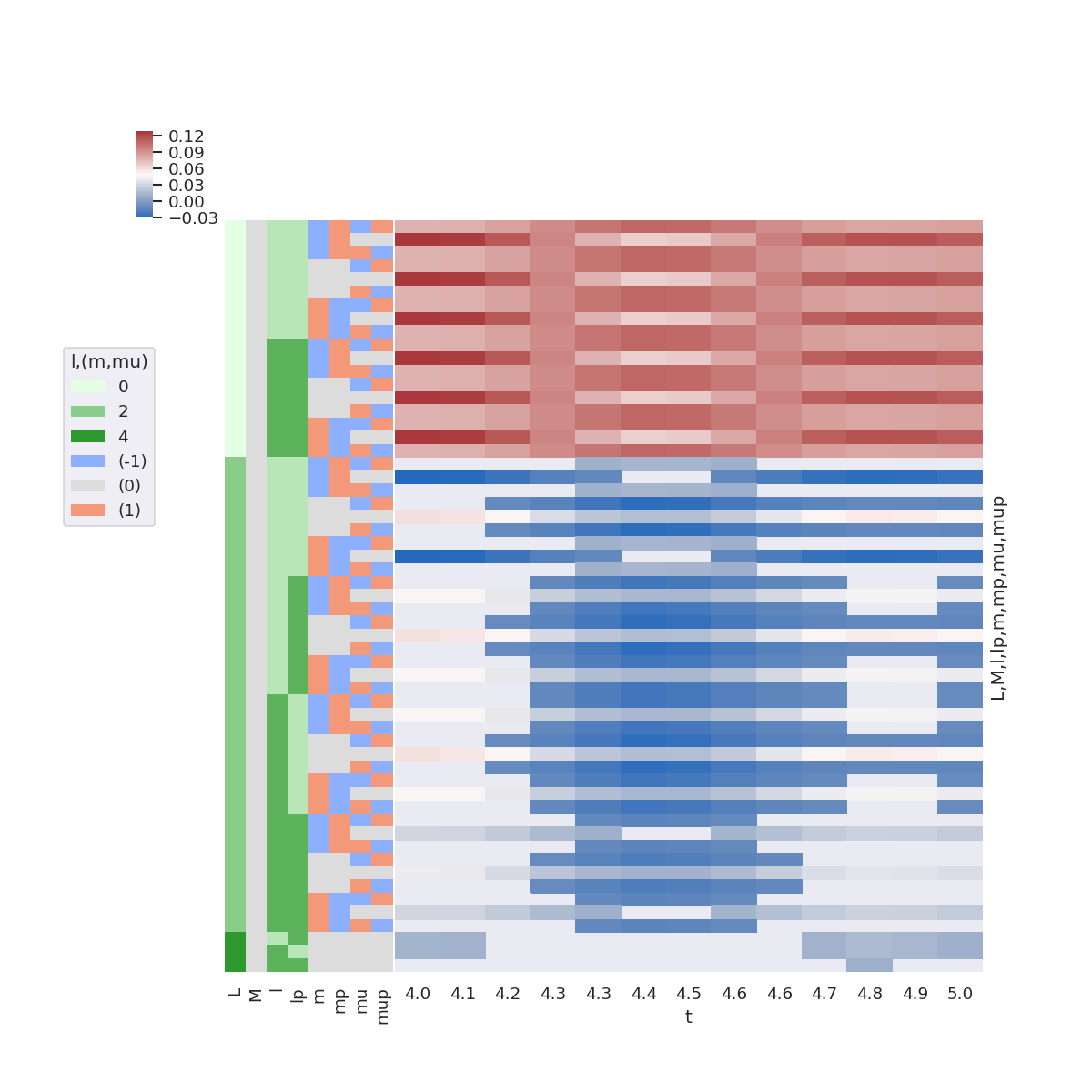}
\caption{Full channel (response) functions \(\varUpsilon_{L,M}^{u,\zeta\zeta'}\).     (Terms \(\ge0.001\) only.)\label{676540}}
\end{center}
\end{figure}

Finally, Fig. \ref{676540} illustrates the full AF channel (response) function $\varUpsilon_{L,M}^{u,\zeta\zeta'}$ (Eqns. \ref{eqn:channel-fns}, \ref{eq:channel-func-defn-AF}), which is essentially the complete geometric basis set,
hence equivalent to the AF-$\beta_{LM}$ (Eqn. \ref{eq:BLM-tensor-AF}) if the ionization matrix elements were set to unity. This illustrates not only the coupling of the geometric terms into the observable $L,M$, but also how the partial wave $|l,m\rangle$ terms map to the observables. 
A few observations in this case:

\begin{itemize}
\item The largest response is in the total cross-section ($\beta_{0,0}$), which is ~2 times larger than any other term; 
\item this response is similar for both the $l=1$ and $l=3$ contributions, and all $m$, hence we expect similar sensitivity to both partial cross-sections (continua) in this case.
\item For $L>0$, $l\neq l'$ are present, indicative of the sensitivity of the PADs to cross-terms (interferences).
\item The response is, as might be expected, distinct for terms with $l=l'$ and $l\neq l'$, and as a function of $L,l,l'$.
\item The parameters indicate enhanced sensitivity to higher-order terms ($L=6$) just after the main revival feature. This was already apparent in Fig. \ref{652406}, and is a consequence of the larger $K=4,6$ terms in the ADMs in this region (see Fig. \ref{720080}). In the current case this is not particularly important for the matrix element retrieval, but in general may be significant in more complex cases with larger $l$ present.

\end{itemize}

These results demonstrate the high level of detail that can be obtained from a set of channel functions, and how such a basis set can aid in both planning and interpretation of experimental measurements in terms of the sensitivity to given channels. Many outstanding questions remain, in particular general metrics of total information content are not yet defined, nor are general methods for determination of a sufficient information content for retrieval in a given case. Such exploration is currently a topic of ongoing research, although it is also of note that some authors have previously explored related issues (see, for example, Refs. \cite{Schmidtke2000,Ramakrishna2012}).


\subsubsection{Further bootstrapping: sub-sample size \& fidelity}

In cases with noisy data, and/or where uncertainties in fitted parameters remain large, fitting to larger and/or alternative sub-sets of the data may be pursued.
An obvious route to improvement of the retrieved matrix elements is via the inclusion of additional data points. In the original experimental demonstration \cite{marceau2017MolecularFrameReconstruction}, after an initial 10-point fit, further data was included, up to 89 data points, to minimise uncertainties in the retrieved matrix elements.\footnote{This procedure was only tested for the X-state in the experimental case, not for the A or B-state datasets. 
and especially the supplementary materials for more details.} Another, similar, route is via traditional statistical bootstrapping, in which fitting is tested for various randomly-selected sub-sets of the data (but the size of the fitted data set is not varied). Additionally, if the channel functions (fitting basis set) are investigated (see Sect. \ref{sec:bootstrapping-info-sensitivity}), additional data points may be selected to enhance the sensitivity to certain partial wave components.

In the current case, work is ongoing, 
but some effort has been made to investigate sampling strategies. In particular, preliminary work using different sample sizes plus additional statistical weightings (statistical bootstrapping) has yielded very promising results. For example, a dataset of 30 t-points equally spaced over the range 3.5~-~4.5~ps (as compared to 4~-~5~ps in the previous demonstration above), with random (Poissonian) statistical weightings additionally applied, yielded:

\begin{itemize}
\item 30~\% convergence (best $\chi^2$) in 100 test fits.
\item Standard deviations on the retrieved parameters of $<10^{-4}$ (cf. Table \ref{tab:matE}).
\item Fidelities on retrieved parameters typically $<10\%$.
\end{itemize}

However, whilst this appears promising, it is of note that the trends in the fidelity of the retrieved matrix element remained similar to those shown in Table \ref{tab:matE}, with the $l=3$ cases less well-defined. This indicates that a more fundamental information-content limitation remained. Additionally, other tests with the same initial dataset but different statistical weightings did not yield the same high percentage of converged fits with low standard deviation (hence retrieval quality), indicating that more careful and methodical work is required here, perhaps with more sophisticated statistical techniques applied.

\subsubsection{Bootstrapping technique outlook \& future directions}

In the current case, it is clear that the bootstrapping methodology for obtaining full photoionization matrix elements from time-domain, AF datasets, works well. The current numerical routines in the PEMtk package (and back-end libraries) are relatively stable and fast, and amenable to detailed inspection. However, a range of outstanding questions and routes of investigation remain, for instance:

\begin{itemize}
\item Scaling to larger problems (larger molecules, more matrix elements). This is currently under investigation, but - based on previous work \cite{hockett2009RotationallyResolvedPhotoelectron, hockett2018QMP2} - it is anticipated that small polyatomic molecules should be tractable to the basic approach. 
\item More sophisticated approaches, for instance careful sub-selection of data based on the channel functions, data obtained for additional polarization geometries or with shaped laser pulses \cite{hockett2014CompletePhotoionizationExperiments, hockett2015CoherentControlPhotoelectron, hockett2015CompletePhotoionizationExperiments,hockett2018QMP1}. Such approaches may be expected to yield higher fidelity reconstructions for small polyatomic systems, and may be required for more complex cases.
\item Faster numerical routines, in particular via GPU-based numerics.
\item Implementation of additional fitting routines, both from standard numerical methods, and also from related specialised domain problems, for example phase-retrieval methods developed for optical interferograms and spectrograms (e.g. general FROG-type retrieval methods \cite{trebino2000FrequencyResolvedOpticalGating}, ptychography/holographic techniques \cite{Spangenberg2015b, Spangenberg2015c}) and homotopy \cite{Sommese2005} approaches may be applicable.
\item Correlations and overlaps with other related/emerging photoionization techniques may also prove fruitful. For instance, photoionization matrix elements are of interest in high-harmonic spectroscopy schemes \cite{Lock2012}, angle-resolved RABBITT \cite{hockett2017AngleresolvedRABBITTTheory,villeneuve2017CoherentImagingAttosecond} and general attosecond ``clocking" and time-delay measurement techniques. In many cases these methods are also directly sensitive to the energy-dependence of the matrix elements, thus providing additional information relative to a basic 1-photon ionization study (albeit with additional complexity), and have recently been of great theoretical interest in the \textit{ab initio} photoionization community \cite{Feist2014,benda2022AnalysisRABITTTime}.

\end{itemize}

Machine learning (ML), and particularly recent ``deep learning" techniques (often also generically referred to as AI), are currently in vogue and evolving rapidly. Such approaches may also present interesting opportunities for MF retrieval problems. Perhaps the main strength of these methods is that, in favourable cases, very complex problems may be treated without complete computation and/or understanding of the underlying physics (e.g. the recent success of DeepMind/AlphaFold in protein folding \cite{eisenstein2021ArtificialIntelligencePowers,jumper2021HighlyAccurateProtein}); the use of ML/AI may, therefore, be particularly interesting for cases which are otherwise intractable to a full \textit{ab initio} analysis, e.g. complicated molecular dynamics or strongly-coupled light-matter systems, where the step-wise and separable approaches detailed herein will currently fail, and closed-form equations/solutions do not exist. A secondary use may be as fast fitting algorithms for cases of the type discussed herein, where the physics is understood, albeit complicated. This is much less interesting scientifically since it does not present a ``new" capability, although may still prove fruitful if the increase in speed (or robustness) is significant compared to current methods, e.g. to allow for real-time analysis during experimental runs. Furthermore, use in this vein, but with the aim of solving a fully coupled problem without the necessity of a bootstrapping or multi-step type of analysis (i.e. combining the various stages illustrated in Figs. \ref{807606}, \ref{731792} into a single ML-based routine) is certainly worth pursuing, with the potential to create a method that simultaneously retrieves both the alignment and photoelectron properties from experimental data without additional researcher intervention. The difficulty in all cases will likely be the generation of a suitable data set for the ML/AI training procedure, which is required before the routines can be deployed on new problems. Significant introductory and general discussion to the (rapidly evolving) topic can be found online and, for example, in Refs. \cite{LeCun2015,carleo2019MachineLearningPhysical,davies2021AdvancingMathematicsGuiding}. Some recent use of ML/AI in the AMO context has proved successful, e.g. Refs. \cite{he2022FilmingMoviesAttosecond,hegazy2022BayesianInferencingDeterministic} tackle specific data-analysis and reconstruction type problems; Refs. \cite{lu2022FastInitioPotential,nandi2021DmachineLearningPotential} examine ML for \textit{ab initio} computation of potential energy surfaces; Ref. \cite{carleo2019MachineLearningPhysical} provides a broad review of ML in the physical sciences, including quantum state reconstruction, as well as particle physics, cosmology and materials science.


\subsection{MF reconstruction via matrix inversion\label{sec:Matrix-inversion-example}}

\begin{figure}[]
\begin{center}
\includegraphics[width=\textwidth,height=\dimexpr\textheight-4\baselineskip-\abovecaptionskip-\belowcaptionskip\relax,keepaspectratio]{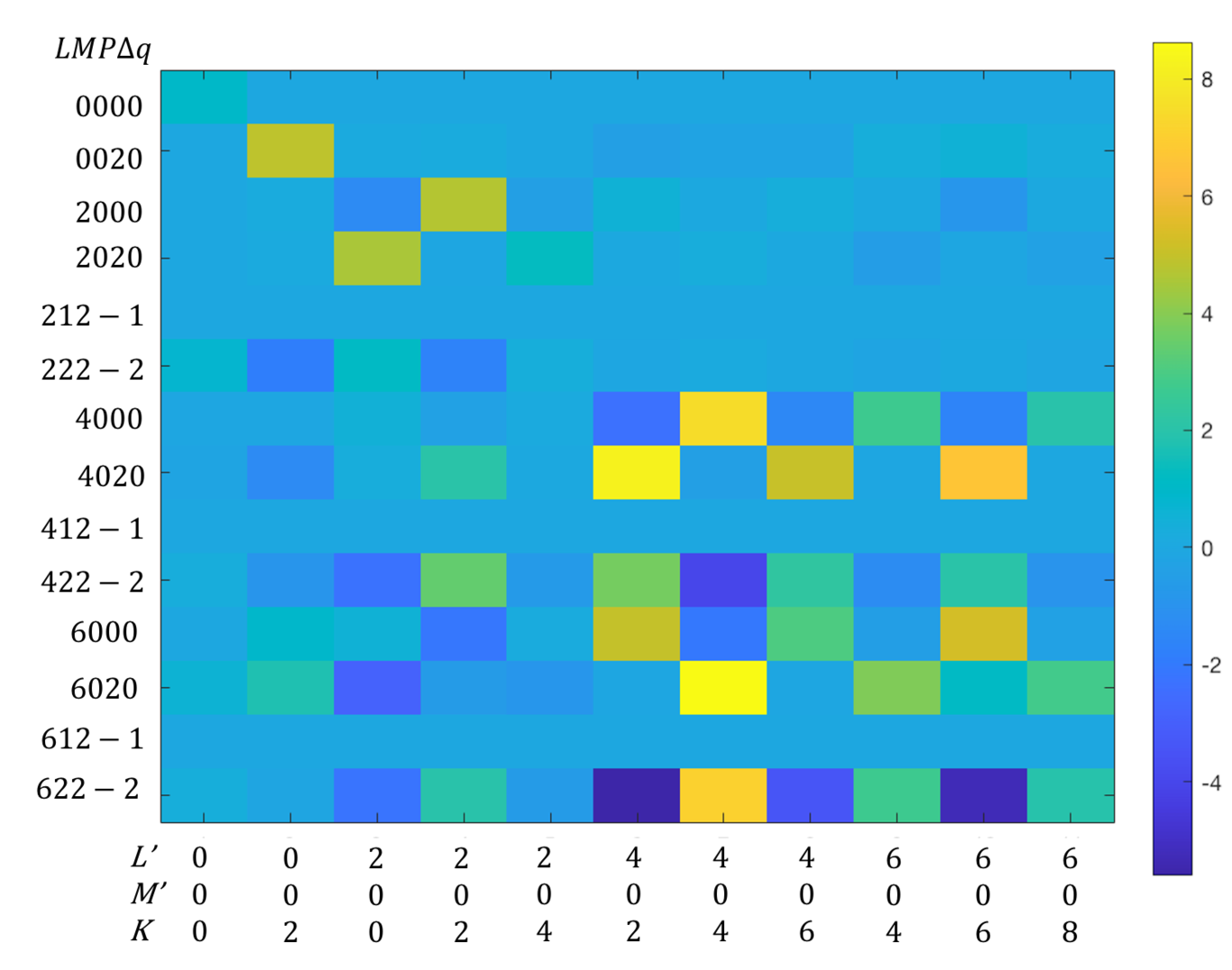}
\caption{\(\) Values of \(\mathbf{G}^{LMP\Delta q }_{L'M'KS}\) for \(N_2(X^{1}\Sigma^{+}_{g}) \rightarrow N^+_2(X^{2}\Sigma^{+}_{g})\). Figure reproduced from Ref. \cite{gregory2021MolecularFramePhotoelectron}.\label{931809}}
\end{center}
\end{figure}

The key aspects of this method involve first fitting out the ADMs $A^K_{QS}(t)$ from measured $\bar{\beta}^u_{LM}(t)$ in Eqn.~\ref{eqn:beta-convolution-C} to retrieve the $\bar{C}^{LM}_{KQS}$ (the vector $\mathbf{C}^{lab}$), and construction of the matrix $\mathbf{{G}}^{LMP\Delta q }_{L'M'KS}$ which facilitates the extraction of the molecular frame coefficients $C^{LM}_{P\Delta q}$ (the vector $\mathbf{C}^{mol}$). The steps in the process are depicted in Fig.~\ref{731792}, which label's primary inputs as the `AF/LF measurements', referring to the $\bar{\beta}^u_{LM}(t)$, and the ADMs $A^K_{QS}(t)$ calculated for a grid of fluence and temperature values by solving the TDSE using the appropriate Hamiltonian for impulsive alignment~\cite{hasegawa2015NonadiabaticMolecularAlignment, koch2019QuantumControlMolecular}. 
The `Alignment Retrieval' process requires solving the linear equations Eqn.~\ref{eqn:beta-convolution-C} for the $\bar{C}^{LM}_{KQS}$ using these primary inputs over the entire gird of ADMs, and comparing to the measured data to find the bets fit solution. This provides the vector $\mathbf{C}^{lab}$ needed for the next step - solving the linear equations Eqn.~\ref{eq:basic}. This also requires that the matrix $\mathbf{{G}}^{LMP\Delta q }_{L'M'KS}$ be invertable. This matrix therefore encodes the uniquely accessible MF information in the experiment. Specifically, rows in the matrix link the LF parameter $\bar{\beta}^u_{LM}(t)$  to an MF beta parameter $\beta_{LM}(\theta,\chi)$ that can be used to construct the MFPADs, $\theta$ and $\chi$ being polar and azimuthal angles of the laser polarization in the MF. Specifically, a particular row, or set of rows, being entirely zero may indicate that certain MF $\beta_{LM}$ are inaccessible by the set of LF measurements. For instance, in~\cite{gregory2021MolecularFramePhotoelectron} the $\mathbf{{G}}^{LMP\Delta q }_{L'M'KS}$ matrix for the $N_2(X^{1}\Sigma^{+}_{g}) \rightarrow N^+_2(X^{2}\Sigma^{+}_{g})$ ionization of N$_2$ with a single photon, from a RWP excited by a linearly polarized pulse was computed. The results are reproduced herein for reference, see Fig. \ref{931809}, and this can be considered as an indicator of sensitivity in a somewhat analogous fashion to the channel functions shown in Figs. \ref{776753} - \ref{676540}. Rows with $M = 1$ were all determined to be identically zero, therefore only allowing the retrieval of MFPADs with the ionizing field parallel and perpendicular to the molecular axis. In this case, the partial wave analysis involved in the previously described method, along with the non-linear fitting necessary to retrieve the radial dipole matrix elements is avoided. However, only two of the MFPADs shown in Fig.~\ref{731792} extracted by bootstrapping are retrieved by this method.\\

\begin{figure}[]
\begin{center}
\includegraphics[width=\textwidth,height=\dimexpr\textheight-4\baselineskip-\abovecaptionskip-\belowcaptionskip\relax,keepaspectratio]{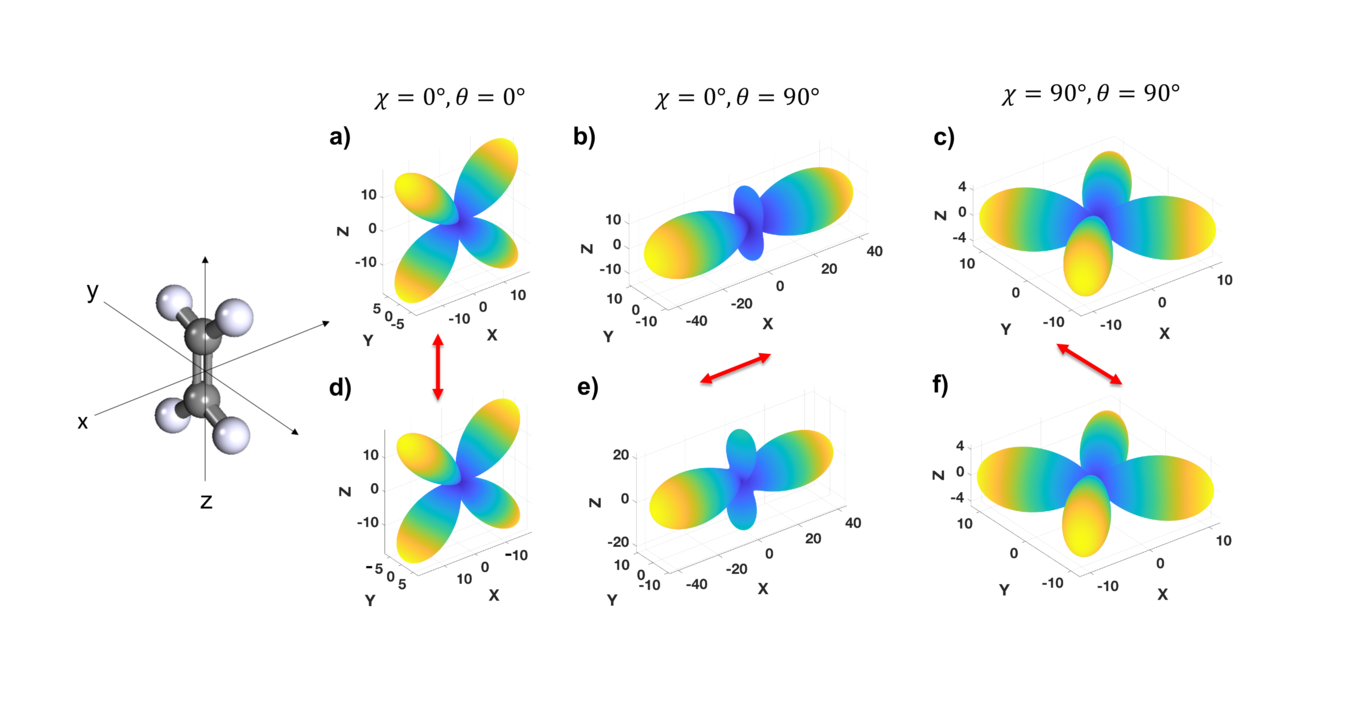}
\caption{Calculated (a,b,c) and Retrieved (d,e,f) MFPADs for \(C_2 H_4(^1A_g) \rightarrow C_2H_4^+(^2B_{3u})\) for \((x,y,z)\) polarization geometries. Figure reproduced from Ref. \cite{gregory2021MolecularFramePhotoelectron}.\label{584598}}
\end{center}
\end{figure}

The utility of this method is better demonstrated by application to a polyatomic molecule of $D_{nh}$ symmetry. Gregory et al.~\cite{gregory2021MolecularFramePhotoelectron} also compute the $\mathbf{{G}}^{LMP\Delta q }_{L'M'KS}$ matrix for $C_2H_4(^1A_g) \rightarrow C_2H_4^+(^2B_{3u})$ ionization of $C_2H_4$, with $D_{2h}$ symmetry, from a RWP excited by linearly polarized light. In contrast with $N_2$, all rows of the $\mathbf{{G}}^{LMP\Delta q }_{L'M'KS}$ are non-zero for $C_2H_4$, but the system of linear equations is inconsistent since there are more MF parameters (in the vector $\mathbf{C}^{mol}$) than LF measurements (in the vector $\mathbf{C}^{lab}$). Nonetheless, the solution selected by Gregory et al. minimizes the retrieved vector $\mathbf{C}^{mol}$ and correctly reproduce the MFPADs for any orientation. The basic $(x,y,z)$ cases are reproduced herein as Fig. \ref{584598}, indicating the high fidelity of the reconstructions (for arbitrary polarization geometries see Fig. 5 of Ref. \cite{gregory2021MolecularFramePhotoelectron}). This `extra' information available in the LF measurement for $C_2H_4$ compared to $N_2$ can be traced back to the fact that the RWP is 2D for an asymmetric top excited by linearly polarized light(cf. Eqn.~\ref{eq:mfrealsig}). While this method does not directly provide radial dipole matrix elements, the MFPADs produced can potentially be used as a constraint, along with additional experimental data, to determine these from a non-linear fit (see Sect. \ref{sec:recon-from-MFPADs}). 

\subsection{Matrix element retrieval from MFPADs\label{sec:recon-from-MFPADs}}

To complete the circle, one can also consider whether MFPADs - either directly measured or reconstructed via matrix inversion - contain sufficient information to retrieve the underlying matrix elements.  The former case has already been the subject of several studies, for instance Refs. \cite{Gessner2002a,Lebech2003,Cherepkov2005, Yagishita2005}, and shown to work well for at least homo and hetero nuclear diatomics. In a sense the latter case can be considered as (yet another) bootstrapping-type scheme - MFPADs are obtained via a minimum effort/information route, then leveraged to obtain underlying properties. 

Fitting for MFPADs is currently implemented in the PEMtk suite in a similar manner to the AF fitting detailed in Sect. \ref{sec:bootstrapping}, based on the tensor formalism detailed in Sect. \ref{sec:tensor-formulation}. An extensive numerical example is not presented herein, but in testing for the $N_{2}$ example case, e.g. using MFPADs as shown in Fig. \ref{454268}, matrix element retrieval was found to be possible in general. However, in line with previous observations for other methods (and the underlying physics), the fidelity and retrievable relations again depend on the exact nature and quality - generally the information content - of the MFPAD dataset (or, equivalently, the set of measured $\beta_{L,M}$). In testing, a number of different datasets were trialed, and some notes and observations are given below.

\begin{itemize}
\item MFPADs were tested at single $\epsilon$ only, as a function of polarization geometry, cf. the MFPADs as shown in Fig. \ref{454268}.
\item Data for fitting therefore consisted of parameters $\beta_{L,M}(\epsilon,R_{\hat{n}})$, with $L_{max}=6$.
\item For fitting intensity-normalised MFPADs (i.e. $\beta_{0,0}=1$), for single or multiple polarization geometries from the ($x,y,z$) set:
\begin{itemize}
\item This was found to be sufficient to determine the corresponding continuum matrix elements in this case, i.e. $z$ polarization corresponds to the $\sigma_u$ continuum, and $x,y$ to the $\pi_u$ continuum. (This separation may not be so clean in other cases, depending on the symmetry of the system.)
\item Matrix elements were constrained as expected, with only \textit{relative} magnitudes and phases obtained (per continuum) in this manner.
\item Similarly, due to a lack of cross-terms between continua in this case, the phase relations between the $\sigma_u$ and $\pi_u$ continua were also undefined in this case.
\item In line with the AF reconstruction procedure, the sign of the phases was also undefined in this case.
\item Similarly, the fidelity on the retrieved matrix elements (for defined relations), was on the order of the uncertainties in the fitted dataset.
\end{itemize}
\item For determination of additional matrix element relations, additional data can be incorporated in the fitting.
\begin{itemize}
\item Incorporating absolute magnitudes ($\beta_{0,0}=\sigma$) allows for the relative magnitudes of the continua to be determined.
\item Incorporating additional suitable interferences, e.g. diagonal polarization (cf. Fig. \ref{454268}), in the dataset allows for additional relative phase relationships to be determined, e.g. between the $\sigma_u$ and $\pi_u$ continua.
\end{itemize}
\item Retrieval of matrix elements vs. $\epsilon$ is also possible; however, it is again constrained by the presence (or otherwise) of interferences. In the basic case, each energy is treated independently, and relative phases as a function of energy are undefined. However, these may be approximately defined by imposing an energy constraint, e.g. defining that the phases are smooth vs. energy or follow a certain functional form (see, for an example, Ref. \cite{Yagishita2005}). 
\end{itemize}

In general, an MF fitting procedure, whether from directly measured MFPADs, or reconstructed MFPADs, may be expected to work generally in principle, with similar caveats to the full AF-bootstrapping methodology of Sect. \ref{sec:bootstrapping} (and the existing literature, see Sect. \ref{sec:CompleteLit}). Both methodologies are constrained by the symmetry of the system, and general information content. Notable differences are the implicit presence of cross-terms between continua in the AF case (Eqn. \ref{eq:BLM-tensor-AF}), which may be missing in the MF case (Eqn. \ref{eq:BLM-tensor-MF}), depending on the choice of polarization geometries.
Outstanding questions remain similar for both cases, namely the fidelity of reconstruction, and the ability to scale-up to more complex cases (larger molecules, dynamic systems) and concomitant information content requirements, and are a subject of ongoing research.




\section{Summary \& Outlook \label{sec:summary-outlook}}

From this topical review, it is hoped that the reader has gained a solid grounding in photoionization physics, and general reconstruction methods, for both MF observable reconstruction and full matrix element retrieval (``complete" photoionization or quantum tomography treatments), and that a suitable toolkit and platform for interested researchers has been developed. 

Over the last decade or two, such experiments, as well as theoretical treatments and unified data-analysis techniques have developed significantly, and become more accessible to non-specialists. It is our hope 
that the new platform and tools discussed herein (Sect. \ref{sec:resources}) will prove useful, and help photoionization (and related) studies involving MF and full matrix element retrieval to (continue to) become more routine and successful. Whilst significant new work has been presented herein, development work is ongoing, with aims to make the code-base more robust and user-friendly, test new cases - particularly for more complex molecular systems, and dynamical cases - and implement new features, such as metrics for information content for a given basis set that will aid experimental planning, and further density-matrix based analysis methods. As well as publication in the literature, new developments and results will also be made available via the open-source platform, and it is hoped that other researchers will also contribute to grow these efforts over time.

Meanwhile, experimental methodologies continue to improve and grow in sophistication, allowing more routes, and often more direct routes, to high information-content observables, whether in the lab or molecular frame, in a range of interactions and scenarios, including for larger molecules and probing dynamical effects. 


Recent work in this vein from the current authors and coworkers includes (atomic) matrix element retrieval from photoelectron imaging measurements with polarization multiplexing (via shaped laser pulses) \cite{hockett2014CompletePhotoionizationExperiments, hockett2015MaximuminformationPhotoelectronMetrology,hockett2015CoherentControlPhotoelectron, hockett2015CompletePhotoionizationExperiments}, (hyperfine) quantum beat spectroscopy from time-resolved photoelectron imaging experiments \cite{forbes2018QuantumbeatPhotoelectronimagingSpectroscopy}, and quantum tomographic determination of (LF) density matrices including electronic dynamics - theory \cite{gregory2022LaboratoryFrameDensity} and application to $NH_3$ (manuscript in preparation). These types of investigations indicate the potential for further general developments, and the utility of the MF retrieval and reconstruction techniques discussed herein as a foundation for a broader range of problems in molecular photoionization and dynamics, ultimately building to a class of molecular quantum state retrieval methods from photoelectron measurements \cite{hockett2018QMP1, hockett2018QMP2}.

\section{Acknowledgements}

The authors would like to thank the JPB editors - particularly A. Stolow - for inviting this submission, their encouragement, and their extreme patience during its (rather lengthy) preparation. Special thanks also to R. Boll, D. Rolles and A. Rudenko for comments \& discussion on an early version of the manuscript, and technical notes on the state-of-the-art in coincidence measurements; M. Schuurman for general discussion; the referees for close-reading of the manuscript, and helpful suggestions for improvements and additional references.
\section{Resources\label{sec:resources}}

The following online resources are available, and it is hoped that interested readers will find these useful, and contribute.

\begin{itemize}
\item Manuscript text, live \href{https://www.authorea.com/users/71114/articles/447808-extracting-molecular-frame-photoionization-dynamics-from-experimental-data}{on Authorea (includes interactive figures)} and \href{https://github.com/phockett/Extracting-Molecular-Frame-Photoionization-Dynamics-from-Experimental-Data}{Github}. Post-publication reader comments \& suggestions can be added directly via the web.
\item \href{https://www.zotero.org/groups/4733878/molecular_frame_pads_measurements_and_reconstruction}{Full bibliography, via a Zotero group \textit{``Molecular Frame PADs Measurements and Reconstruction"}} \cite{hockettZoteroGroupsMolecular}. Additional references and suggestions can be added to the group, and it is hoped that this will grow to provide a useful community resource.
\item Code for AF computation and matrix element retrieval, as illustrated in Sect. \ref{sec:Recon}. 
\begin{itemize}
\item Full Jupyter notebooks and source data are \href{http://dx.doi.org/10.6084/m9.figshare.20293782}{available on Figshare, DOI: 10.6084/m9.figshare.20293782}. This also includes a complete archive of the python software releases and environment used during this work \cite{hockett2022MFreconFigshare}.
\item The open-source python libraries \href{https://epsproc.readthedocs.io}{ePSproc} \cite{ePSprocAuthorea,ePSprocGithub,ePSprocDocs} and \href{https://pemtk.readthedocs.io}{PEMtk} \cite{hockett2021PEMtkDocs, hockett2021PEMtkGithub} developed for these problems. (See Sects. \ref{sec:numerics-intro}, \ref{sec:numerical-notes} for further discussion.)
\end{itemize}
\item Discussion forum at \href{https://amoopenscience.femtolab.ca/}{AMO Open Science}. We hope interested readers will use this forum for general discussion on the topic, if only to suggest other venues for discussion.
\item A Docker-based distribution of various codes for tackling photoionization problems is also available from the \href{https://github.com/phockett/open-photoionization-docker-stacks}{Open Photoionization Docker Stacks} project, which aims to make these tools more accessible to interested researchers \cite{hockettOpenPhotoionizationDocker}.
\end{itemize}

\section{Appendix A - Further reading\label{sec:Appendix-A}}

The following sections provide further reading on specific topics for interested readers. In line with the aims of this manuscript, these sections are (very) far from comprehensive, but aim to provide a jumping-off point for the topics; it is hoped that a more comprehensive, crowd-sourced and ongoing listing will be generated via the \href{https://www.zotero.org/groups/4733878/molecular_frame_pads_measurements_and_reconstruction}{\textit{``Molecular Frame PADs Measurements and Reconstruction"}} online bibliography on Zotero \cite{hockettZoteroGroupsMolecular}.

\subsection{MF experimental literature sample\label{appendix:MF-expt}}

Reviews covering this class of experiment can be found in  \cite{Yagishita2005,Reid2012,dowek2012PhotoionizationDynamicsPhotoemission,Yagishita2015, jahnke2022PhotoelectronDiffraction, dowek2022TrendsAngleresolvedMolecular}; some (representative) examples from the literature include:

\begin{itemize}
\item MFPADs for N2 core ionization \cite{Shigemasa1995}
\item MFPADs from various diatomics with coincidence velocity-map imaging \cite{Eland2000} \cite{Hikosaka2000}
\item MFPADs from CO K-shell photoionization \cite{motoki2000KshellPhotoionizationCO} and matrix element retrieval \cite{Cherepkov2000}.
\item MFPADs and matrix element retrieval for N2 \cite{Gessner2002a}
\item MFPADs and matrix element retrieval for NO inner-valence ionization \cite{Lucchese2002,Lebech2003} (this makes use of a general $F_{LN}(\theta_e)$ parameterized treatment of the MFPADs (see also Ref. \cite{Lafosse2002}), which provides a suitable reduction method for experimental data analysis for RF and MFPADs, and the potential for matrix element retrieval, in a somewhat similar manner to the tensor approach discussed herein). 
\item NO2 RFPADs \cite{Toffoli2007}.
\item NO MFPADs (1s channels) \cite{Li2007}.
\item Interference and entanglement in H2 double ionization in the MF \cite{Akoury2007}.
\item General discussion of multi-particle imaging with COLTRIMS \cite{Trinter2012a}.
\item Imaging methane MFPADs with electron-ion-ion coincidences \cite{Williams2012, Williams2012a}.
\item MFPADs from naphthalene ($C_{10}H_{8}$) via alignment and tomographic imaging \cite{Maurer2012}.
\item MFPADs from H2 and D2 at ~30eV photon energies (HHG source) \cite{Billaud2012a}.
\item MFPADs via alignment for Auger-Meitner electrons \cite{Cryan2010,Cryan2012a}.
\item MFPADs from core ionization of carbon tetrafluoride ($CF_4$), ethane ($C_2H_6$) and 1,1-difluoroethylene ($C_2 H_2 F_2$)\cite{Menssen2016}.
\end{itemize}

\subsection{Theory literature sample\label{sec:theory-lit}}

A number of authors have developed relevant theory for photoionization problems. Whilst the fundamentals are similar, different formalisms have been derived for various specific cases, or to emphasize particular aspects of the problem. A few notable examples are listed here.

\begin{itemize}
\item Early derivations for atomic PADs \cite{Cherepkov1979,Cooper1968,Cooper1969}
\item Early derivations for molecular (LF)PADs \cite{Tully1968}, rotationally-resolved cases \cite{Buckingham1970}, and optically-active cases \cite{Ritchie1976}
\item Angular momentum transfer in LF and MFPADs \cite{Fano1972} 
\item General MFPAD (``fixed-molecule") derivation \cite{Dill1976}
\item Atomic PADs and spin polarization \cite{Klar1982}
\item Resonant multiphoton PADs \cite{Dixit1983}
\item Molecular orbital continuum decomposition \cite{Park1996}
\item A simple-model description for MFPADs, including $F_{LN}$ function decomposition \cite{Lucchese2004}
\item Matrix element retrieval from time-resolved photoelectron imaging experiments \cite{Suzuki2007}.
\item Time-resolved PADs and dynamics (including molecular alignment) \cite{Underwood2000,Seideman2001,Stolow2008}
\item LF, RF and MFPADs theory and determination (review article) \cite{dowek2012PhotoionizationDynamicsPhotoemission}
\item Photoionization with a focus on alignment/RWP effects, including numerical studies \cite{Ramakrishna2012, Ramakrishna2013,hockett2015GeneralPhenomenologyIonization}, and MFPAD case-study \cite{reid2018AccessingMolecularFramea}.
\item LF and MFPADs theory review and tutorials (with numerical examples) \cite{hockett2018QMP1}
\end{itemize}
\subsection{Complete experiments literature sample\label{sec:CompleteLit}}


The topic of complete photoionization experiments has been recently reviewed in \cite{kleinpoppen2013perfect, hockett2018QMP2}, see also older review articles \cite{Becker1998,Reid2003,Kleinpoppen2005}. Some representative and noteworthy examples are also listed below (see also Sect. \ref{appendix:MF-expt} above for cases involving MF measurements).

\begin{itemize}
\item First experimental demonstration for photoionization of an atomic target by Berry and coworkers \cite{Duncanson1976}, who studied
$Na(^{2}P_{1/2},\,^{2}P_{3/2})$.
\item First molecular demonstration for $NO$, from the Zare group, with state-resolved LF measurements including linear and circularly polarized fields \cite{Reid1992} (see also prior work \cite{Allendorf1989,Leahy1991,Reid1991}).
\item General theoretical discussion on complete experiments in atoms and molecules \cite{Cherepkov2005}.
\item Time-resolved rotational-wavepacket methods applied to $NO$ (narrow wavepacket) \cite{Tsubouchi2004, Tang2010}, see also Refs. \cite{Suzuki2006} (review) and \cite{Suzuki2007} (theory \& analysis).
\item Non-linear polyatomic demonstration with state-resolved LF measurements ($NH_3$) \cite{hockett2009RotationallyResolvedPhotoelectron}.
\item Theoretical investigation of matrix element retrieval for photoelectron imaging experiments \cite{Ramakrishna2012} (see also \cite{Ramakrishna2013} for rotational wavepacket reconstruction from complementary methods including high-harmonic generation).  
\item Complete experiments with polarization shaping \cite{hockett2014CompletePhotoionizationExperiments, hockett2015CompletePhotoionizationExperiments}.
\item Time-resolved rotational-wavepacket method applied to $N_2$ (broad wavepacket) and bootstrap technique development \cite{marceau2017MolecularFrameReconstruction} (as per Sect. \ref{sec:bootstrapping} herein.)
\end{itemize} 
\section{Appendix B - Further formalism\label{sec:Appendix-B}}

A complete accounting of the formalism used herein is given in the following sections, expanding on the brief introduction of Sect. \ref{sec:tensor-formulation}, along with some additional notes for interested readers.

\subsection{Full Photoionization Formalism \label{appendix:formalism}}


The equations for the \(\beta_{LM}\) parameters, in the molecular and
lab frames (MF \& AF respectively), written in terms of geometric
tensor parameters, are given in Sect. \ref{sec:full-tensor-expansion}. The necessary tensor components are further detailed below. For further discussion and derivations see Refs. \cite{Reid2000,Underwood2000,Stolow2008}, and Ref. \cite{hockett2018QMP1} for a summary including various different formalisms; further examples can also be found in the ePSproc documentation \cite{ePSprocDocs}. For general discussion on tensor methods in atomic and molecular spectroscopy and related problems, see Refs. \cite{BlumDensityMat,zareAngMom}. 

Note, also, that several equivalent formalisms have been presented in the literature (Sect. \ref{sec:theory-lit}), often specialised to a particular problem or choice of phase conventions. In the following the equations are general, although the phase conventions are chosen to match those used in ePolyScat (Sect. \ref{sec:mat-ele-conventions}), hence the numerics for the analysis presented herein. In practice the ePSproc codes allow the user to set these conventions as desired. 

\subsubsection{Electric field term}\label{electric-field-term}

The
\href{https://epsproc.readthedocs.io/en/latest/methods/geometric_method_dev_260220_090420_tidy.html\#E_\%7BP,R\%7D-tensor}{coupling
of two 1-photon terms can be written as a tensor contraction} \cite{BlumDensityMat,zareAngMom}:

\begin{equation}
E_{P,R}(\hat{e})=[e\otimes e^{*}]_{R}^{P}=[P]^{\frac{1}{2}}\sum_{p}(-1)^{R}\left(\begin{array}{ccc}
1 & 1 & P\\
p & R-p & -R
\end{array}\right)e_{p}e_{R-p}^{*}
\label{eq:EPR-defn-1}
\end{equation}

Where \(e_{p}\) and \(e_{R-p}\) define the field strengths for the
polarizations \(p\) and \(R-p\), which are coupled into the spherical
tensor \(E_{PR}\). For the simplest case of a linearly-polarized field, $p=0$, and only terms $P=0,2$ with $R=0$ (i.e. $E_{0,0}$ and $E_{2,0}$) are non-zero.

Note this notation implicitly describes only the time-independent photon angular momentum coupling,
but time-dependent/shaped laser fields can be readily incorporated by allowing for time-dependent fields $e_{p}(t)$ (see, for instance, Ref. \cite{hockett2015CompletePhotoionizationExperiments}).

\subsubsection{\texorpdfstring{\(B_{L,M}\)
term}{B\_\{L,M\} term}}\label{b_lm-term}

The coupling of the partial wave pairs, \(|l,m\rangle\) and
\(|l',m'\rangle\), into the observable set of \(\{L,M\}\) is
\href{https://epsproc.readthedocs.io/en/latest/methods/geometric_method_dev_260220_090420_tidy.html\#B_\%7BL,M\%7D-term}{defined
by a tensor contraction with two 3j terms} \cite{zareAngMom}.

\begin{equation}
B_{L,M}=(-1)^{m}\left(\frac{(2l+1)(2l'+1)(2L+1)}{4\pi}\right)^{1/2}\left(\begin{array}{ccc}
l & l' & L\\
0 & 0 & 0
\end{array}\right)\left(\begin{array}{ccc}
l & l' & L\\
-m & m' & M
\end{array}\right)
\label{eq:BLM-func-defn}
\end{equation}

Note for the AF case the terms may be reindexed by $M=S-R'$ (see below). This allows for all MF projections to contribute, rather than a single specified polarization geometry.

\subsubsection{\texorpdfstring{\(\Lambda\)
Term}{\textbackslash{}Lambda term}}\label{lambda-term}

A general geometric projection term can be defined in the MF and AF.
For the \href{https://epsproc.readthedocs.io/en/latest/methods/geometric_method_dev_260220_090420_tidy.html\#/Lambda-Term}{MF
projection term}, \(\Lambda_{R',R}(R_{\hat{n}})\):

\begin{equation}
\Lambda_{R',R}(R_{\hat{n}})=(-1)^{(R')}\left(\begin{array}{ccc}
1 & 1 & P\\
\mu & -\mu' & R'
\end{array}\right)D_{-R',-R}^{P}(R_{\hat{n}})
\label{eq:lambda-func-defn-MF}
\end{equation}

This is similar to the $E_{P,R}$ term, and essentially rotates the field defined in the LF into
the MF by a set of rotations (Euler angles) defined by $R_{\hat{n}}=\{\chi,\Theta,\Phi\}$.

For
\href{https://epsproc.readthedocs.io/en/latest/methods/geometric_method_dev_pt3_AFBLM_090620_010920_dev_bk100920.html\#/beta_\%7BL,M\%7D\%5E\%7BAF\%7D-rewrite}{the
AF case}, a simplified form can be used, since there is no single
orientation/rotation defined in relation to the MF, and the relations are defined by the
molecular axis distribution (see below). 

\begin{equation}
\bar{\Lambda}_{R'}=(-1)^{(R')}\left(\begin{array}{ccc}
1 & 1 & P\\
\mu & -\mu' & R'
\end{array}\right)\equiv\Lambda_{R',R'}(R_{\hat{n}}=0)
\label{eq:lambda-func-defn-AF}
\end{equation}

The notation here implies that the electric field and axis distributions are expanded about the same axis ($R_{\hat{n}}=0$), which corresponds to the simplest parallel align-probe field geometry. Additional frame rotation(s) may be applied in some cases (e.g. for crossed-polarization of the alignment and probe fields).

\subsubsection{Alignment term\label{alignment-term}}

The axis distribution moments (ADMs) define the LF in this case, and are given above as a set of parameters $A_{Q,S}^{K}(t)$. These give rise to couplings which can be written as:

\begin{equation}
\Delta_{L,M}(K,Q,S)=(2K+1)^{1/2}(-1)^{K+Q}\left(\begin{array}{ccc}
P & K & L\\
R & -Q & -M
\end{array}\right)\left(\begin{array}{ccc}
P & K & L\\
R' & -S & S-R'
\end{array}\right)
\label{eq:delta-func-defn}
\end{equation}

Hence the coupling between the LF and MF is, effectively, defined by the final term in the AF observable (Eqn. \ref{eq:BLM-tensor-AF}):

\begin{equation}
\sum_{K,Q,S}\Delta_{L,M}(K,Q,S)A_{Q,S}^{K}(t)
\end{equation}

And the observable is restricted to $L_{max}=|P+K|$, and $M=0$ only if  $Q=S=R=0$ (the usual cylindrically symmetric case).

\subsubsection{Dipole matrix elements and conventions\label{sec:mat-ele-conventions}}


Herein the numerical form of the dipole matrix elements is chosen to match the definitions used by ePolyScat \cite{Gianturco1994,Natalense1999,Toffoli2007}. In the notation of Ref. \cite{Toffoli2007}:

\begin{equation}
I_{l,m,\mu}^{p_{i}\mu_{i},p_{f}\mu_{f}}(\epsilon)=\langle\Psi_{i}^{p_{i},\mu_{i}}|\hat{d_{\mu}}|\Psi_{f}^{p_{f},\mu_{f}}\varphi_{klm}^{(-)}\rangle\label{eq:eps-I}
\end{equation}

\begin{equation}
T_{\mu_{0}}^{p_{i}\mu_{i},p_{f}\mu_{f}}(\theta_{\hat{k}},\phi_{\hat{k}},\theta_{\hat{n}},\phi_{\hat{n}})=\sum_{l,m,\mu}I_{l,m,\mu}^{p_{i}\mu_{i},p_{f}\mu_{f}}(\epsilon)Y_{lm}^{*}(\theta_{\hat{k}},\phi_{\hat{k}})D_{\mu,\mu_{0}}^{1}(R_{\hat{n}})\label{eq:eps-TMF}
\end{equation}

\begin{equation}
I_{\mu_{0}}(\theta_{\hat{k}},\phi_{\hat{k}},\theta_{\hat{n}},\phi_{\hat{n}})=\frac{4\pi^{2}\epsilon}{cg_{p_{i}}}\sum_{\mu_{i},\mu_{f}}|T_{\mu_{0}}^{p_{i}\mu_{i},p_{f}\mu_{f}}(\theta_{\hat{k}},\phi_{\hat{k}},\theta_{\hat{n}},\phi_{\hat{n}})|^{2}\label{eq:eps-MFPAD}
\end{equation}

In this formalism:
\begin{itemize}
\item $I_{l,m,\mu}^{p_{i}\mu_{i},p_{f}\mu_{f}}(\epsilon)$ is the radial part of the dipole matrix element (as per Eqns. \ref{eq:r-kllam}, \ref{eqn:I-zeta} herein), determined from the initial and final state electronic wavefunctions $\Psi_{i}^{p_{i},\mu_{i}}$and $\Psi_{f}^{p_{f},\mu_{f}}$,
and the radial part of the photoelectron wavefunction $\varphi_{klm}^{(-)}$ (cf. $\phi_{lm}(k)$ in Eqn. \ref{eq:r-kllam}, and where $^{(-)}$ denotes outgoing wave normalisation) and dipole operator $\hat{d_{\mu}}$. Here the wavefunctions are indexed by irreducible representation (i.e. symmetry) by the labels $p_{i}$ and $p_{f}$, with components $\mu_{i}$ and $\mu_{f}$ respectively; $l,m$ are angular momentum components, $\mu$ is the projection of the polarization into the MF. Each energy and irreducible representation corresponds to a calculation in ePolyScat.
\item $T_{\mu_{0}}^{p_{i}\mu_{i},p_{f}\mu_{f}}(\theta_{\hat{k}},\phi_{\hat{k}},\theta_{\hat{n}},\phi_{\hat{n}})$
is the full matrix element (expanded in polar coordinates) in the
MF, where $\hat{k}$ denotes the direction of the photoelectron $\mathbf{k}$-vector, and $\hat{n}$ the direction of the polarization vector $\mathbf{n}$ of the ionizing light. This is equivalent to the full photoelectron wavefunction denoted $\Psi_e(\mathbf{k})$ in Sect. \ref{sec:dynamics-intro}.
\item $Y_{lm}^{*}(\theta_{\hat{k}},\phi_{\hat{k}})$ is a spherical harmonic. Note the conjugate form here.
\item $D_{\mu,-\mu_{0}}^{1}(R_{\hat{n}})$ is a Wigner rotation matrix element, with a set of Euler angles $R_{\hat{n}}=(\phi_{\hat{n}},\theta_{\hat{n}},\chi_{\hat{n}})$, which rotates/projects the polarization into the MF .
\item $I_{\mu_{0}}(\theta_{\hat{k}},\phi_{\hat{k}},\theta_{\hat{n}},\phi_{\hat{n}})$ is the final (observable) MFPAD, for a polarization $\mu_{0}$ and summed over all symmetry components of the initial and final states, $\mu_{i}$ and $\mu_{f}$. Note that this sum can be expressed as an incoherent summation, since these components are (by definition) orthogonal.
\item $g_{p_{i}}$ is the degeneracy of the state $p_{i}$.
\end{itemize}

As noted previously, in general there are multiple equivalent definitions for these terms used by different authors; the ePolyScat definitions above apply to the test matrix elements used herein (e.g. Table \ref{tab:inputMatE}), which lists $I_{l,m,\mu}^{p_{i}\mu_{i},p_{f}\mu_{f}}(\epsilon)$ for continuum symmetry components $p_f=\sigma_u,\pi_u$. These matrix elements are normalised to the total cross-section (in Mb), and include symmetrization. 



For cases where symmetrization is not included in the matrix elements directly, it can be addressed via the use of symmetrized (or generalised) harmonics, which essentially provide correctly symmetrized expansions of spherical harmonics for a given irreducible representation, $\Gamma$. These can be defined by linear combinations of spherical harmonics (see Refs.\cite{Altmann1963a,Altmann1965,Chandra1987} for more):

\begin{equation}
X_{hl}^{\Gamma\mu*}(\theta,\phi)=\sum_{\lambda}b_{hl\lambda}^{\Gamma\mu}Y_{l,\lambda}(\theta,\phi)\label{eq:symm-harmonics}
\end{equation}

where: 

\begin{itemize}
\item $\Gamma$ is an irreducible representation, 
\item ($l$, $\lambda$) define the usual spherical harmonic indicies (rank, order)
\item $b_{hl\lambda}^{\Gamma\mu}$ are symmetrization coefficients, 
\item index $\mu$ allows for indexing of degenerate components,
\item $h$ indexes cases where multiple components are required with all other quantum numbers identical. 
\end{itemize}
    
The exact form of these coefficients will depend on the point-group of the system, see, e.g. Refs. \cite{Chandra1987,Reid1994}; for numerical implementation notes in PEMtk see Sect. \ref{sec:numerical-notes}.

\subsection{Channel functions expansion\label{sec:channel-funs-full}}

Following the tensor components detailed above, the full form of the channel functions can be written as:








\begin{equation}
\varUpsilon_{L,M}^{u,\zeta\zeta'}=(-1)^{M}(2P+1)^{\frac{1}{2}}E_{P-R}(\hat{e};\mu_{0})(-1)^{(\mu'-\mu_{0})}\Lambda_{R',R}(R_{\hat{n}};\mu,P,R,R')B_{L,-M}(l,l',m,m')
\label{eq:channel-func-defn-MF}
\end{equation}

\begin{equation}
\bar{\varUpsilon}_{L,M}^{u,\zeta\zeta'}=(-1)^{M}[P]^{\frac{1}{2}}E_{P-R}(\hat{e};\mu_{0})(-1)^{(\mu'-\mu_{0})}\bar{\Lambda}_{R'}(\mu,P,R')B_{L,S-R'}(l,l',m,m')\Delta_{L,M}(K,Q,S)A_{Q,S}^{K}(t)
\label{eq:channel-func-defn-AF}
\end{equation}

\subsection{Final state density matrix\label{sec:density-mat-full}}







As introduced in Sect. \ref{sec:density-mat-basic}, the (radial) density matrix can be expressed as the outer-product of the (radial) matrix elements. Following the channel function notation, it is also trivial to write the radial matrix elements in density matrix form in the $\zeta\zeta'$ representation:

\begin{equation}
\mathbf{\rho}^{\zeta\zeta'} = |\zeta\rangle\langle\zeta'| \equiv \mathbb{I}^{\zeta,\zeta'}
\end{equation}

And the full final continuum state as a density matrix in the $\zeta\zeta'$ representation (with the observable dimensions $L,M$ explicitly included in the density matrix), which will also be dependent on the choice of channel functions ($u$):

\begin{equation}
\mathbf{\rho}_{L,M}^{u,\zeta\zeta'}=\varUpsilon_{L,M}^{u,\zeta\zeta'}\mathbb{I}^{\zeta,\zeta'}
\end{equation}

Here the density matrix can be interpreted as the final, LF/AF or
MF density matrix (depending on the channel functions used), incorporating both the intrinsic and extrinsic
effects (i.e. all channel couplings and radial matrix elements for
the given measurement), with dimensions dependent on the unique sets of quantum numbers required - in the simplest case, this will just be a set of partial waves $\zeta = (l,m)$. Note that this final state is distinct from the ``radial" density matrix (Eqn. \ref{eqn:radial-density-mat}), which encodes purely intrinsic (molecular scattering) photoionization dynamics (thus characterises the scattering event). The $L,M$ notation indicates here that these dimensions should not be summed over, hence the tensor coupling into the $\beta_{L,M}^{u}$ parameters can also be written in terms of the density matrix:

\begin{equation}
\beta_{L,M}^{u}=\sum_{\zeta,\zeta'}\mathbf{\rho}_{L,M}^{u,\zeta\zeta'}
\end{equation}

In fact, this form arises naturally since the $\beta_{L,M}^{u}$ terms are the state multipoles (geometric tensors) defining the system, which can be thought of as a coupled basis equivalent of the density matrix representations (see, e.g., Ref. \cite{BlumDensityMat}, Chpt. 4.).

In a more traditional notation (cf. Eqn. \ref{eq:cstate}, see also Refs. \cite{gregory2021MolecularFramePhotoelectron} 
), the density operator can be expressed as:


\begin{equation}
\rho(t) =\sum_{LM}\sum_{KQS}A^{K}_{QS}(t)\sum_{\zeta\zeta^{\prime}}\varUpsilon_{L,M}^{u,\zeta\zeta'}|\zeta,\Psi_+\rangle\langle\zeta,\Psi_+|\mu_q\rho_i\mu_{q\prime}^{*}|\zeta^{\prime},\Psi_+\rangle\langle\zeta^{\prime},\Psi_+|
\end{equation}

with $\rho_i = |\Psi_i\rangle\langle\Psi_i|$. This is, effectively, equivalent to an expansion in the various tensor operators defined above, in a state-vector notation.


\subsection{Matrix inversion formalism\label{app:mat-inversion}}

As discussed in Sect. \ref{sec:matrix-inv-intro}, and following Gregory et. al. \cite{gregory2021MolecularFramePhotoelectron}, the general formalism can also be rewritten with the ADMs separate (see also Reid \& Underwood convolution form, \cite{Reid2000}), and the LF/AF given as per Eqn. \ref{eqn:beta-convolution-C} (cf. Eqn. \ref{eq:BLM-tensor-AF}). Herein, the notation from Gregory et. al. \cite{gregory2021MolecularFramePhotoelectron} is rewritten slightly, as per Eqns. \ref{eq:basic} - \ref{eq:MPinversion}.

In the current notation, the full expressions are written as per Eqn. \ref{eqn:beta-convolution-C} for the LF/AF (cf. Eqn. \ref{eq:BLM-tensor-AF}):

\begin{equation}
\bar{\beta}_{L,M}(E,t)=\sum_{K,Q,S}C_{K,Q,S}^{L,M}(E)A_{Q,S}^{K}(t)
\end{equation}


And for the MF (cf. Eqn. \ref{eq:BLM-tensor-MF}):

\begin{equation}
\beta_{L,M}(E,\Omega)=\sum_{P,R,\Delta q}C_{P,R}^{L,M}(E,\Delta q)D_{R,\Delta q}^{P}(\Omega)
\end{equation}

And the required coefficients defined as:

\begin{equation}
\bar{C}_{KQS}^{LM}(\epsilon)=\sum_{\zeta\zeta'}\mathbb{I}_{\zeta\zeta'}^{\Gamma,\Gamma'}(\epsilon)\Gamma_{KQS}^{\zeta\zeta'LM}
\end{equation}

\begin{eqnarray}
\Gamma_{KQS}^{\zeta\zeta'LM} & = & (-1)^{M}[P]^{\frac{1}{2}}E_{P-R}(\hat{e};\mu_{0})(-1)^{(\mu'-\mu_{0})}\bar{\Lambda}_{R'}(\mu,P,R')B_{L,S-R'}(l,l',m,m')\Delta_{L,M}(K,Q,S)\\
 & = & \bar{\varUpsilon}_{L,M}^{u,\zeta\zeta'}/A_{Q,S}^{K}(t)
\end{eqnarray}

\begin{equation}
C_{PR}^{LM}(\epsilon,\Delta q)=\sum_{\zeta\zeta'}\mathbb{I}_{\zeta\zeta'}^{\Gamma,\Gamma'}(\epsilon)\Gamma_{PR\Delta q}^{\zeta\zeta'LM}
\end{equation}

\begin{equation}
\Gamma_{PR\Delta q}^{\zeta\zeta'LM}=(-1)^{M}(2P+1)^{\frac{1}{2}}E_{P-R}(\hat{e};\mu_{0})(-1)^{(\mu'-\mu_{0})}\bar{\Lambda}_{\Delta q}(\mu,P,\Delta q)B_{L,-M}(l,l',m,m')
\end{equation}

Where it is assumed that $R'=\Delta q$, and that the rotational matrix element $D_{R,\Delta q}^{P}(\Omega)$
is computed independently (note that the current numerical function $\Lambda_{R',R}(R_{\hat{n}};\mu,P,R,R')$
could possibly be used directly here, but this version keeps the angle-dependence
separate as per the matrix inversion formalism).

Note also that $Q=0$ only for the derivations in Gregory et. al. \cite{gregory2021MolecularFramePhotoelectron}, and the the inverse matrix as given in Eqn. \ref{eq:MPinversion}, although the results should generalise - see discussion in Ref. \cite{gregory2021MolecularFramePhotoelectron} for details, particularly Sect. 3.2 and the appendices.
\subsection{Numerical implementation in ePSproc and PEMtk\label{sec:numerical-notes}}

\subsubsection{Photoionization calculations with ePSproc}

The ePSproc codebase \cite{ePSprocAuthorea, ePSprocGithub, ePSprocDocs} aims to provide methods for post-processing with ePolyScat matrix elements (or equivalent matrix elements from other sources), including computation of AF and MF observables. The numerical implementation thus follows the conventions of ePolyScat, as given in Sect. \ref{sec:mat-ele-conventions}. Additionally, various switches can also be set to define alternative choices, e.g. conjugate forms, use of real harmonics etc., for computation of observables. Since the code is open-source Python, users may also swap libraries/conventions to their preference. By default the following conventions/libraries are used:

\begin{itemize}
\item Angular momentum functions (Wigner D and 3js) are currently implemented directly, or via the Spherical Functions library \cite{boyle2022SphericalFunctions}, and have been tested for consistency with the definitions in Zare (for details see \href{https://epsproc.readthedocs.io/en/latest/tests/Spherical_function_testing_Aug_2019.html}{the ePSproc docs} \cite{ePSprocDocs}).\cite{ePSprocDocs}).\cite{ePSprocDocs}).
\item Spherical harmonics are defined with the usual physics conventions: orthonormalised, and including the Condon-Shortley phase. Numerically they are implemented directly or via SciPy's \verb+sph_harm+ function (see \href{https://docs.scipy.org/doc/scipy/reference/generated/scipy.special.sph_harm.html}{the SciPy docs for details} \cite{SciPyDocumentation}. Further manipulation and conversion between different normalisations can be readily implemented with the SHtools library \cite{wieczorek2018SHToolsToolsWorking,SHtoolsGithub}.
\item General tensor handling and manipulation makes use of the Xarray library \cite{hoyer2017XarrayNDLabeled,XarrayDocumentation}.
\end{itemize}

\subsubsection{Data handling and fitting with PEMtk}

The Photoelectron Metrology Toolkit (PEMtk) codebase \cite{hockett2021PEMtkDocs, hockett2021PEMtkGithub} aims to provide various general data handling routines for photoionization problems. At the time of writing, only fitting routines are implemented, along with some basic utility functions, and backend functionality from ePSproc. Further details can be found in the \href{https://pemtk.readthedocs.io/en/latest/about.html}{PEMtk documentation} \cite{hockett2021PEMtkDocs}.

The results presented in Sect. \ref{sec:bootstrapping} make use of PEMtk routines, including functions provided to wrap matrix elements and ePSproc observable calculations for fitting, and analysis routines for identifying candidate matrix elements. The full analysis notebooks are available in the \href{http://dx.doi.org/10.6084/m9.figshare.20293782}{Figshare repository for this article} \cite{hockett2022MFreconFigshare}.

Non-linear optimization (fitting) is handled via the \href{https://lmfit.github.io/lmfit-py/index.html}{lmfit library}, which implements and/or wraps a range of non-linear fitting routines in Python \cite{LMFITDocumentation, newville2014LMFITNonLinearLeastSquare}; for the Levenberg-Marquardt least-squares minimization method used herein this wraps 
\href{https://docs.scipy.org/doc/scipy/reference/generated/scipy.optimize.least_squares.html}{Scipy's \texttt{least\_squares} functionality}, which therefore constituted the core minimization routine \cite{SciPyDocumentation} for the demonstration case.

Although not demonstrated herein, computation of $X_{hl}^{\Gamma\mu*}(\theta,\phi)$ (Eqn. \ref{eq:symm-harmonics}) is also currently implemented in the PEMtk codebase, making use of libmsym \cite{johansson2017AutomaticProcedureGeneratinga, johansson2022LibmsymGithub} (symmetry coefficients) and SHtools \cite{wieczorek2018SHToolsToolsWorking,SHtoolsGithub} (general spherical harmonic handling and conversion). For worked examples, see \href{https://pemtk.readthedocs.io/en/latest/sym/pemtk_symHarm_demo_160322_tidy.html}{the PEMtk docs} \cite{hockett2021PEMtkDocs}. It is hoped that this will be a useful tool for tackling photoionization problems more generally, without \textit{a priori} knowledge of the matrix elements for a given system.


\printbibliography[heading=bibintoc]

\end{document}